\documentclass[twoside,12pt]{article}
\usepackage{graphicx}
\usepackage{color}
\usepackage{cite}
\def\Journal#1#2#3#4{{#1} {#2} (#4) #3 }

\def\NPA{{\em Nucl. Phys.} A}

\def\NPB{{\em Nucl. Phys.} B}

\def\PLB{{\em Phys. Lett.} B}

\def\PRL{\em Phys. Rev. Lett.}
\def\PREV{\em Phys. Rev.}
\def\PREP{\em Phys. Rep.}
\def\PRA{{\em Phys. Rev.} A}
\def\PRD{{\em Phys. Rev.} D}
\def\PRC{{\em Phys. Rev.} C}

\def\ZPC{{\em Z. Phys.} C}

\def\ANNP{\em Ann. Phys. (N.Y.)}
\def\RMP{{\em Rev. Mod. Phys.}}

\newcommand{\dissum}[2]{\displaystyle \sum_{#1}^{#2}}
\newcommand{\abs}[1]{\left| #1\right|}
\newcommand{\bsm}[1]{{\boldmath$#1$}}
\newcommand{\be}{\begin{equation}}
\newcommand{\ee}{\end{equation}}
\newcommand{\bea}{\begin{eqnarray}}
\newcommand{\eea}{\end{eqnarray}}

\newcommand{\rsub}[1]{\mbox{\scriptsize #1}}
\newcommand{\ssz}{${}^{1\!}S_0$}
\newcommand{\tso}{${}^{3\!}S_1$}
\newcommand{\tdo}{${}^{3\!}D_1$}
\newcommand{\tpz}{${}^{3\!}P_0$}
\newcommand{\tpo}{${}^{3\!}P_1$}
\newcommand{\tpt}{${}^{3\!}P_2$}
\newcommand{\sdt}{${}^{1\!}D_2$}
\newcommand{\tdt}{${}^{3\!}D_2$}
\newcommand{\tft}{${}^{3\!}F_2$}
\newcommand{\otpo}{$1\,{}^{3\!}P_1$}
\newcommand{\ttpo}{$2\,{}^{3\!}P_1$}
\newcommand{\spo}{${}^{1\!}P_1$}
\newcommand{\ospo}{$1\,{}^{1\!}P_1$}
\newcommand{\tspo}{$2\,{}^{1\!}P_1$}
\newcommand{\otpt}{$1\,{}^{3\!}P_2$}
\newcommand{\ttpt}{$2\,{}^{3\!}P_2$}
\newcommand{\otpz}{$1\,{}^{3\!}P_0$}
\newcommand{\ttpz}{$2\,{}^{3\!}P_0$}

\newcommand{\otso}{$1\,{}^{3\!}S_1$}
\newcommand{\tssz}{$2\,{}^{1\!}S_0$}
\newcommand{\hssz}{$3\,{}^{1\!}S_0$}
\newcommand{\ttso}{$2\,{}^{3\!}S_1$}
\newcommand{\htso}{$3\,{}^{3\!}S_1$}
\newcommand{\ftso}{$4\,{}^{3\!}S_1$}
\newcommand{\otdo}{$1\,{}^{3\!}D_1$}
\newcommand{\ttdo}{$2\,{}^{3\!}D_1$}
\newcommand{\osdt}{$1\,{}^{1\!}D_2$}
\newcommand{\tsdt}{$2\,{}^{1\!}D_2$}
\newcommand{\otdt}{$1\,{}^{3\!}D_2$}
\newcommand{\ttdt}{$2\,{}^{3\!}D_2$}
\newcommand{\otft}{$1\,{}^{3\!}F_2$}
\newcommand{\kbw}{k_{\mbox{\scriptsize BW}}}
\newcommand{\kp}{k_{\mbox{\scriptsize pole}}}

\newcommand{\ep}{E_{\mbox{\scriptsize pole}}}
\newcommand{\mbw}{M_{\mbox{\scriptsize BW}}}
\newcommand{\mpo}{M_{\mbox{\scriptsize pole}}}
\newcommand{\gbw}{\Gamma_{\mbox{\scriptsize BW}}}
\newcommand{\gp}{\Gamma_{\!\mbox{\scriptsize pole}}}
\newcommand{\Imag}[1]{\Im {\it m}(#1 )}
\newcommand{\One}{1\!\!1}
\newcommand{\rse}{\mathcal{R}}
\newcommand{\bes}[1]{j^{#1}_{L_{#1}}}
\newcommand{\han}[1]{h^{(1)#1}_{L_{#1}}}

\newcommand{\tmat}[2]{T_{#1#2}^{(L_{#1},L_{#2})}}

\begin{document}
\title{ \vspace{1cm}
Modern meson spectroscopy: \\ the fundamental role of unitarity
}
\author{
E.~van Beveren$^1$ and G.~Rupp$^2$ \\ \\
$^1$Centro de F\'{\i}sica da UC,  Departamento de F\'{\i}sica \\
Universidade de Coimbra, P-3004-516 Coimbra, Portugal \\[1mm]
$^2$Centro de F\'{\i}sica e Engenharia de Materiais Avan\c{c}ados,
Instituto Superior T\'{e}cnico \\
Universidade de Lisboa, P-1049-001 Lisbon, Portugal
}
\maketitle
\begin{abstract}
The importance of $S$-matrix unitarity in realistic meson spectroscopy
is reviewed, both its historical development and more recent
applications. First the effects of imposing $S$-matrix unitarity on
meson resonances are demonstrated in both the elastic and the inelastic
case. Then, the static quark model is revisited and its theoretical as
well as phenomenological shortcomings are highlighted. A detailed
account is presented of the mesons in the tables of the Particle Data
Group that cannot be explained at all or only poorly in models
describing mesons as pure quark-antiquark bound states. Next the
earliest unitarised and coupled-channel models are revisited, followed
by several examples of puzzling meson resonances and their
understanding in a modern unitarised framework. Also, recent and fully
unquenched lattice descriptions of such mesons are summarised. Finally,
attention is paid to production processes, which require an unconventional
yet related unitary approach. Proposals for further improvement are
discussed.
\end{abstract}
\section{Introduction}
\label{intro}
Knowledge of low-energy QCD is encoded in the observable
properties of hadrons, that is, mesons and baryons. Most importantly,
hadronic mass spectra should provide detailed information on the forces
that keep the quarks and/or antiquarks in such systems permanently confined,
inhibiting their observation as free particles. However, since QCD is
not tractable through perturbative calculations at low energies, owing to
a large running coupling in that regime, quark confinement is usually
dealt with employing a confining potential in the context of some
phenomenological quark model. The shape of this potential is largely
empirical, though its short-distance behaviour can be reasonably determined
from one-gluon exchange, resulting in a Coulomb-like interaction, usually
endowed with an $r$-dependent coupling constant in order to simulate
asymptotic freedom. At large distances, the potential is mostly supposed
to grow linearly, on the basis of flux-tube considerations, which have been
observed \cite{bali95} in lattice simulations of string formation for static
quarks. The most cited quark model of mesons with such a Coulomb-plus-linear
confining potential, sometimes also called ``funnel potential'', is due to
Godfrey \& Isgur (GI) \cite{GI85}, which also accounts for
kinematically relativistic effects. The enormous popularity of the model is
understandable, in view of its exhaustive description of practically all
imaginable $q\bar{q}$ systems, including those with one or two top quarks,
which had not yet even been discovered then. Experimentalists as well as
model builders often invoke GI predictions as a touchstone for their
observations or results. However, the GI model does not reproduce
the excitation spectra of mesons made of light quarks, as already analysed
in Ref.~\cite{rupp12}. Its principal shortcoming is the prediction of much
too large radial splittings for mesons in the range of roughly 1--2 GeV,
resulting in several experimentally observed states that do not fit in the
GI level scheme. Also the lowest-lying scalar mesons, below 1 GeV, are not
at all reproduced in the GI model. In Sec.~\ref{static} we shall come
back to this particular case in much more detail.

Logically, there can be two reasons for the problems of the GI
model in the mentioned energy region, namely possible deficiencies of the
employed confining potential and/or certain approximations inherent in the
model.
Let us first consider the employed funnel-type confining potential
\be
V_{\rsub{conf}} \; = \; -\frac{\alpha_s(r)}{r} + \lambda r \; ,
\label{funnel}
\ee
where the constant parameter $\lambda$ is the so-called string tension
and $\alpha_s(r)$ a configuration-space parametrisation of the running
strong coupling. Ignoring for the moment the $r$-dependence of $\alpha_s$,
we see that $V_{\rsub{conf}}$ is independent of (quark) mass and therefore
also flavour-independent, in accordance with the QCD Lagrangian.
Consequently, the mass spectrum of a Schr\"{o}dinger or related
relativistic equation with such a potential will inexorably be mass-dependent,
that is, radial splittings will increase according as the quark mass
decreases. This had already been realised by the Cornell group when developing
their potential model of charmonium \cite{eichten78}, in which the strength
$\kappa$ of the Coulombic part was fitted to the few then known charmonium
 observables (quote):
\begin{quote}  \em
``The recent discovery of the $\mu^+\mu^-$ enhancement $\Upsilon$ 
probably implies the existence of another $Q\bar{Q}$ 
family, where $Q$ is a quark carrying a new flavor
and having a mass of 4--5 GeV. The variation of
the spectrum with quark mass $m_Q$ is very sensitive
to the form of $V(r)$, and present indications are
that our ansatz (1.1) may not pass this test.'' \em
\end{quote}
Two years later, with more data available, the same authors \cite{eichten80a}
adjusted their model parameters so as to try to accommodate charmonium as well
as bottomonium states, both in the static and the coupled-channel version of
the model, though resulting in a clearly too heavy \htso\ state in the latter
case. Also, one of the authors \cite{eichten80b} used a slightly 
smaller value of $\kappa$ when applying the coupled-channel model only to
bottomonium states, by indeed arguing on the basis of a running coupling in the
Coulombic part. We shall come back to unitarisation effects in the Cornell
model in Sec.~\ref{coupled}. An empirical way to account for a running
strong coupling $\alpha_s(r)$ in the context of the static quark model 
simultaneously applied to charmonium and bottomonium states was 
presented in Ref.~\cite{richardson79}.

Now, the mechanism that allows to reproduce the approximate equal radial
spacings in charmonium and bottomonium for the funnel potential, despite the
quark-mass independence of its confining part, is a very delicate balance
between the Coulombic behaviour and running coupling at short distances, as
well as the linear rising at larger distances. However, this is impossible
to sustain for light quarkonia, resulting in considerably increased
spacings in the energy region of 1--2 GeV, in conflict with experiment.
A model proposed by us almost four decades ago \cite{nijmegen80,nijmegen83}
does reproduce the observed radial spacings for light, charm, and bottom
quarks, being based on a flavour-dependent harmonic-oscillator confining
potential, while also accounting for non-perturbative coupled-channel effects
in a manifestly unitary $S$-matrix formalism. In Sec.~\ref{coupled} the model
will be described in detail.

The other possible reason why the GI and related quark models often fail in
light-meson spectroscopy is the usual neglect of unitarisation effects. By
this we mean that, as most mesons are resonances and not stable or quasi-stable
$q\bar{q}$ states, they must strictly speaking be described as poles in some
unitary $S$-matrix, and not as bound states in a static potential.
This becomes all the more true if one realises that many mesonic resonances
are broad or very broad, some of which having widths of the same order of
magnitude as the observed radial spacings. Picking just one typical example
from the PDG \cite{PDG2020} Meson Summary Tables, we see that the mass
difference between the ground-state $s\bar{s}$ tensor meson $f_2^\prime(1525)$
and its first radial excitation $f_2(1950)$ is about 420 MeV, while the full
width of the latter resonance is $(464\pm24)$~MeV \cite{PDG2020}. So it is
clear that a reliable determination of radial splittings that originate
exclusively in the underlying confining potential demands to account for
unitarisation effects, which inevitably will give rise to both the observed
decay widths and real mass shifts hidden in the spectrum. Note that the size
of such a mass shift may very well depend on the specific radial (or angular)
quantum number and/or the vicinity of some decay threshold. This will be
comprehensively discussed in Sec.~\ref{coupled}.

Before outlining the organisation of the present review, we should stress
that this is not intended to be an exhaustive overview of research on meson
spectroscopy. We rather aim at showing where the traditional static quark
model fails to reproduce the observed mass spectra and how unitarisation
can explain several discrepancies. For a not very recent yet highly
insightful review of mesons from an experimentalist's point of view, we
recommend Ref.~\cite{bugg04}. A detailed review of truly exotic charmonium-
and bottomonium-like candidate states, not treated here, can be found in
Ref.~\cite{xiang16}.
For reviews on glueballs and baryons, we recommend
Refs.~\cite{crede09,cotanch03} and \cite{eichmann16}, respectively.

Coming back to our work, in Sec.~\ref{unitarity} a simple unitary ansatz for
the $S$-matrix in the elastic case is shown to lead to significant deviations
from common Breit-Wigner (BW) parametrisations, when applied to some light
resonances. Section~\ref{static} revisits the static quark model, in which the
dynamical effects of strong decay are neglected. Several serious weaknesses
will be highlighted. In Sec.~\ref{coupled} we briefly review a number of
unitarised or coupled-channel quark models of mesons with their respective
predictions for mass shifts, discussing some of the differences. This
includes the original model developed by the present authors, in collaboration
with others at the University of Nijmegen, as well as its more recent
formulation in momentum space. The latter ``Resonance Spectrum Expansion'' 
(RSE) formalism is presented in Sec.~\ref{rse}, both in its simplest form and
the fully unitary multichannel version, with several applications to 
controversial meson resonances. The charmed axial-vector mesons and the
puzzling charmonium-like state $\chi_{c1}(3872)$ are described in detail
in Sec.~\ref{enigmatic} as typical examples of employing momentum-space and
coordinate-space formulations of the RSE model. Section~\ref{lattice} presents
selected results of fully unquenched lattice-QCD computations of some mesons
that are often modelled as tetraquarks. These show that standard
quark-antiquark configurations are adequate provided that two-meson
interpolators are included in the simulations as well. In
Sec.~\ref{production} the issue of resonances produced in production processes
instead of elastic scattering is discussed. An unconventional production
formalism strongly related to the RSE model is shown to lead to non-resonant
threshold enhancements that may be mistaken for genuine resonances. Finally,
Sec.~\ref{conclusions} summarises the main results and outlines avenues of
possible future research.

\section{Unitarity and Breit-Wigner amplitudes}
\label{unitarity}
The most fundamental cornerstone of the PDG tables is the uniqueness of
$S$-matrix pole positions of unstable particles, as a consequence of
quantum-field-theory principles. Therefore, the unitarity property of the
$S$-matrix should ideally be respected in whatever description of mesonic
resonances in experiment, quark models, and lattice-QCD simulations. In this
section we use a minimal yet manifestly unitary $S$-matrix ansatz to compare
pole and BW masses in the case of three broad to very broad mesons. Simplistic
BW parametrisations continue to be widely used in data analyses of mesonic
processes, in spite of often not satisfying unitarity, as e.g.\ in the isobar
model for overlapping resonances. Moreover, even for an isolated resonance and
so unitary BW parametrisation of the amplitude, the BW mass will be different
from the real part of the complex-energy pole in an $S$-matrix parametrisation,
as demonstrated below. Thus, as an illustration (also see
Refs.~\cite{kaminski19,rupp20}), we study the resulting
discrepancies for the meson resonances $\rho(770)$, $f_0(500)$ (alias
$\sigma$), and $K_0^\star(700)$ (alias $\kappa$). Let us consider the simplest
possible parametrisation of an elastic resonance that respects $S$-matrix
unitarity. The regularity of the independent wave-function solutions at the
origin and infinity, respectively, leads to a $1\times1$ partial-wave
$S$-matrix as a function of the on-shell relative momentum $k$ expressed as
\cite{taylor}
\begin{equation}
S_l(k) \; = \; \frac{J_l(-k)}{J_l(k)} \; ,
\label{smatrix}
\end{equation}
where $J_l(k)$ is the so-called Jost function. A resonance then corresponds
to a simple pole in $S_l(k)$ for complex $k$ with positive real part
and negative imaginary part, that is, a pole lying in the fourth
quadrant of the complex-$k$ plane. So the simplest ansatz for the $S$-matrix
and thus for the Jost function is to write
\begin{equation}
J_l(k) \; = \; k - \kp \; = \; k-(c-id) \; , \;\;\mbox{with}\;\;
c>0 \; , \;d>0 \; .
\label{josts}
\end{equation}
Note that Eq.~(\ref{smatrix}) requires $S_l(k)$ to have a zero in the
second quadrant, viz.\ for $k=-c+id$.
However, then the $S$-matrix cannot be unitary, for real $k$, i.e.,
\begin{equation}
S_l^\star(k) \; \neq \; S_l^{-1}(k) \;\;\;\Leftrightarrow\;\;\;
J_l^\star(k)\,J_l(k) \; \neq \; J_l^\star(-k)\,J_l(-k) \; .
\label{notunitary}
\end{equation}
Thus, $S_l(k)$ will only satisfy unitarity if \cite{taylor}, for real $k$,
\begin{equation}
J_l^\star(k) \; = \; J_l(-k) \; .
\label{unitary}
\end{equation}
Consequently, instead of Eq.~(\ref{josts}), the Jost function should read
\begin{equation}
J_l(k) \; = \; (k-\kp)(k+\kp^\star)  \; = \; (k-c+id)(k+c+id) \; .
\label{jostu}
\end{equation}
So $S_l(k)$ has a symmetric pair of poles in the 3rd and 4th quadrants,
corresponding to an equally  symmetric pair of zeros in the 1st and 2nd
quadrants. Note that in the corresponding complex-energy plane, given by
\begin{equation}
E \; = \; 2\sqrt{k^2+m^2}
\label{energy}
\end{equation}
in the case of two equal-mass particles, this amounts to one pole and one
zero lying symmetrically in the 4th and 1st quadrants, respectively. Since
a $1\times1$ $S$-matrix can generally be written as
\begin{equation}
S_l(k)\;=\;\exp2i\delta_l(k)\;=\;\frac{1+i\tan\delta_l(k)}{1-i\tan\delta_l(k)}\;,
\label{phase}
\end{equation}
and from Eqs.~(\ref{smatrix},\ref{jostu}) also as
\begin{equation}
S_l(k) \; = \; \frac{(-k-\kp)(-k+\kp^\star)}{(k-\kp)(k+\kp^\star)} \; = \;
\frac{\displaystyle 1+\frac{2ik\,\mbox{Im}(\kp)}{k^2-|\kp|^2}}
     {\displaystyle 1-\frac{2ik\,\mbox{Im}(\kp)}{k^2-|\kp|^2}} \; ,
\label{sjost}
\end{equation}
we get, using $\kp=c-id$ from Eq.~(\ref{josts}),
\begin{equation}
\tan\delta_l(k) \; = \; \frac{2k\,\mbox{Im}(\kp)}{k^2-|\kp|^2} \; = \;
\frac{2dk}{c^2+d^2-k^2} \; .
\label{tandelta}
\end{equation}
With the partial-wave amplitude given by
$T_l(k)=e^{i\delta_l(k)}\sin\delta_l(k)$, we thus obtain 
\begin{equation}
|T_l(k)|^2 \; = \; \sin^2\delta_l(k) \; = \;
\frac{\tan^2\delta_l(k)}{1+\tan^2\delta_l(k)} \; = \;
\frac{4d^2k^2}{(k^2-c^2+d^2)^2+4c^2d^2} \; .
\label{tsquared}
\end{equation}
So we see that the modulus of the amplitude is maximum, i.e., equal to 1, for
\begin{equation}
k^2_{\mbox{\scriptsize max}} \; = \; c^2+d^2 \; ,
\label{kmax}
\end{equation}
where the phase shift passes through $90^\circ$. Note that the corresponding
energy
\begin{equation}
E_{\mbox{\scriptsize max}} \; = \; 2\sqrt{k^2_{\mbox{\scriptsize max}}+m^2}
\; = \; 2\sqrt{c^2+d^2+m^2} 
\label{emax}
\end{equation}
usually corresponds to the mass $\mbw$ in a typical Breit-Wigner (BW)
amplitude
\begin{equation}
T_l(E) \; \propto \; \frac{1}{E-\mbw+i\gbw/2} \; ,
\label{tbw}
\end{equation}
at least in the elastic case. So we take the (unitary) BW mass as
\begin{equation}
\mbw\;=\;2\sqrt{k^2_{\mbox{\scriptsize max}}+m^2}\;=\;2\sqrt{c^2+d^2+m^2} \; .
\label{ebw}
\end{equation}
Next we compare this BW mass with the pole mass in the complex-energy plane for
several cases.

\subsection{\boldmath$\rho(770)$}
\label{rho770}
Now we illustrate the consequences of these considerations in the
simple case of the very well-known meson $\rho(770)$ \cite{PDG2020}, which
is an elastic $P$-wave resonance in $\pi\pi$ scattering. The PDG 
lists its mass and total width as \cite{PDG2020}
\begin{equation}
M_{\rho^0}=(775.26\pm0.25) \; \mbox{MeV}\; , \;\;
\Gamma_{\!\!\rho^0}=(147.8\pm0.9) \; \mbox{MeV} \; ,
\label{rho}
\end{equation}
where the width follows almost exclusively ($\approx100$\%) from the decay mode
$\rho^0\to\pi^+\pi^-$, with $m_{\pi^\pm}=139.57$~MeV. The PDG makes the following
remark ahead of the listed $\rho^0$ masses, from which the average value given in
Eq.~(\ref{rho}) is determined:
\begin{quote} \em
``We no longer list $S$-wave Breit-Wigner fits, or data with high combinatorial
  background.''  \em
\end{quote}
However, for several of these analyses it remains unclear whether unitarity is 
respected.

In the following and as mentioned above, we shall refer to BW mass ($\mbw$)
for the energy where the resonance's phase shift passes through $90^\circ$ and
so the modulus of the amplitude is maximum
(cf.\ Eqs.~(\ref{tsquared},\ref{emax})). This indeed also holds for the standard
BW amplitude in Eq.~(\ref{tbw}).
Here, we want to determine for the $\rho(770)$ the discrepancy between
pole mass and BW mass 
(cf.\ Eq.~(\ref{ebw}))
\begin{equation}
\mbw \; = \; 2\sqrt{c^2+d^2+m^2} \; .
\label{mbw}
\end{equation}
In order to find the corresponding pole mass, we write
\begin{equation}
\ep \; = \; 2\sqrt{\kp^2+m^2}  =  2\sqrt{c^2-d^2+m^2-2icd}  = 
\mpo-i\gp/2 \; .
\label{mpole}
\end{equation}
Thus,
\begin{equation}
\mbox{Re}(\ep^2) \; = \; 4(c^2-d^2+m^2) \; = \; \mpo^2-\gp^2/4 
\label{reps}
\end{equation}
and
\begin{equation}
\mbox{Im}(\ep^2) \; = \; -8cd \; = \; -\mpo \gp \;.
\label{ieps}
\end{equation}
So combining Eqs.~(\ref{mbw}) and (\ref{reps}) we get
\begin{equation}
\mbw^2+\mbox{Re}(\ep^2) \; = \; 8(c^2+m^2) \; = \mbw^2+\mpo^2-\gp^2/4
\label{mbwspreps}
\end{equation}
and
\begin{equation}
\mbw^2-\mbox{Re}(\ep^2) \; = \; 8d^2 \; = \mbw^2-\mpo^2+\gp^2/4 \; .
\label{mbwsmreps}
\end{equation}
From Eqs.~(\ref{mbwspreps}) and (\ref{mbwsmreps}), $c$ and $d$ follow
straightforwardly, viz.\
\begin{equation}
c \; = \; \sqrt{\mbw^2+\mpo^2-\gp^2/4-8m^2}/\sqrt{8}
\label{c}
\end{equation}
and
\begin{equation}
d \; = \; \sqrt{\mbw^2-\mpo^2+\gp^2/4}/\sqrt{8} \; ,
\label{d}
\end{equation}
respectively.
Combining Eqs.~(\ref{c}) and (\ref{d}) with Eq.~(\ref{ieps}), we arrive,
after some basic algebra, at the final expression
\begin{equation}
\mpo \; = \; \sqrt{\sqrt{(\mbw^2-4m^2)^2-4m^2\,\gp^2}+4m^2-\gp^2/4} \; .
\label{mp}
\end{equation}
Note that it is not possible to write $\mpo$ as a simple closed-form expression
in terms of both $\mbw$ and $\gbw$. Nevertheless, assuming for the moment that
$\gp\simeq\gbw$, we substitute in Eq.~(\ref{mp}) the PDG values given in
Eq.~(\ref{rho}) for $\mbw$ and $\gp$, which yields
\begin{equation}
\mpo \; = \; 770.67 \; \mbox{MeV} \; .
\label{rhopolemass}
\end{equation}
Thus, we see that even in the case of the very well-known $\rho(770)$
resonance, whose mass is given with 5 significant digits in the PDG tables, 
the difference between the (unitary) BW mass and the pole mass is as large
as 4.5~MeV! Now we check whether indeed $\gp\simeq\gbw$, by calculating the
half-width of the $\rho(770)$ peak from the modulus-squared of the amplitude
$T_l(k)$. Using Eq.~(\ref{tsquared}) we have
\begin{equation}
\frac{1}{2} \; = \; |T_l(k_\pm)|^2 \; = \;
\frac{4d^2k_\pm^2}{(k_\pm^2-c^2+d^2)^2+4c^2d^2} \; ,
\label{tshalf}
\end{equation}
yielding
\begin{equation}
k^2_\pm \; = \; c^2+3d^2 \pm \sqrt{8d^4+4c^2d^2} \; .
\label{kpm}
\end{equation}
Taking now the expressions for $c$ and $d$ in Eqs.~(\ref{c}) and (\ref{d}),
respectively, we can evaluate $\gbw$ as
\begin{equation}
\gbw \; = \; E_+ - E_- \; = \; 2\sqrt{k_+^2+m^2}-2\sqrt{k_-^2+m^2} \; ,
\label{gpm}
\end{equation}
with $\mpo=770.67$~MeV and $\gp=147.8$~MeV, resulting in
\begin{equation}
\gbw \; = \; 147.83 \; \mbox{MeV} \; .
\label{gbw}
\end{equation}
So indeed, in this case $\gbw$ is an excellent approximation for $\gp$,
to be contrasted with the significant difference between $\mbw$ and $\mpo$.

At this point we should return to the peak amplitude given by
Eqs.~(\ref{tsquared}) and (\ref{kmax}). As a matter of fact, the maximum
in the partial-wave cross section \cite{taylor}
\begin{equation}
\sigma_l(k) \; \equiv \; \frac{4\pi}{k^2}(2l+1)\sin^2\delta_l(k) \; = \;
\frac{16\pi(2l+1)d^2}{(k^2-c^2+d^2)^2+4c^2d^2} 
\label{cross}
\end{equation}
does not occur at the same $k$ value as in $|T_l(k)|^2$,
but rather at $k^2=c^2-d^2$ instead of $k^2=c^2+d^2$. Accordingly,
the cross-section half-width follows from
\begin{equation}
k^2_\pm \; = \; c^2-d^2\pm2cd \;
\label{kpmcs}
\end{equation}
instead of Eq.~(\ref{kpm}). As a result, for the PDG $\rho(770)$
mass of 775.26~MeV and width of 147.8~MeV, the cross-section peak comes out
at 767.13~MeV, with a half-width of 149.19~MeV, which values are 8.13~MeV
lower and 1.39~MeV higher than the PDG ones, respectively. This further
reinforces the need for a consistent unitary approach.

A final remark is due, as $\rho(770)$ is an elastic resonance and
so suitable for our simple study here. However, its first radial 
excitation, listed as $\rho(1450)$ \cite{PDG2020} in the PDG tables,
is highly inelastic and also considerably broader. So one naturally
expects potentially larger effects due to unitarity. Yet a discrepancy
of 170~MeV between a unitary and a non-unitary analysis of the very
same resonance, as found in Ref.~\cite{PRD96p113004}, may still come as a
shock to some, because it could wreak havoc with mainstream meson spectroscopy,
in particular the $\rho(1450)$ PDG assignment. As a matter of fact, a very
recent reanalysis \cite{hammoud20} of $P$-wave $\pi\pi$ phase shifts
and inelasticities employing a manifestly unitary multichannel $S$-matrix
parametrisation of resonance poles, with crossing-symmetry constraints, shows
strong evidence of a $\rho(1250)$ instead of $\rho(1450)$ as the first radial
excitation of $\rho(770)$. We shall come back to this controversy in
Sec.~\ref{static}. Let us just mention here the inevitability of resorting to
a coupled-channel $S$-matrix description, if one insists not only on a fully
unitary description of the data, but also on proper analyticity.

\subsection{\boldmath$f_0(500)$ and \boldmath$K_0^\star(700)$}
\label{sigmakappa}
Next we focus our attention on two resonances that have much larger
widths than $\rho(770)$, namely the scalar mesons $f_0(500)$ (alias
$\sigma$) and $K_0^\star(700)$ (alias $\kappa$). These two states have been
very controversial over the years \cite{tornqvist96,beveren99a}, but are now
included as established resonances in the 2020 PDG Meson Summary Tables
\cite{PDG2020}. For extensive reviews of the $\sigma$ and $\kappa$ meson, see
Refs.~\cite{pelaez16} and \cite{pelaez20}, respectively.
In order to be able to deal with $K_0^\star(700)$, we must consider the
unequal-mass case, as its (only) hadronic decay mode is $K\pi$.
Of course, the unequal-mass formulae can also be applied to
$f_0(500)\to\pi\pi$, possibly with a negligibly small correction so
as to avoid numerical problems.

The relativistic relative momentum of two unequal-mass particles reads
\begin{equation}
k=\frac{E}{2}\sqrt{\left\{1-\frac{(m_1+m_2)^2}{E^2}\right\}
                   \left\{1-\frac{(m_1-m_2)^2}{E^2}\right\}} \; ,
\label{krelat}
\end{equation}
with the total energy
\begin{equation}
E = \sqrt{k^2+m_1^2} + \sqrt{k^2+m_2^2} \; .
\label{erelat}
\end{equation}
Now it is not feasible anymore to derive a simple closed-form expression
for the pole mass as in Eq.~(\ref{mp}). Instead, we shall derive $\mbw$
and $\gbw$ for a given $\mpo$ and $\gp$, though not in an explicit form.
Thus, we write
\begin{equation} 
c \; = \; \mbox{Re}(\kp) \; = \; \mbox{Re}\left[\frac{\ep}{2}
\sqrt{\left\{1-\frac{(m_1+m_2)^2}{\ep^2}\right\}
\left\{1-\frac{(m_1-m_2)^2}{\ep^2}\right\}}\:\right] 
\label{creal}
\end{equation}
and similarly for the corresponding imaginary part
\begin{equation} 
d \; = \; -\mbox{Im}(\kp) \; .
\label{dimag}
\end{equation}
Using 
\begin{equation} 
\ep \; = \mpo - i\gp/2 
\label{emgpole}
\end{equation}
and
\begin{equation} 
\mbw \; = \; \sqrt{\kbw^2+m_1^2}+\sqrt{\kbw^2+m_2^2} \; , \; \; 
\mbox{with} \;\; \kbw \; = \; \sqrt{c^2+d^2} \; ,
\label{mkbw}
\end{equation}
we can numerically compute $\mbw$ as a function of $\mpo$ and $\gp$
with Eqs.~(\ref{creal},\ref{dimag}). In the same way, $\gbw$ can be
calculated employing Eqs.~(\ref{kpm},\ref{gpm}).

Let us now check what this means for the very broad scalar mesons $f_0(500)$
and $K_0^\star(700)$. Their pole positions as well as BW masses and widths are
listed in the PDG Meson Tables as \cite{PDG2020}
\begin{equation}
\hspace*{-5mm}
f_0(500): \; \;
\left\{\begin{array}{l}
\ep\;=\;\left\{\mbox{(400--550)}-i\mbox{(200--350)}\right\} \mbox{MeV}\;,\\[1mm]
\mbw \; = \; (475 \pm 75) \; \mbox{MeV} \;,\;\;
\gbw \; = \; (550 \pm 150) \; \mbox{MeV}
\end{array}\right.
\label{sigmapdg}
\end{equation}
and
\begin{equation}
K_0^\star(700): \;\;
\left\{\begin{array}{l}
\ep\;=\;\left\{\mbox{(630--730)}-i\mbox{(260--340)}\right\} \mbox{MeV}\;,\\[1mm]
\mbw \; = \; (824 \pm 30) \; \mbox{MeV} \;,\;\;
\gbw \; = \; (478 \pm 50) \; \mbox{MeV} \; .
\end{array}\right.
\label{kappapdg}
\end{equation}
If on the other hand we use our Eqs.~(\ref{kpm},\ref{gpm}) and
(\ref{krelat}--\ref{mkbw}), we obtain
\begin{equation}
\begin{array}{rl}
f_0(500): & \mbw \; = \; (591 \pm 70) \; \mbox{MeV} \; , \;\;
\gbw \; = \; (560 \pm 153) \; \mbox{MeV} \; ; \\[1mm]
K_0^\star(700): & \mbw \; = \; (906 \pm 37) \; \mbox{MeV} \; , \;\;
\gbw \; = \; (707 \pm 97) \; \mbox{MeV} \; .
\label{skunitary}
\end{array}
\end{equation}
The errors in Eq.~(\ref{skunitary}) are determined with the variance formula,
given the data spreading for the real and imaginary parts of the pole
positions shown in Eqs.~(\ref{sigmapdg},\ref{kappapdg}), instead of
simply \cite{rupp20} from those PDG limits.
The conclusion is that the PDG seems to underestimate the BW masses of both
$f_0(500)$ and $K_0^\star(700)$, as well as the BW width of
$K_0^\star(700)$. As a word of caution, we repeat that our label `BW' refers
to the energy where the phase shift passes through $90^\circ$, in the context
of the present simple pole model. Note that reality is more complicated,
as the $f_0(500)$ resonance overlaps with $f_0(980)$ and $K_0^\star(700)$ 
overlaps with $K_0^\star(1430)$, apart from the influence of Adler zeros on
the amplitudes \cite{bugg03}. 
We also recall our alternative definition of $\mbw$ and $\gbw$ based
on the cross-section peak and its half-width (cf.\ Eqs.~(\ref{cross}) and
(\ref{kpmcs})). When applied to the light scalars, we find an $f_0(500)$ BW
mass below the pole mass, while no BW width can be consistently computed.
In the $K_0^\star(700)$ case, not even a BW mass can be defined in part
of the PDG's range of pole values, while a BW width is again not possible
to calculate. This lends further support to the generally accepted non-BW
nature of the $f_0(500)$ and $K_0^\star(700)$ resonances, making the use of
BW ansatzes in their description highly questionable.
\section{Shortcomings of the Static Quark Model}
\label{static}
Traditional quark models treat hadrons as $q\bar{q}$ or $qqq$ bound states,
in spite of the fact that the vast majority of mesons and baryons are
resonances, many of which have large to very large decay widths. In some
cases these are of the same order of magnitude as average level spacings
(see e.g.\ the example of excited $f_2$ mesons above). The rationale behind
these models is essentially a poor man's approach, since one cannot
convincingly argue that decay processes giving rise to hadronic widths of
hundreds of MeV will not also normally lead to comparably large real mass
shifts. Such effects are always governed by analyticity of some $S$-matrix
in which those resonances show up as simple poles. Since the couplings
that produce strong-decay widths are large, the resulting complex mass shifts
are generally non-perturbative and non-linear and so there is no reason to
believe that unitarisation effects will keep real mass splittings in the bare
spectrum intact. To make things worse, even states lying below their respective
lowest OZI-allowed decay threshold may undergo sizeable mass shifts due to
virtual meson loops, which in this case are always real and negative. A famous
example is the $D_{s0}^\star(2317)$ \cite{PDG2020} charmed-strange scalar
meson, whose bare state was found on the lattice \cite{mohler13b} to shift
from above the $DK$ threshold to clearly below it owing to the inclusion of
meson-meson interpolators in the simulation, besides the usual $c\bar{s}$
ones. This result was confirmed very recently in the independent lattice
computation of Ref.~\cite{wagner20}. We shall extensively revisit this meson
in Secs.~\ref{rse} and \ref{lattice}.

Nevertheless, the success of the Cornell model in reasonably reproducing the
then known $c\bar{c}$ \cite{eichten78,eichten80a} and $b\bar{b}$
\cite{eichten80b} states in the model's static as well as the coupled-channel
version, seemed to indicate that unitarisation could be accommodated through
a modest change of parameters. In the following we shall try to make clear that
this is not the case at least in the light-meson spectra, on the basis of the
GI model \cite{GI85}. Note that this choice is exclusively motivated by the
exhaustiveness of the GI predictions of meson spectra for all flavours and is
in no way intended to downplay the importance of this pioneering work.

\subsection{Observed meson spectrum and Godfrey-Isgur \cite{GI85} model}
The GI \cite{GI85} quark model of mesons is still referred to very
frequently for comparison when new mesons are observed in experiment
or other models make predictions. This is understandable in view of the GI
model's completeness in predicting meson spectra for almost any desired flavour
combination and quantum numbers, besides the employment of the widely
accepted funnel potential, with a running coupling constant in the
colour-Coulombic part similarly to the approach in Ref.~\cite{richardson79}.
Also, the model employs relativistic kinematics, which should make its 
predictions for light meson more reliable. Other typical ingredients are
colour spin-spin and spin-orbit interactions, as well as the use of constituent
quark masses. Note that this makes the model's application to especially the
pion somewhat questionable, in view of the neglect of dynamical chiral-symmetry
breaking \cite{orsay85,bicudo90a}, which may account for even more than half of
the $\rho(770)$-$\pi$ mass splitting \cite{orsay85}. In spite of this, the pion
mass is reproduced in the GI parameter fit, which is possible owing to
the small non-strange constituent quark mass of 220~MeV, usually of
the order of 350~MeV in non-relativistic models.

\begin{figure}[!ht]
\begin{center}
\includegraphics[trim = 20mm 153mm 18mm 37mm,clip,width=18cm,angle=0]
{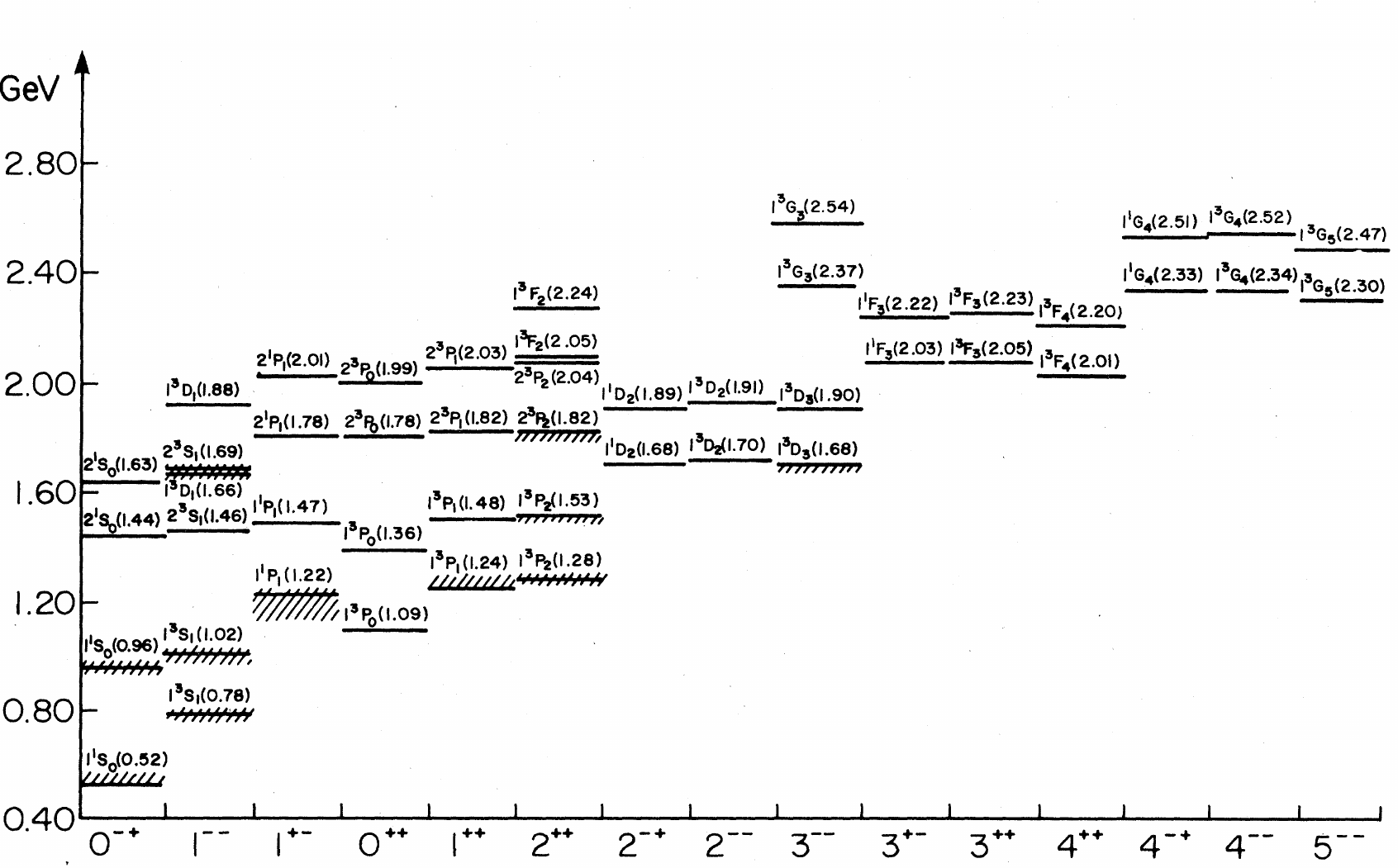}
\end{center}
\caption{Light-quark isoscalar mesons as predicted in the GI model
\cite{GI85}.}
\label{GIisoscalars}
\end{figure}

In the following, we compare the predictions of the GI model for meson masses
with those of the latest PDG tables \cite{PDG2020}, focusing especially on
radial meson spectra for light and strange quarks. A similar analysis was
already carried out in Ref.~\cite{rupp12}, on the basis of the 2012 PDG tables
\cite{PDG2012}.
\subsubsection{Light-quark isoscalar mesons \cite{GI85}
 (Fig.~\ref{GIisoscalars})}
\begin{figure}[!h]
\begin{center}
\includegraphics[trim = 18mm 62mm 20mm 128mm,clip,width=18cm,angle=0]
{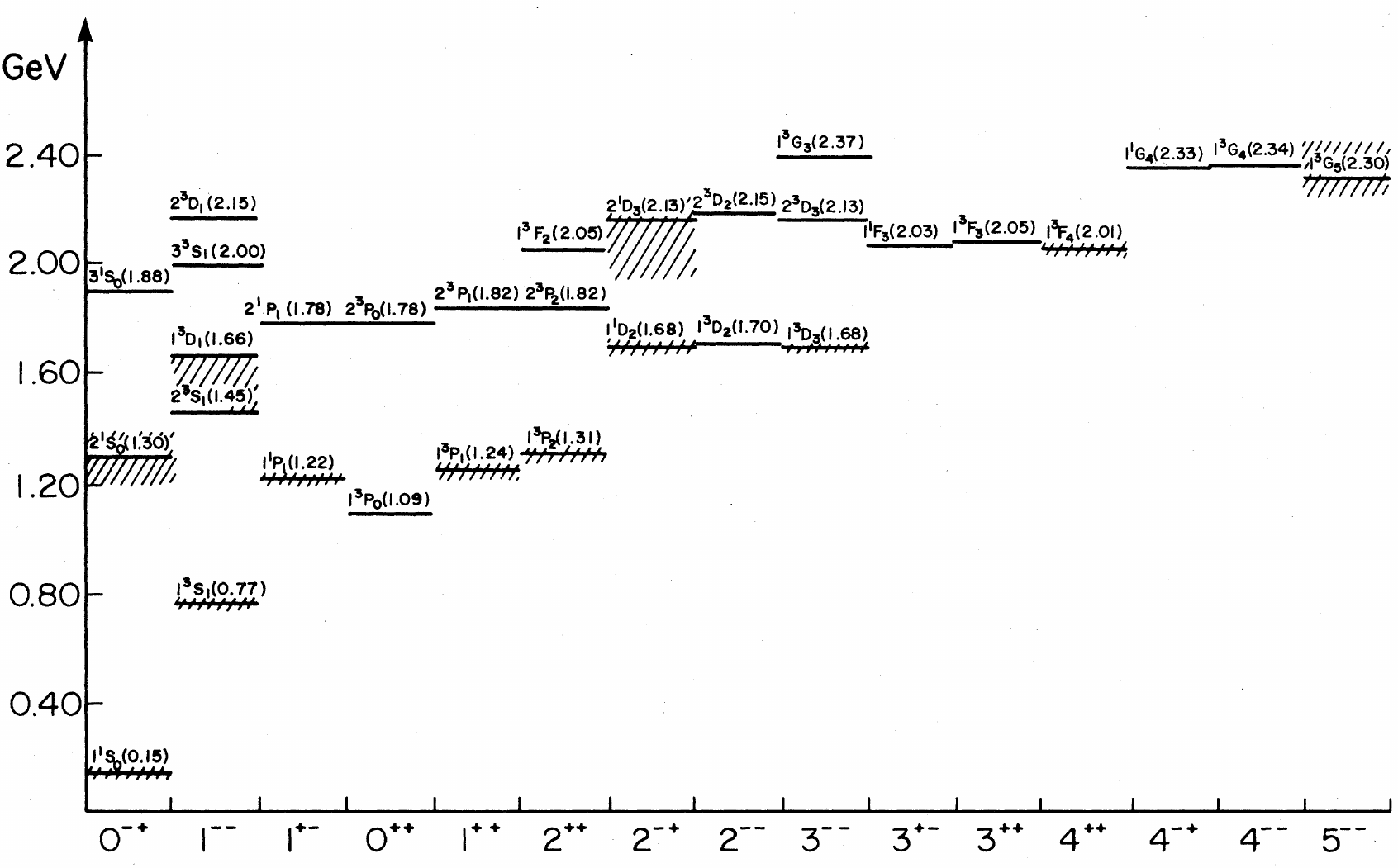}
\end{center}
\caption{Light-quark isovector mesons as predicted in the GI model
\cite{GI85}.}
\label{GIisovectors}
\end{figure}

\begin{itemize}
\item
$0^{++}$/\tpz: \\
Lowest GI scalar $\sim\!600$ MeV heavier than \bsm{f_0(500)}\footnote
{Henceforth, we shall print the states included in the PDG Summary Tables
\cite{PDG2020} in boldface.} (alias \bsm{\sigma}); \\
GI $s\bar{s}$ scalar almost 400 MeV heavier than
\bsm{f_0(980)}. \\[1mm]
Note: we shall come back to these light scalar mesons in much more detail
in Secs.~\ref{coupled} and \ref{rse}.
\item
$2^{++}$/\tpt-\tft: \\
PDG listings report 6 likely $n\bar{n}$ ($n=u,d$) states up to
$\approx\!2.15$~GeV, viz.\ \bsm{f_2(1270)}, $f_2(1565)$,
$f_2(1640)$, $f_2(1810)$, $f_2(1910)$, and $f_2(2150)$, whereas GI only
predict 3. In the probably dominant $s\bar{s}$ sector, PDG also lists 6 states
up to $\approx\!2.35$ GeV: $f_2(1430)$, \bsm{f_2^\prime(1525)},
\bsm{f_2(1950)}, \bsm{f_2(2010)}, \bsm{f_2(2300)}, and
\bsm{f_2(2340)}, while GI again only predict 3. \\[1mm]
Note: some PDG $f_2$ states may not be resonances \cite{bugg04},
but $f_2(1565)$ looks reliable. Also,
PDG: $m(\mbox{\ttpt})-m(\mbox{\otpt})\approx300$~MeV; 
GI: $m(\mbox{\ttpt})-m(\mbox{\otpt})=540$~MeV. \\
For unclear reasons, the PDG has been omitting $f_2(1565)$ from the
Summary Tables.
\item
$1^{+-}$/\spo: \\
PDG $n\bar{n}$ entries: \bsm{h_1(1170)},
$h_1(1595)$; \\
GI predict: $h_1(1220)$ (\ospo), $h_1(1780)$ (\tspo).
\end{itemize}
\subsubsection{Light-quark isovector mesons \cite{GI85}
(Fig.~\ref{GIisovectors})}
\begin{figure}[!t]
\begin{center}
\includegraphics[trim = 20mm 138mm 18mm 37mm,clip,width=18cm,angle=0]
{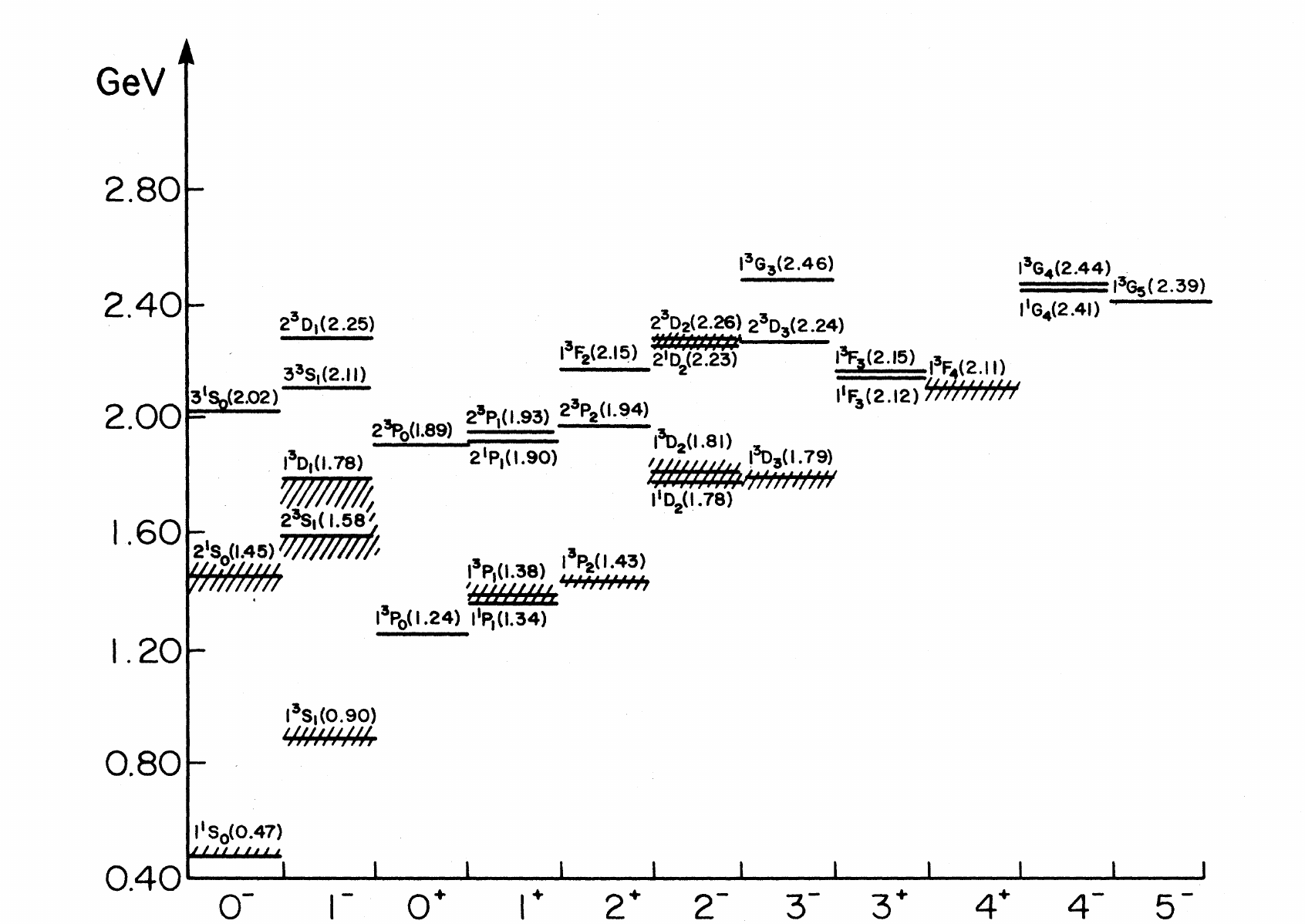}
\end{center}
\caption{Strange mesons as predicted in the GI model \cite{GI85}.}
\label{GIstrange}
\end{figure}

\begin{itemize}
\item
$0^{++}$/\tpz: \\
PDG entries: \bsm{a_0(980)}, \bsm{a_0(1450)}; \\
GI: $a_0(1090)$ (\otpz), $a_0(1780)$ (\ttpz).
\item
$1^{++}$/\tpo: \\
PDG entries: \bsm{a_1(1260)}, \bsm{a_1(1640)}; \\
GI: $a_1(1240)$ (\otpo), $a_1(1820)$ (\ttpo).
\item
$2^{++}$/\tpt: \\
PDG entries: \bsm{a_2(1320)}, \bsm{a_2(1700)}; \\
GI: $a_2(1310)$ (\otpt), $a_2(1820)$ (\ttpt).
\item
$1^{--}$/\tso-\tdo: \\
PDG entries: \bsm{\rho(1450)}, $\rho(1570)$,
\bsm{\rho(1700)}, $\rho(1900)$;  \\
GI: $\rho(1450)$ (\ttso), $\rho(1660)$ (\otdo),
$\rho(2000)$ (\htso), $\rho(2150)$ (\ttdo). \\
Note: as already mentioned above, a very recent \cite{hammoud20}  multichannel
unitary $S$-matrix reanalysis of $P$-wave $\pi\pi$ phase shifts and
inelasticities has confirmed $\rho(1250)$, besides finding evidence of
the further resonances $\rho(1450)$, $\rho(1600)$, and $\rho(1800)$.
\end{itemize}
\subsubsection{Strange mesons \cite{GI85} (Fig.~\ref{GIstrange})}
\begin{itemize}
\item
$0^{-}$/\ssz: \\
PDG entries: $K(1460)$, $K(1830)$; \\
GI: $K(1450)$ (\tssz), $K(2020)$ (\hssz).
\item
$0^{+}$/\tpz: \\
PDG entries: \bsm{K_0^\ast(700)}, \bsm{K_0^\ast(1430)}, $K_0^\ast(1950)$; \\
GI: $K_0^\ast(1240)$ (\otpo), $K_0^\ast(1890)$ (\ttpo).  \\[1mm]
For $K_0^\star(700)$, also see Secs.~\ref{coupled} and \ref{rse}.
\item
$1^{-}$/\tso-\tdo:
PDG entries: \bsm{K^\ast(1410)}, \bsm{K^\ast(1680)}; \\
GI: $K^\ast(1580)$ (\ttso), $K^\ast(1780)$ (\otdo).
\item
$1^{+}$/\tpo-\spo: \\
PDG entries: \bsm{K_1(1270)}, \bsm{K_1(1400)},
$K_1(1650)$; \\
GI: $K_1(1340)$ (\ospo), $K_1(1380)$ (\otpo), $K_1(1900)$ (\tspo),
$K_1(1930)$ (\ttpo).
\item
$2^{-}$/\sdt-\tdt: \\
PDG entries: $K_2(1580)$, \bsm{K_2(1770)},
\bsm{K_2(1820)}, $K_2(2250)$; \\
GI: $K_2(1780)$ (\osdt), $K_2(1810)$ (\otdt),
$K_2(2230)$ (\tsdt), $K_2(2260)$ (\ttdt).
\end{itemize}
\subsubsection{Summary of light mesons \cite{GI85}}
As we have seen above, The GI model predicts radial level splittings in
the light and strange meson sectors that are considerably larger than the
listed \cite{PDG2020} ones. Lattice QCD simulations that ignore the 
dynamical effects of strong decay predict radial splittings that are even
larger \cite{morningstar14} (also see Sec.~\ref{lattice} below). Moreover,
there is no indication that some of the observed resonances might be
crypto-exotics, i.e., states with non-exotic quantum numbers yet not having
the standard $q\bar{q}$ configuration, like tetraquarks ($qq\bar{q}\bar{q}$)
and hybrids ($q\bar{q}g$). Therefore, no excess of regular mesons can be
claimed on the basis of the experimental data \cite{beveren15b}.
On the other hand, several missing states in e.g.\ the strange and vector
$\phi$ sectors further complicate the picture. Nevertheless, there can be no
doubt that this part of meson spectroscopy poses a huge challenge to the usual
quark model.
\subsubsection{Charm and bottom mesons }
Especially the scalar $D_{s0}^\star(2317)$ but also the axial-vector
$D_{s1}(2460)$ come out too heavy \cite{GI85} in the GI model. However, this
may be strongly related to the failure of the static quark model in describing
the light scalar mesons as $q\bar{q}$, as demonstrated in
Ref.~\cite{beveren03} (also see Sec.~\ref{rse} below). Apart from these clear
discrepancies, there are insufficient data on radial excitations of open-charm
and open-bottom mesons for any definite conclusions.
\subsubsection{Charmonium and bottomonium}
Up to the 2002 PDG edition \cite{PDG2002}, the list of charmonia and bottomonia
had remained largely unaltered, with all states more or less compatible
\cite{GI85} with the GI model and similar ones. Things changed dramatically in
2003 with the discovery \cite{x3872} of the charmonium-like state $X(3872)$,
which by now has been firmly established as a \tpo\ state and is therefore
called \cite{PDG2020} $\chi_{c1}(3872)$. Its mass is about 80 MeV below the one
predicted \cite{GI85} in the GI model, which is a lot when compared with the
accuracy of the model in reproducing the masses of the \otpo\ $c\bar{c}$
states. Moreover, at 3872~MeV, $\chi_{c1}(3872)$ lies on top of the lowest
strong decay threshold, viz.\ $D^0\bar{D}^{\star0}$, within a hundredth of an
MeV when considering the central values of the 2020 fitted PDG masses of the
$D^0$ and $D^{\star0}$ charmed mesons. Therefore, $\chi_{c1}(3872)$ is often
considered a kind of very weakly bound $D^0\bar{D}^{\star0}$ molecule. However,
in Refs.~\cite{coito11a,coito13,cardoso15} it was shown that this state can be
successfully described as a strongly unitarised \ttpo\ $c\bar{c}$ meson, with
a naturally very large $D^0\bar{D}^{\star0}$ component in its wave function
\cite{coito13,cardoso15}. For more details, see Sec.~\ref{rse} below.

Starting in 2005, many more enhancements have been found in the charmonium
spectrum and included in the PDG listings, besides a few new ones for
bottomonium, too. As for charmonium-like states, several have only been
observed in OZI-violating strong-decay modes like $J/\psi\pi\pi$, most notably
the vector $X(4260)$, now called \cite{PDG2020} $\psi(4260)$. This state has a
very long list of open-charm decay modes listed in the PDG tables as ``not
observed'' \cite{PDG2020}, which has led to a plethora of suggestions
involving different crypto-exotic configurations for $\psi(4260)$, viz.\ a
hybrid, tetraquark, charm-baryonium molecule, \ldots\ . For more details, see
e.g.\ the review on hidden-charm pentaquarks and tetraquarks in
Ref.~\cite{xiang16}. An alternative explanation \cite{beveren10b} of the
$\psi(4260)$ peak in $e^+e^-\to J/\psi\pi\pi$ data amounts to a non-resonant,
merely apparent enhancement owing to strong inelasticities from competing
OZI-allowed channels and true $\psi$ vector resonances. We shall come back to
this modelling \cite{beveren10b} in Sec.~\ref{production} below.

Even several charged hidden-charm and hidden-bottom states have been observed
\cite{PDG2020} in recent years. Clearly, if these indeed correspond to genuine
resonances, they must be of an exotic non-$q\bar{q}$ nature, but there are
alternative explanations in terms of purely kinematical ``triangle''
singularities. Discussion of such models lies outside the scope of the present
review, so let us again refer to Ref.~\cite{xiang16} for a detailed discussion
of all these exotic candidates. For an overview of the rapidly expanding and
changing PDG listings of hidden-charm and hidden-bottom states between 2002
and 2018, see Ref.~\cite{rupp19}.
\subsection{Other relativistic static quark models}
In order to show that problems foremostly in the light-meson spectrum are not
unique to the GI model, we now briefly resume the formalisms and predictions of
two static quark models that treat relativity in a more complete fashion. The
first model \cite{tjon94} concerns a relativistically covariant quasipotential
calculation in configuration space, applied to most light, heavy-light, and
heavy mesons, It excludes light isoscalars as e.g.\ the scalars $f_0(500)$,
$f_0(980)$, and the several $J^{PC}=2^{++}$ $f_2$ states, owing to the
difficulty of dealing with the inevitable $n\bar{n} \leftrightarrow s\bar{s}$
mixing. Although the starting point is the fully relativistic four-dimensional
(4D) Bethe-Salpeter \cite{BS51} equation (BSE) with the complete Dirac spin
structure, the enormous complexity of properly dealing with all singularities
involved in the Wick rotation and the confining potential \cite{tiemeijer93},
a three-dimensional (3D) reduction of the BSE is carried out, by integrating
out the relative-energy variable. This can be done in many different ways,
but special care is required in order to preserve relativistic covariance.
In Ref.~\cite{tjon94}, two different approaches are tested empirically in
fits to the meson spectrum. The first 3D approximation is the
Blankenbecler-Sugar-Logunov-Tavkhelidze \cite{bslt63_65} (BSLT)
formalism, which basically uses delta functions to put the two particles
equally off mass shell in the intermediate state when they have the same mass,
while correctly reducing to the Dirac equation if one of the two becomes
infinitely heavy \cite{CJ89}. The other approach employed in Ref.~\cite{tjon94}
is the so-called equal-time (ET) equation \cite{ET87}, which is based on the
assumption that the potential does not depend on the relative energy 
in the centre-of-mass system. This facilitates integrating out this variable,
thus reducing the equation from 4D to 3D. However, in order to ensure that
it reduces to the Dirac equation for the lighter particle if the other mass
goes to infinity, an extra term is added to the two-body propagator in the
intermediate state which stands for propagation inside the crossed-box diagram.
This is also guarantees to approximately account for all crossed diagrams when
iterating the equation. For further details, see Ref.~\cite{ET87}.
In both 3D approaches \cite{tjon94}, the $q\bar{q}$ potential in $r$-space
consists of a Coulombic part from one-gluon exchange, similarly to
Ref.~\cite{richardson79}, plus a linearly rising confining part with a constant
term as well. The former part has a natural Dirac vector structure, while the
confining piece is taken as dominantly scalar with a small vector admixture.
The resulting coupled integro-differential equations for the bound-state wave
functions in coordinate space, including both positive- and negative-energy
components, are solved numerically using an expansion in terms of cubic Hermite
spline functions \cite{tjon94}. Finally, the solutions of the equations are
gauge dependent as far as the vector parts in the potential are concerned, due
to the 3D nature of the formalism and the fixing of the relative energy.
Therefore, two different gauges are chosen in order to test the corresponding
sensitivity of the results, viz.\ the Feynman gauge and the Coulomb gauge, but
only in the BSLT approximation, since the ET formalism does not allow a
satisfactory fit to the spectra at all.

Coming now to the fit results in Ref.~\cite{tjon94}, our first observation is
that the constituent quark masses $m_{u,d}$, $m_s$, $m_c$, and $m_b$ are close
to those found in the GI model, especially $m_{u,d}$ and $m_s$. As for the thus
computed spectra, the radial spacings are clearly much too large, both in
the light-quark sector and for the heavy quarkonia $c\bar{c}$, $b\bar{b}$ 
(see Table~III in Ref.~\cite{tjon94}). Therefore, the discrepancies are even
worse than in the GI model. This may not have been so evident back in 1994,
but is now undeniable when comparing the predictions \cite{tjon94} to PDG
\cite{PDG2020} mass values. A typical example:
$m_{\rho^{''''}}($\htso$)-m_{\rho^{''}}($\ttso$)\simeq600$--700~MeV, 
depending mildly on the BSLT or ET approach and also on the chosen gauge (for
BSLT), to be compared to 300--350~MeV in the very recent analysis of
Ref.~\cite{hammoud20}. Note that the $(n\!+\!1)\,$\tso\ states lie above the
corresponding $n\,$\tdo\ states in the light-quark sector, as opposed to the
GI and non-relativistic models, but not for charmonium and bottomonium.
Another light-quark example:
$m_{\pi{''}}($\hssz$)-m_{\pi^{'}}($\tssz$)\simeq600$--800~MeV \cite{tjon94},
cf.\ 510~MeV according to the PDG. The radial splittings also come out much too
large for $c\bar{c}$ and $b\bar{b}$ states, giving rise to mass predictions for
the \ftso\ bottomonium and charmonium\footnote
{Note that there is a typographical error in Table~III of Ref.~\cite{tjon94}:
\htso\ for $\psi^v$ should read \ftso.}
states that are 300--400~MeV too high.

The other \cite{hersbach94} relativistic model calculation of meson spectra
is of a completely different nature, though the employed interaction is very
similar to the one used in Ref.~\cite{tjon94}. The so-called RdG 
formalism \cite{hersbach92} is not derived from the BSE. Its principal
characteristic is that all particles are always kept on their mass shells,
even in the intermediate state, where total three-velocity is conserved and
not total four-momentum. Nevertheless, in asymptotic states total
four-momentum is naturally conserved. Also, since there is no relative-energy
variable, the resulting equations are automatically of a 3D type. Furthermore,
as the RdG approach is not derived from quantum field theory but rather from
ordinary quantum mechanics, no negative-energy states occur in the equations,
despite the use of Dirac spinors for the individual quark and antiquark 
composing a meson. Finally, Lorentz invariance of the formalism is guaranteed,
as the potential is a function of scalars constructed from the quark and
antiquark four-momenta. The RdG equations are solved numerically in momentum
space, again with cubic Hermite splines, after regularising the potential's
singularities in both the Coulombic part with running strong coupling and the
linear-plus-constant part. Also, different Dirac structures in the confining
potential are tested in the fits, namely as purely scalar or with a sizeable
vector admixture. For further details, see Ref.~\cite{hersbach94}.

The results of this RdG model of mesons are better than those of the previous
quasipotential approach when focusing on the radial excitations in charmonium
and bottomonium. In the light-quark sector, this aspect is more difficult to
evaluate, in view of the scarcity of reported higher radial excitations in 
Table~III of Ref.~\cite{hersbach94}. Nevertheless, the radial spacings for the
excited $\rho$ states, though considerably smaller than in Ref.~\cite{tjon94},
still seem to be somewhat on the large side when compared with
the analysis of Ref.~\cite{hammoud20}, albeit agreeing on interpreting
$\rho(1450)$ as the \otdo\ state instead of the \ttso\ favoured by the PDG.
However, these good fit results come at a high price,
namely much too large constituent quark masses, in all 4 flavour sectors 
(see Table~II \cite{hersbach94}), when compared to other quark models. 
The values $m_{u/d}=512$--966~MeV, $m_s=766$--1072~MeV, $m_c=2066$--2249~MeV,
and $m_b=5474$--5593~MeV, depending on the specific fit, are very unusual and
quite worrying, notwithstanding the fact that quark masses are not directly
observable. As the author explains \cite{hersbach94}, these very heavy
constituent quarks are a consequence of the large negative constant
$C\sim-1$~GeV in the confining potential needed to obtain reasonable radial and
angular splittings for light and strange quarks. The author also adds that
the quality of the fits in charmonium and bottomonium depends only weakly on
$C$, but that a large negative value is absolutely required for good results
in the lighter sectors.
Now, it seems obvious to ask why a negative constant $c$ of the same order of
magnitude ($c\sim-1$~GeV) in the confining potential as employed in
Ref.~\cite{tjon94} does not give rise to such huge quark masses. However, one
must realise that the quasipotential equations in the latter work are
time-reversal invariant, so that for each positive-energy solution there is
also a negative-energy solution with equal mass modulus. A further lowering of
$c$ would eventually lead to two equal massless solutions, beyond which point
the equations do not support a bound-state solution for the ground state
anymore and only for excitations. This is to be contrasted with the RdG
equations in Ref.~\cite{hersbach94}, which do not have negative-energy
solutions. So the role that the constant term in the confining potential plays
is very different in the two approaches.

The purpose of discussing these two different relativistic models of mesons in
quite some detail was to show that a much more complete treatment of relativity
than in the GI model does not necessarily lead to better or even unambiguous
predictions. In particular, either the radial spacings in all meson spectra
become much too large, as is the case of the quasipotential model of
Ref.~\cite{tjon94}, or one ends up with unrealistically heavy constituent
quarks, with possibly still somewhat too large radial splittings in the light
sector. Thus, it appears safe to conclude that either the commonly used
static $q\bar{q}$ potential must be reconsidered or other mechanisms lead to
very significant deformations of the static meson spectrum, which might allow
to bridge the gap with experiment. In the remainder of this review, we shall
explore the latter hypothesis.
\section{Unitarity and coupled-channel mass shifts in quark
models}
\label{coupled}
As mentioned in Sec.~\ref{intro}, even nowadays many experimentalists still
confront any enhancement in their meson data with predictions of the
GI model \cite{GI85} in order to arrive at a $q\bar{q}$ assignment or otherwise
claim to have found evidence of some exotic state. Yet, models going beyond
the static quark model have been around for more than four decades, the
pioneering ones dating back to several years before the GI model. These were
the already mentioned Cornell model for charmonium \cite{eichten78,eichten80a}
and bottomonium \cite{eichten80b}, the Helsinki model for light pseudoscalars
and vectors \cite{tornqvist79,tornqvist80}, and the also referred Nijmegen
model for heavy quarkonia \cite{nijmegen80} and all pseudoscalar as well as
vector mesons \cite{nijmegen83}. Despite the at times huge mass shifts
predicted by these models, for many years the effects of decay, also called
coupled-channel contributions or unitarisation, were largely ignored. The
success of the Cornell model for heavy quarkonia and the vast scope of
the GI model no doubt contributed to this state of affairs.

Quark models that try to account for the effects of $q\bar{q}$
pair creation and/or hadronic decay are often called ``unquenched''
\cite{PRL79p1998,JPG31p845,bijker07,gonzalez11,burns14,cardoso15}.
Now, this is actually a very sloppy name, as the term ``unquenched'' originates
in lattice calculations with dynamical instead of static quarks, via a fermion
determinant. We shall nevertheless use this inaccurate name when generically
referring to such quark models, because the various approaches are very
different. For instance, Refs.~\cite{bijker07} and \cite{burns14}
evaluate real mass shifts from lowest-order hadronic loops.
Now, Ref.~\cite{bijker07} introduced an unquenched quark model
for baryons (also see Ref.~\cite{bijker09}), based on the flux-tube-breaking
model of Ref.~\cite{geiger91}, but the authors and their collaborators later
applied it to mesons as well, as e.g. $\chi_{c1}(3872)$ \cite{ferretti13} and
bottomonium states \cite{ferretti14}. In Ref.~\cite{geiger91} it was claimed,
on the basis of overlap integrals of (real) harmonic-oscillator wave functions,
that cancellations among  different sets of hadronic loops will lead to
violations of the OZI rule that are much smaller than typical widths of meson
resonances. In the same spirit, Ref.~\cite{burns14} argued that the hyperfine
spin-orbit splittings among $P$-wave states remain surprisingly small, in
spite of often large mass shifts. In particular, the ordering of these states
remains unaltered. These features were shown to be a consequence of negative
real mass shifts of comparable size for the different states, depending 
critically on Clebsch-Gordan coefficients and on the created \tpz\ light
$q\bar{q}$ pair having spin 1. However, the conclusions of both
Refs.~\cite{geiger91} and \cite{burns14} do not necessarily hold above the
physical decay thresholds. Indeed, as we shall see below, in the case of 
genuine decay and complex meson-meson loops, resonance pole positions are also
strongly influenced by unitarity, analyticity, phase space, and possibly the
proximity of other thresholds as well as poles. This makes any predictions
about the real parts of mass shifts highly unreliable without a full-fledged
$S$-matrix approach.

An alternative method to simulate effects from $q\bar{q}$ pair creation is
via a screened confining potential (see e.g.\
Refs.~\cite{gonzalez11,segovia10}), which is in fact not confining anymore
above a certain energy. The problem with this approach is that thereabove it
will give rise to spurious free quarks instead of physical free mesons.
Moreover, the threshold for e.g.\ charmonium decay to open-charm mesons lies
far below this ``deconfinement'' energy, so that the method will not naturally
account for hadronic widths, let alone unitarity.

Finally, the unquenched quark model of Ref.~\cite{cardoso15} for different
stable charmonium states is formulated as a multichannel problem in coordinate
space, with a confining potential in the $c\bar{c}$ sector and several 
open-charm meson-meson channels. Despite the fact that this calculation is
limited to states below all open-charm thresholds, the formalism is based on a
fully unitary $S$-matrix, analytically continued below threshold and so 
including meson loops to all orders. Applications of a simpler version of this
$r$-space model yet also in the scattering region can be found in
Ref.~\cite{coito13} (for both versions, see Subsecs.~\ref{x3872r2} and
\ref{x3872r1} below, respectively). Also the
original models of Refs.~\cite{eichten78,nijmegen80,nijmegen83} were truly
unitarised, but there are enormous differences as well in the computed mass
shifts from unquenching, even among in principle similar models. In
Table~\ref{massshifts} (also see Refs.~\cite{rupp16,rupp17})
\begin{table}[ht]
\caption{Negative mass shifts from unquenching. Abbreviations: BT = bootstrap,
$\chi$ = chiral, QM = quark model, RGM = Resonating Group Method, RSE =
Resonance Spectrum Expansion, CC = coupled channels, HO = harmonic oscillator,
WF = wave function, PT = perturbation theory, $q$ = light quark; $P,V,S$ =
pseudoscalar, vector, scalar meson, respectively.}
\begin{center}
\begin{tabular}{|l|l|l|c|}
\hline\hline
Refs.\ & Approach &  Mesons & $-\Delta M$ (MeV) \\ \hline\hline
\cite{eichten78} & $S$-matrix, $r$-space & charmonium &
48--180  \\
\cite{tornqvist79,tornqvist80,tornqvist84} & one-loop BT & $q\bar{q}$,
$c\bar{c}$, $b\bar{b}$; $P,V$ & 23--1300, 11--500 \\
\cite{nijmegen80,nijmegen83} & $S$-matrix, $r$-space &
$q\bar{q}$, $c\bar{q}$, $c\bar{s}$, $c\bar{c}$, $b\bar{b}$; $P,V$ &
180--700, 3--350 \\
\cite{beveren86} & $S$-matrix, $r$-space &
light, intermediate $S$ & 510--830, $\sim0$\\
\cite{bicudo90c} & $\chi$ QM, RGM &
$\rho(770)$, $\phi(1020)$ & 328, 94\\
\cite{beveren03}  & RSE, $p$-space &
$D_{s0}^\star(2317)$, $D_0^\star(2300)$ & 260, 410 \\
\cite{simonov04} & CC, $\chi$ Lagrangian &
$D_{s0}^\star(2317)$, $D_s^\star(2632)$ & 173, 51 \\
\cite{kalashnikova05} & CC, PT & charmonium & 165--228 \\
\cite{barnes08} & CC, HO WF & charmonium & 416--521 \\
\cite{coito11a} & RSE, $p$-space & $X(3872)$ & $\approx\!100$ \\
\cite{coito11b}  & RSE, $p$-space &
$c\bar{q}$, $c\bar{s}$; $J^P=1^+$ & 4--13, 5--93  \\ \hline\hline
\end{tabular}
\end{center}
\label{massshifts}
\end{table}
we show the corresponding predictions of a number of unquenched quark models
for mesons, including the Nijmegen model and its more recent momentum-space
version (RSE), to be dealt with comprehensively in Sec.~\ref{rse}. Note that
the mass shifts in
Refs.~\cite{nijmegen80,nijmegen83,beveren86,beveren03,coito11a,coito11b}
are in general complex, in some cases \cite{beveren86,beveren03,coito11b}
with huge imaginary parts, corresponding to pole positions in an exactly solved
$S$-matrix. As for the disparate shifts among the various approaches, they are
due to differences in the assumed decay mechanism, included channels, and
possibly drastic approximations. Another crucial point should be to properly
account for the nodal structure of the bare $q\bar{q}$ wave functions.

Let us now look in more detail at some results shown in Table~\ref{massshifts}.
First of all, we emphasise again that mass shifts for resonances must be
complex, amounting to pole positions in the complex-energy plane with respect
to the inherently bound states in a quenched model calculation. These poles
may show up in a fully unitary and non-perturbative $S$-matrix, or just
correspond to the real and imaginary parts of complex mass shifts computed
perturbatively. In Table~\ref{massshifts} we only specify the real parts of
such mass shifts, both for bound states and resonances. Concerning the
charmonium results, we see large differences among the different models
referred in Table~\ref{massshifts}, both in size and pattern of the obtained
mass shifts. Starting with the two earliest approaches, namely by the
Cornell \cite{eichten78,eichten80a} and Nijmegen \cite{nijmegen80,nijmegen83}
groups, we observe negative mass shifts that on average are of the same order of
magnitude. However, in the Nijmegen model the shift is largest for the
\otso\ state ($J\!/\!\psi(3097)$) and in the Cornell model for the \otdo\
($\psi(3770)$). This can at least qualitatively be understood by noting
that in the Cornell model the decay process is thought to be directly
linked to the confining potential, whereas in the Nijmegen model it is
supposed to be governed by the empirically successful \tpz\ model
\cite{micu69,carlitz70}. Since the
Cornell potential is instantaneous, the created $q\bar{q}$ pair in the decay
process is then necessarily in a \ssz\ state, which implies that the
transition potential between the confined charm-anticharm sector and the final
state with two free open-charm mesons essentially probes the $c\bar{c}$
wave function's inner region. In contrast, the \tpz\ model assumes that a
new $q\bar{q}$ pair is created with vacuum quantum numbers, that is,
$J^{PC}=0^{++}$, which necessarily implies $S=L=1$ for the created pair,
being in a \tpz\ spectroscopic state. The corresponding transition potential
has necessarily a peaked structure at some interquark distance, as it must
vanish both at the origin, for being in a $P$-wave, and at large
separations because of quark confinement. Such a peaked behaviour, resulting
from string breaking at a certain distance and confirmed \cite{bali05}
in lattice-QCD calculations, was represented phenomenologically in the Nijmegen
model by a spherical delta-shell \cite{nijmegen80} or a Gaussian-type peaked
function \cite{nijmegen83}. As a consequence, in this approach it is the
ground state that shifts most, because its wave function has no nodes.
Since the Cornell and Nijmegen models do not use the same confining potential
and so predict disparate bare spectra, both models are capable of fitting the
physical spectra, despite the significant differences in the pattern of mass
shifts due to unitarisation. This emphasises the need to apply unitarised
models to a wide range of mesons, in order to allow to disentangle confinement
and decay effects on mass spectra. Note that in Ref.~\cite{nijmegen83} the
model parameters were fitted to the vector and pseudoscalar spectra, with
good results for the vector mesons in the light, charmonium, and bottomonium
sectors, both for masses and hadronic widths. Moreover, exactly the same
parameters and Gaussian-like transition potential were employed in
Ref.~\cite{beveren86}, predicting a complete nonet of light scalar mesons,
besides the regular \tpz\ $q\bar{q}$ resonances in the energy region of
1.3--1.5~GeV (also see Subsec.~\ref{scalars}).

The other two papers in Table~\ref{massshifts} on coupled-channel mass shifts
in charmonium, viz.\ Refs.~\cite{kalashnikova05} and \cite{barnes08}, both
show little sensitivity to the radial or angular excitation of the bare
$c\bar{c}$ state. So this roughly amounts to an overall downward mass shift of
the spectra, which could be compensated for by accordingly increasing the charm
quark mass \cite{barnes08}. However, in view of the enormous quantitative
discrepancy between the mass shifts in these two works, namely 165--228~MeV
vs.\ 416--521~MeV, these results should be interpreted with a lot of caution.

Next we compare the predictions of the Helsinki
\cite{tornqvist79,tornqvist80,tornqvist84} and Nijmegen
\cite{nijmegen80,nijmegen83} models for light as well as heavy pseudoscalar
and vector mesons (cf.\ second and third line in Table~\ref{massshifts}).
{\it Grosso modo}, the results are quite comparable. This is not surprising,
since both models employ the \tpz\ mechanism for decay and satisfy multichannel
unitarity. However, there are two significant differences. First of all, the
Helsinki model, based on a Cornell-type potential for the bare spectra,
predicts much smaller mass shifts for bottomonium than for
charmonium, whereas in the Nijmegen model these shifts are comparable. Note,
though, that the Nijmegen model was intentionally formulated in a universal
way, with the same confinement and decay parameters for light, medium-heavy,
and heavy mesons. Anyhow, both models manage to describe the spectra reasonably
well without resorting to many adjustable parameters. This is a reminder that
just looking at the observed resonance spectra does not allow to draw
straightforward conclusions on the kind of confining potential. Another
difference concerns the \tso\ and \tdo\ states in charmonium and bottomonium.
In the Helsinki model, these levels are generally well separated and undergo
comparable shifts. On the other hand, in the Nijmegen model, based on HO
confinement, the $(n\!+\!1)\,$\tso\ and $n\,$\tdo\ bare states are degenerate.
Now, upon unitarisation, the degeneracy is lifted, with the \tso\ level
becoming an $S$-matrix pole that shifts much more than the \tdo\ partner pole.
A similar phenomenon occurs through unitarisation of the degenerate \tpo\ and
\spo\ bare $c\bar{q}$ or $c\bar{s}$ state, when describing the open-charm
axial-vector mesons in the momentum-space formulation of the Nijmegen model
\cite{coito11b} (eleventh line in Table~\ref{massshifts}).
Since such mesons have no definite $C$-parity, the physical states are
linear combinations of \tpo\ and \spo. One combination then results in a
strongly shifted pole, whereas the other combination almost decouples from
the decay channel(s), giving rise to a quasi-bound state in the continuum
\cite{coito11b}.

Other very intriguing results in Table~\ref{massshifts} concern the already
mentioned scalar mesons, as described in the Nijmegen model \cite{beveren86}
(fourth line in Table~\ref{massshifts}).
Here, we see some enormous mass shifts, but also others that
are close to zero, which clearly requires further explanation. The point is
that one is not actually dealing with mass-shifted states here, but rather with
emergence of extra states, of a more dynamical origin and not present in the
unquenched spectrum. The real parts of the latter resonances indeed lie many
hundreds of MeV below the bare \tpz\ $q\bar{q}$ states. However, the latter ones
also shift, but mostly in the imaginary direction upon unquenching, thus
acquiring significant decay widths yet with their masses changed much less. So
it is more correct to refer to this phenomenon as duplication instead of mass
shift of states, since two complete $SU(3)_{\mbox{\scriptsize flavour}}$
nonets result from this doubling. The unusual mass pattern in the lightest
nonet reflects the very strong influence of hadronic decay thresholds and is
not an indication of a non-$q\bar{q}$ substructure as postulated in
Ref.~\cite{jaffe77} (see Subsec.~\ref{scalars} for more details).

To conclude this section, let us look in more detail at the light pseudoscalar
mesons, in particular the pion. Checking again Table~\ref{massshifts}, second
and third line, we see huge mass shifts of up to 700~MeV in the Nijmegen model
and even about 1300~MeV in the Helsinki model. However, one should realise that
a significant fraction of these shifts is due to a coupling constant made
proportional to the lowest-order hyperfine interaction in QCD, due to
one-gluon exchange. In the Helsinki model this was done as a fundamental
ingredient of the approach, whereas in the Nijmegen model it was a numerical
procedure of making this interaction proportional to the unitarised wave
function at the origin. Anyhow, in both models the result is that
pseudoscalar mesons shift roughly three times as much as vectors, because
of the eigenvalues of the colour-spin operator
$\vec{F}_q\!\cdot\!\vec{F}_{\bar{q}}\:\vec{S}_q\!\cdot\!\vec{S}_{\bar{q}}$,
which are -1 for \ssz\ and +1/3 for \tso.
The same is done in traditional static quark models of mesons. However, it
must be noted that the very low mass of the pion is not only due to one-gluon
exchange, but equally or even dominantly to dynamical chiral-symmetry breaking.
The Helsinki and Nijmegen models, which use effective, constituent quark
masses, are not well-suited to described the pion, as it is much lighter than
such constituent quarks. Instead, starting from massless current quarks and a
chirally symmetric confining interaction with the appropriate Dirac spin
structure, the models of Refs.~\cite{orsay85,bicudo90a} indeed obtain a
massless pion in the chiral limit, via a gap equation describing dynamical
chiral-symmetry breaking. Moreover, in the approach of \cite{bicudo90a} and by
modelling the vector-meson resonances $\rho(770)$ and $\phi(1020)$ via
Salpeter \cite{bicudo90b} and Resonating-Group-Method (RGM) \cite{bicudo90c}
equations with the very same interaction, reasonable masses and widths are
found for these mesons. Also note that the real mass shift thus obtained
\cite{bicudo90c} for $\rho(770)$ is -328~MeV, so very near the value found
in the Nijmegen model \cite{nijmegen83}.

Summarising, we have seen that unitarised and other coupled-channel models of
mesons can give rise to enormous real mass shifts, which may completely
obfuscate the underlying bound-state spectrum resulting from the confining
potential only and so the very nature of this potential. The challenge then is
to determine the most realistic approach, in view of the quantitatively very
different predictions of the here described models. 
For a minireview on unquenching hadrons and some more references, see
Ref.~\cite{pennington15}.
Next we shall revisit the special case of the light scalar mesons in the
Nijmegen model \cite{nijmegen83}.

\subsection{Light scalar mesons in the Nijmegen model}
\label{scalars}
Now we look in more detail at the predictions of the Nijmegen model
for light and heavy pseudoscalar and vector mesons \cite{nijmegen83},
as applied \cite{beveren86} to the light scalar mesons. It must be
emphasised that this did not involve any fit, as the parameters 
determined in Ref.~\cite{nijmegen83} were left completely unaltered.
Recapitulating, the resulting modelling of light isoscalar, isovector, and
strange scalar mesons amounts to coupling a confined quark-antiquark pair
with \tpz\ quantum numbers to all OZI-allowed decay channels with pairs
of ground-state pseudoscalar or vector mesons, in relative $S$-waves or
$D$-waves (only for vector-meson pairs with intrinsic spin 2). The isoscalar
$q\bar{q}$ states require two channels, namely $n\bar{n}$ and $s\bar{s}$,
owing to their unavoidable mixing through the common $KK$ and
$K^\star K^\star$ channels.  All included two-meson channels can be found in
Table~1 of Ref.~\cite{beveren86}, together with the corresponding channel
couplings derived in Ref.~\cite{beveren84} employing a microscopic formalism
of wave-function overlaps in a harmonic-oscillator basis.

The kinematically relativistic coupled-channel Schr\"{o}dinger
equations in coordinate space are solved numerically for the $S$-matrix
as in Ref.~\cite{nijmegen83}, using an increasing number of delta functions
to approximate the transition potential between the $q\bar{q}$ and
meson-meson sectors. Thus, regular scalar resonances show up in the
energy region of about 1.3--1.5~GeV, through $S$-matrix poles with
real parts not far from the discrete energy levels of \tpz\ $q\bar{q}$
bound states and imaginary parts that are also roughly compatible with
the observed widths of the PDG \cite{PDG2020} states $f_0(1370)$, $a_0(1450)$,
$K_0^\star(1430)$, and $f_0(1500)$. Note that these experimentally determined
masses are about 200--300~MeV higher than those found in the GI \cite{GI85}
model. However, the big surprise --- as it certainly was almost four decades
ago when the phenomenon was first observed \cite{dullemond83} --- is the
appearance of additional $S$-matrix poles, viz. at much lower energies and
larger imaginary parts. The nature of these extra poles is clearly different
from the ones above 1~GeV, as they do not approach the corresponding $q\bar{q}$
bound states on the real energy axis when the overall coupling constant is
reduced to zero. In contrast, these poles disappear in the scattering continuum
with ever increasing negative imaginary part. So they are clearly of a dynamical
origin, as opposed to the other, ``intrinsic'' \cite{rupp02} mesons, which can
be simply linked to the bare \tpz\ quark-antiquark spectrum.
\begin{figure}[!h]
\label{pipi}
\begin{center}
\includegraphics[trim = 50mm 175mm 50mm 45mm,clip,width=18cm,angle=0]
{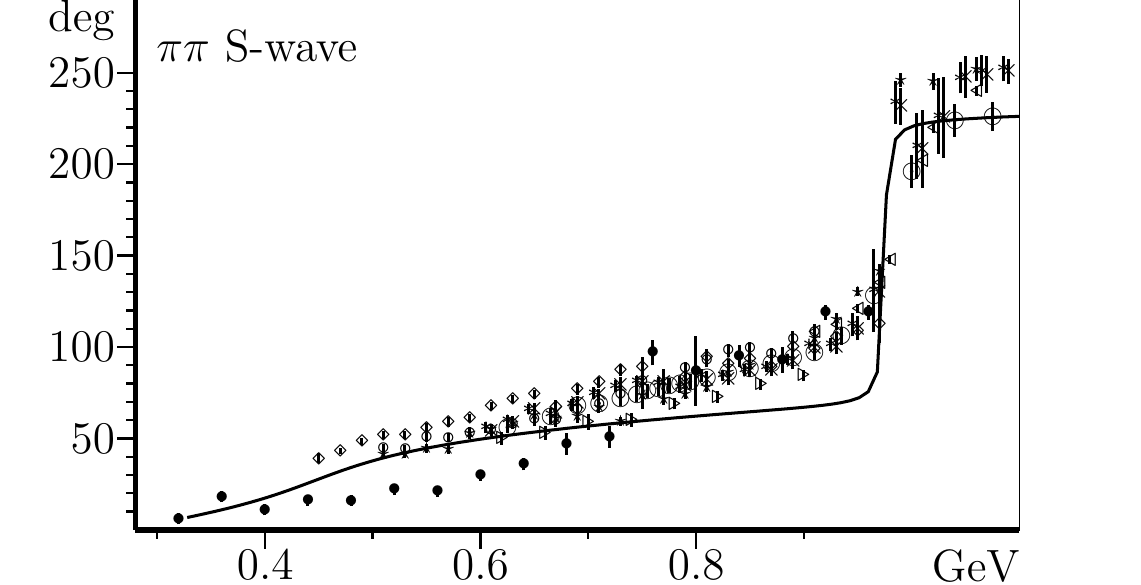}
\end{center}
\caption{$S$-wave $\pi\pi$ phases predicted in
Ref.~\cite{beveren86} (see this reference for the different data).}
\end{figure}

Coming now to the actual findings of light scalar-meson resonances as published
in Ref.~\cite{beveren86}, the lowest-lying isoscalar, isodoublet, and isovector
poles are, with the old names $\epsilon$ for $f_0(500)$, $S$ for $f_0(980)$,
$\kappa$ for $K_0^\star(700)$, and $\delta$ for $a_0(980)$:
\begin{equation}
\epsilon(470-i208)\:\mbox{MeV}, \;S(994-i20)\:\mbox{MeV},\footnote
{Note that there is a typographical error in Ref.~\cite{beveren86}, with the
imaginary part of the $S$ resonance pole given as $-2i$~MeV instead of the
value $-20i$~MeV as corrected in the posterior arXiv version 0710.4067v1
[hep-ph].}
\;\kappa(727-i263)\:\mbox{MeV},\;\delta(968-i28)\:\mbox{MeV} \; .
\label{scalarpoles}
\end{equation}
It is worthwhile noting that the corresponding masses and widths are still
compatible with present-day PDG \cite{PDG2020} limits.

Since the Nijmegen model allows to obtain a closed-form expression for the
full multichannel $S$-matrix, it is straightforward to predict a variety of
observables, as e.g.\ partial-wave phase shifts for any of the included
meson-meson channels. Thus, in Ref.~\cite{beveren86} $S$-wave $\pi\pi$ phase
shifts were computed up to 1.2~GeV, shown in
Fig.~4 with the then
available data. Considering that no fit was carried out, with the theoretical
phase shifts being \em bona fide \em \/predictions, the global agreement with
experiment was and is still remarkable.

In conclusion, we must stress that the pole doubling discovered in 
Refs.~\cite{dullemond83,beveren86} does not depend on details of the
confining potential, only on the existence of bare $q\bar{q}$ states
that couple strongly to $S$-wave meson-meson channels. As a matter of
fact, the same phenomenon was observed \cite{tornqvist96} about a
decade later, in the context of the already mentioned Helsinki 
\cite{tornqvist79,tornqvist80,tornqvist84} model. However, in 
Ref.~\cite{tornqvist96} no pole doubling\footnote
{In an earlier \cite{tornqvist82} application of the Helsinki model
to the light scalar mesons, not even an $f_0(500)$ (``$\sigma$'') was
found.}
was found for the isodoublet scalar meson (``$\kappa$'') and it was
even argued why in this case the doubling mechanism should not occur.
Thus, no $K_0^\star(700)$ was predicted and merely a $K_0^\star(1430)$.
This failure was explained as possibly due to the use of channel couplings
that are not fully flavour symmetric \cite{beveren99a} or the inclusion of
an \em ad hoc \em \/negative Adler zero in $K\pi$ $S$-wave scattering
\cite{rupp05}.
In much more recent work, the pole-doubling phenomenon for one $q\bar{q}$
seed in the scalar-meson sector was also observed in an effective-Lagrangian
model to one-loop order, namely for the isovector pair $a_0(980)$, $a_0(1450)$
\cite{wolkanowski16a} and the isodoublet pair $K_0^\star(800)$ 
(cf.\ $K_0^\star(700)$), $K_0^\star(1430)$ \cite{wolkanowski16b}.

\section{Resonance Spectrum Expansion and Applications}
\label{rse}
\subsection{Simple unitarised momentum-space model}
\label{rse1}
Despite the surprising results for the light scalar mesons in
Ref.~\cite{beveren86}, this work was largely ignored for 15 years. Things
changed in 2001, when we received an email \cite{appel01} from a
co-spokesperson of the E791 Collaboration at Fermilab, informing us about
preliminary evidence \cite{gobel00} of a light strange scalar meson
($K_0^\star$, ``$\kappa$''). He also mentioned that our Comment
\cite{beveren99a} on Ref.~\cite{tornqvist96}, in which we insisted on the
existence of a light $K_0^\star$ meson, had been encouraging to E791 and
motivated his sending us the email \cite{appel01}. The final E791 result was
eventually published in Ref.~\cite{aitala02}, confirming a light $K_0^\star$
with mass $(797\pm19\pm43)$~MeV and width $(410\pm43\pm87)$~MeV, i.e., values
fully compatible with our pole position of $(727-i263)$~MeV predicted 
\cite{beveren86} in 1986 (cf.\ Eq.~(\ref{scalarpoles})).

In the ensuing email exchange \cite{appel01} with the same E791
co-spokesperson, he encouraged us to provide easy-to-use formulae to fit the
light scalars to data, as these mesons had been awkward not only to model
builders but also to experimentalists. Indeed, our original work predicting
\cite{beveren86} a complete light scalar nonet was based on a very complicated
multichannel quark model in coordinate space \cite{nijmegen83} and
impracticable for experimental analysis. Thus, we immediately developed
\cite{beveren01} an exactly solvable yet fully unitary and analytic general
ansatz in momentum space for non-exotic meson resonances. The only assumption
here is that, like in Ref.~\cite{nijmegen80}, transitions between the confined
$q\bar{q}$ and free meson-meson sectors are governed by string breaking at a
sharp distance, modelled with a spherical delta function. In the case of only
one confined quark-antiquark channel and only one free meson-meson channel,
which is appropriate to describe e.g.\ $S$-wave $K\pi$ scattering in the
elastic region, the resulting expression for the phase shift takes the simple
form \cite{beveren01}
\begin{equation}
\mbox{cotg}\left(\delta_{\ell}(p)\right)\; =\;
\frac{n_{\ell}(pa)}{j_{\ell}(pa)}\; -\;
\left[ 2\lambda^{2}\;\mu\; pa\; j^{2}_{\ell}(pa)
\sum_{n=0}^{\infty}\frac{\left|{\cal F}_{n\ell_{c}}(a)\right|^{2}}
{E-E_{n\ell_{c}}}\right]^{-1}
\; ,
\label{cotgd}
\end{equation}
where $j_\ell$ and $n_\ell$ are spherical Bessel and Neumann functions,
respectively, with $\ell$ the orbital angular momentum in the meson-meson
channel, $p$ and $\mu$ are the relativistically defined momentum and reduced
mass in the latter channel, $a$ is the string-breaking radius, $\lambda$ is
an overall coupling constant, $\ell_c$ is the orbital angular momentum in the
$q\bar{q}$ sector,
${\cal F}_{n\ell_c}(a)$ is the $n$-th radial solution of whatever confining
potential is employed in the $q\bar{q}$ channel, $E_{n\ell_c}$ is the
corresponding bare radial spectrum, and $E$ is the total energy of the system,
which becomes complex for physical meson resonances. For a detailed derivation
of Eq.~(\ref{cotgd}) in momentum space via a step-wise solution of the
Lippmann-Schwinger Born series, see Ref.~\cite{beveren06a}. Furthermore, by
solving the same Hamiltonian directly in coordinate space, a spectral 
representation of the bare $q\bar{q}$ Green's function was derived, viz.\
\begin{equation}
\sum_{n=0}^{\infty}\frac{\left|{\cal F}_{n\ell_c}(a)\right|^{2}}
{E-E_{n\ell_c}}\; =\;\frac{2\mu}{a^{4}}\;
\frac{F_{c,\ell_c}(E,a)G_{c,\ell_c}(E,a)}
{W\left(F_{c,\ell_c}(E,a),G_{c,\ell_c}(E,a)\right)}
\; ,
\label{spectral}
\end{equation}
where $F_{c,\ell_c}$ and $G_{c,\ell_c}$ are the two solutions of the
Schr\"{o}dinger equation with the unspecified confining potential that are
regular at the origin and at infinity, respectively. The identity in 
Eq.~(\ref{spectral}) gave rise to the designation \em Resonance Spectrum
Expansion \em \/(RSE) for the formalism developed in
Refs.~\cite{beveren01,beveren06a}.

Now, contrary to BW-type expressions, Eq.~(\ref{cotgd}) does not exhibit
resonance pole positions explicitly. Nevertheless, the complex resonance
energies can be easily found by solving the equation numerically for
cotg$\left(\delta_\ell(p)\right)=i$.  Moreover, the
$\left|{\cal F}_{n\ell_c}(a)\right|^2$ are just a set of real numbers for
different $n$, which makes Eq.~(\ref{cotgd}) extremely flexible. Namely,
theorists can use their favourite model and corresponding bare spectrum for
the confining potential, being left with only two free parameters ($\lambda$,
$a$) to show the predictive power of that model. On the other hand,
experimentalists may just consider those real numbers and energies fit
parameters when analysing their data, while restricting the infinite sum to
merely a few terms, with the  argument that much higher energies will hardly
influence the fit in a certain region. In the following we shall demonstrate
the potentiality of either approach.

\subsubsection{\boldmath$K_0^\star(700)$ from a fit to \boldmath$S$-wave
\boldmath$K\pi$ phase shifts}
\label{kappa-rse}
In Ref.~\cite{beveren01}, Eq.~(\ref{cotgd}) serves as the basis to fit elastic
$K\pi$ scattering in $P$- and $S$-waves, by taking only a few terms in the
infinite sum and fit the real constants to the data, together with the 
parameters $\lambda$ and $a$. In the $P$-wave case, two terms and one bare
energy are sufficient to obtain a good fit and a very reasonable value for the
$K^\star(892)$ pole position, whose real part comes out about 60~MeV below the
fitted bare $K^\star(892)$ mass. Note that this relatively small negative mass
shift from unitarisation when compared to those found in
Refs.~\cite{nijmegen80,nijmegen83} is easy to understand, as in this simple model
only one meson-meson channel is coupled. All the other, kinematically closed
channels, which are not considered here, will contribute with extra negative
yet purely real mass shifts and so not affect in principle the imaginary part
of the pole. Inclusion of such channels will lead to a significantly higher
bare ground-state mass, though, of the order of 1.2~GeV in the case of
$K^\star(892)$, as found in the model of Ref.~\cite{nijmegen83}.

The more complicated $S$-wave case requires an additional term and two
instead of one bare energies when truncating Eq.~(\ref{cotgd}), in order 
to be able to extract resonance information above 1~GeV as well. Moreover,
the different experimental $S$-wave $K\pi$ data are rather conflicting
with each other below 1~GeV (see
Fig.~5).
In the fit to the
\begin{figure}[!t]
\label{kappa}
\begin{tabular}{cc}
\includegraphics[trim = 60mm 177mm 50mm 15mm,clip,width=8.5cm,angle=0]
{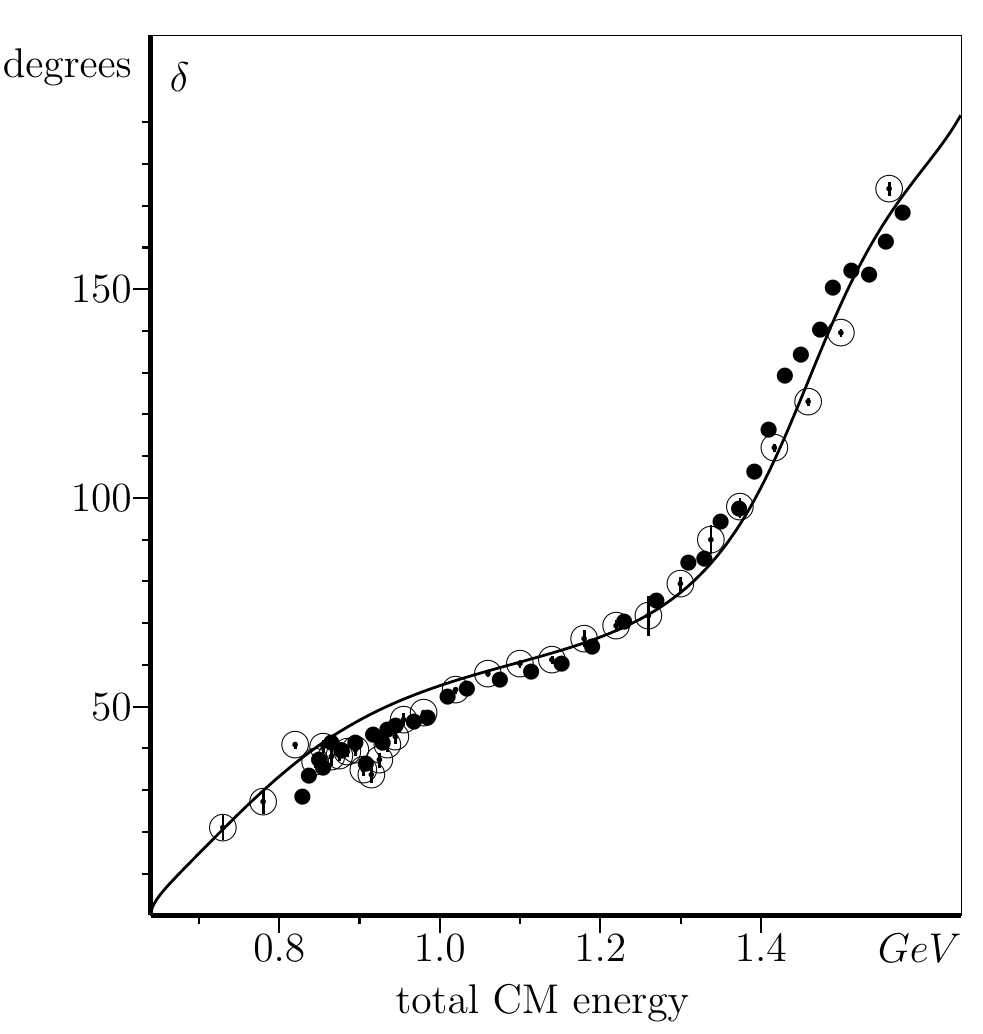}
&
\includegraphics[trim = 60mm 177mm 50mm 15mm,clip,width=8.5cm,angle=0]
{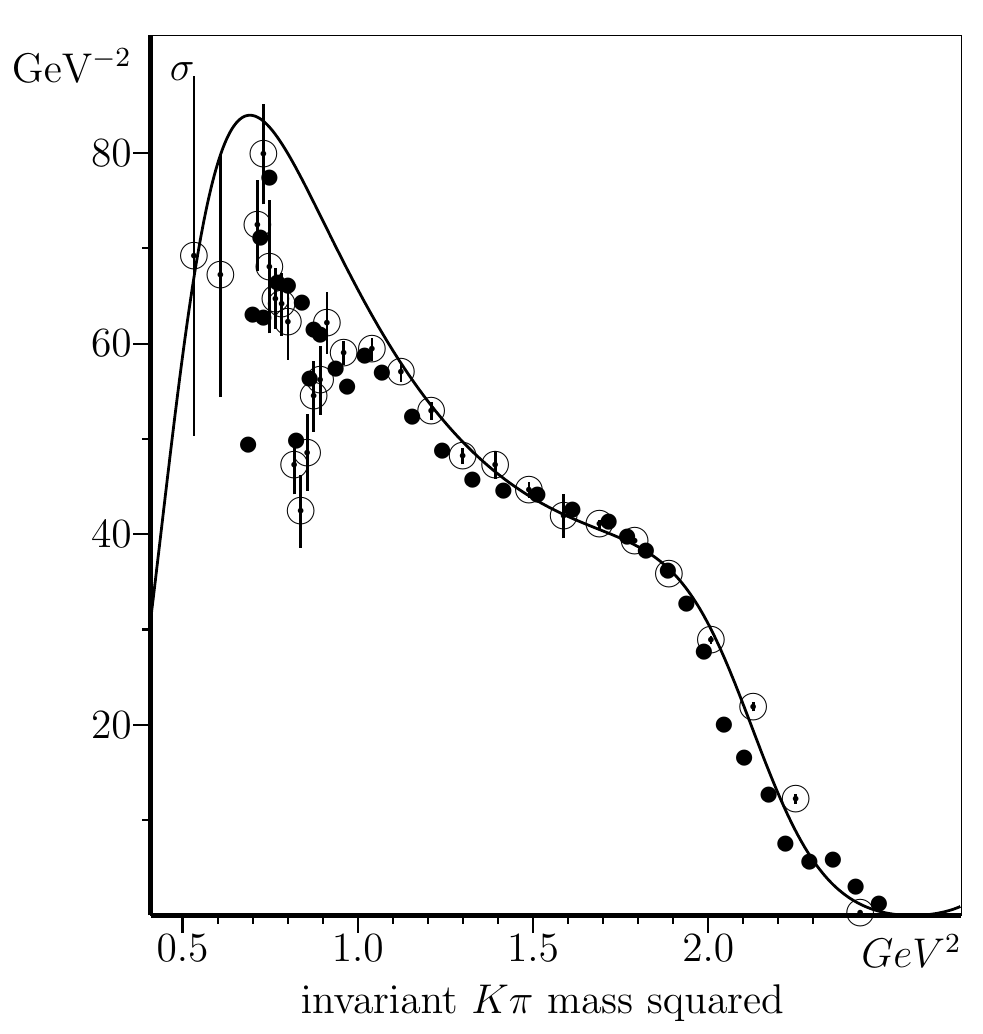}
\end{tabular}
\caption{Fitted $S$-wave $K\pi$ phase shift (left) and corresponding cross
section (right) from truncating the infinite sum in Eq.~(\ref{cotgd}) (see
text).  For more details and the experimental data, see Ref.~\cite{beveren01}.}
\end{figure}
$S$-wave phase shifts, the $\lambda$ value found for the $P$-wave at
0.75~GeV$^{-3/2}$ is kept fixed, whereas the decay radius $a$ is reduced from
5~GeV$^{-1}$ to 3.2~GeV$^{-1}$. This is related to the different
behaviour of spherical Bessel functions for small argument in the $\ell\!=\!1$
and the $\ell\!=\!0$ case. The results of the fit are shown in
Fig.~5,
both for the phase shift and the extracted cross section. 
The broad cross-section peak below 1~GeV must be due to a pole with a relatively
large imaginary part in the corresponding energy region, despite the fact that
the phase shift only reaches $90^\circ$ above 1.3~GeV. As already mentioned
above, this is due to the presence of an Adler zero in the $S$-wave $K\pi$
amplitude \cite{bugg03,rupp05} below yet very close to threshold, which slows
down the phase-shift's rise at low energies, and the overlapping of the
$K_0^\star(700)$ and $K_0^\star(1430)$ resonances \cite{bugg03}. Searching for
poles, we indeed we find two of them below 1.5~GeV, namely at $(714-i228)$~MeV
for $K_0^\star(700)$ and $(1458-i118)$~MeV for $K_0^\star(1430)$. These results
are excellent for such a simple model, without the inelastic $K\eta^\prime$
channel. Note that the $K\eta$ channel largely decouples \cite{beveren07},
because of a partial cancellation between the $n\bar{n}$ and $s\bar{s}$
components of the $\eta$ mesons, via $n\bar{n}$ and $s\bar{s}$ pair creation,
respectively.

\subsubsection{Charmed scalar mesons \boldmath$D_{s0}^\star(2317)$ and
\boldmath$D_0^\star(2300)$}
\label{scalar-charmed}
In April 2003, the BABAR Collaboration announced the observation of a 
narrow meson with a mass of about 2.32~GeV decaying to $D_s^+\pi^0$,
which was published in Ref.~\cite{babar03}. They baptised the new state
as $D_{sJ}^\star(2317)^+$ (now called $D_{s0}^\star(2317)^+$ \cite{PDG2020})
and argued that the low mass of this charmed-strange
meson as well as its natural parity favour a $J^P=0^+$ assignment.
If indeed confirmed as the lowest scalar $c\bar{s}$ state, its small width
would be the consequence of lying below the $DK$ threshold, with $D_s^+\pi^0$
being an isospin-violating decay mode and therefore suppressed. However, such
a low mass is in conflict with the GI \cite{GI85} and similar static quark
models, as already mentioned above.

After learning about the discovery, we immediately realised \cite{beveren03}
that this enigmatic new meson, if confirmed as a scalar, couples strongly to
the $S$-wave $DK$ channel, which may significantly alter its mass resulting
from the confinement potential alone. Moreover, with its $c\bar{s}$ quark
content and only one nearby OZI-allowed decay channel, it has strong
similarities with the coupled $K_0^\star(700)$-$K\pi$ system. The problem is
that in the $D_{sJ}^\star(2317)$ case no $DK$ phase shifts are and probably
never will be available. So we decided to leave several parameters resulting
from the RSE fit to the $S$-wave $K\pi$ phase shifts unaltered, namely
$\lambda$, $a$, and the three constants replacing the infinite sum in
Eq.~(\ref{cotgd}). The only obvious changes were replacing the $K\pi$ threshold
by the $DK$ threshold at 2363~MeV, besides taking two fixed bare energy levels
instead of fitting them as in the $K\pi$ case. To choose these, we resorted to
the Nijmegen model for pseudoscalar and vectors mesons \cite{nijmegen83},
taking the charm quark mass at 1562~MeV, the strange quark mass at 508~MeV, and
the first two radial levels of the harmonic-oscillator spectrum, with
frequency $\omega=190$~MeV. This results in bare $c\bar{s}$ energy levels at
2545~MeV and 2925~MeV. Furthermore, we also included the isodoublet
charmed-light scalar meson \cite{beveren03}, with bare $c\bar{n}$ energy levels
at 2443~MeV and 2823~MeV, from $m_n=406$~MeV \cite{nijmegen83}, as there was
already some indication \cite{belle02} of a very broad charmed scalar resonance
decaying to $D\pi$ in an $S$-wave.

The thus found lowest charmed-strange and charmed-light scalar poles are
\cite{beveren03}
\begin{equation}
c\bar{s}: (2.28+i0)\;\mbox{GeV}\;,\;\;c\bar{n}: (2.03-i0.075)\;\mbox{GeV}\;.
\label{cscn}
\end{equation}
The $c\bar{s}$ bound state pole clearly represents $D_{s0}^\star(2317)$, with
the pole coming out remarkably close to 2.317~GeV in view of the lack of
any fit and just the use of parameters fixed in previous work
\cite{nijmegen83,beveren01}. As for the $c\bar{n}$ pole, it seems to lie 
much too low to stand for the $D_0^{\star0}$ resonance first observed
\cite{belle02} and then confirmed \cite{belle04} by the Belle Collaboration,
with mass $2308\pm17\pm15\pm28$~MeV and width $276\pm21\pm18\pm60$~MeV, now
listed by the PDG \cite{PDG2020} with average mass 2300~MeV and width 274~MeV.
However, one should realise that the precise position of this resonance pole
is extremely sensitive to the employed parameters, in particular $\lambda$
and $a$. As a matter of fact, using the flavour-symmetry \cite{beveren99b}
arguments underlying the Nijmegen model \cite{nijmegen80,nijmegen83}, $\lambda$
and $a$ should scale as \cite{beveren04b} 
\begin{equation}
\lambda\sqrt{\mu_{q_1q_2}} = C_1 \; , \;\; a\sqrt{\mu_{q_1q_2}} = C_2 \; ,
\label{scaling}
\end{equation}
where $\mu_{q_1q_2}$ is the reduced mass of a meson composed of quarks $q_1$
and $q_2$, and the constants $C_1$, $C_2$ are fixed by
$\lambda=0.75$~GeV$^{-3/2}$ and $a=3.2$~GeV$^{-1}$ for the scalar $n\bar{s}$
system (see above), with $m_n=406$~MeV and $m_s=508$~MeV \cite{nijmegen83}.
When the values of $\lambda$ and $a$ resulting from Eq.~(\ref{scaling}) are 
used to determine the lowest scalar $c\bar{s}$ and $c\bar{n}$ poles, we find
\cite{beveren04b}
\begin{equation}
c\bar{s}: (2.33+i0)\;\mbox{GeV}\;,\;\;c\bar{n}: (2.14-i0.16)\;\mbox{GeV}\;,
\label{cscnscaling}
\end{equation}
which is a further improvement as compared to Eq.~(\ref{cscn}), especially for
the $D_0^\star(2300)$ resonance. In a more complete model calculation
\cite{beveren06d}, with all relevant pseudoscalar-pseudoscalar and
vector-vector channels included, the $S$-wave $D\pi$ cross section
representing $D_0^\star(2300)$ peaks close to 2.19~GeV (also see below).

In Fig.~6 we depict the complex-energy pole trajectories 
\begin{figure}[!t]
\label{ds2317}
\includegraphics[trim = 10mm 155mm 10mm 10mm,clip,width=18cm,angle=0]
{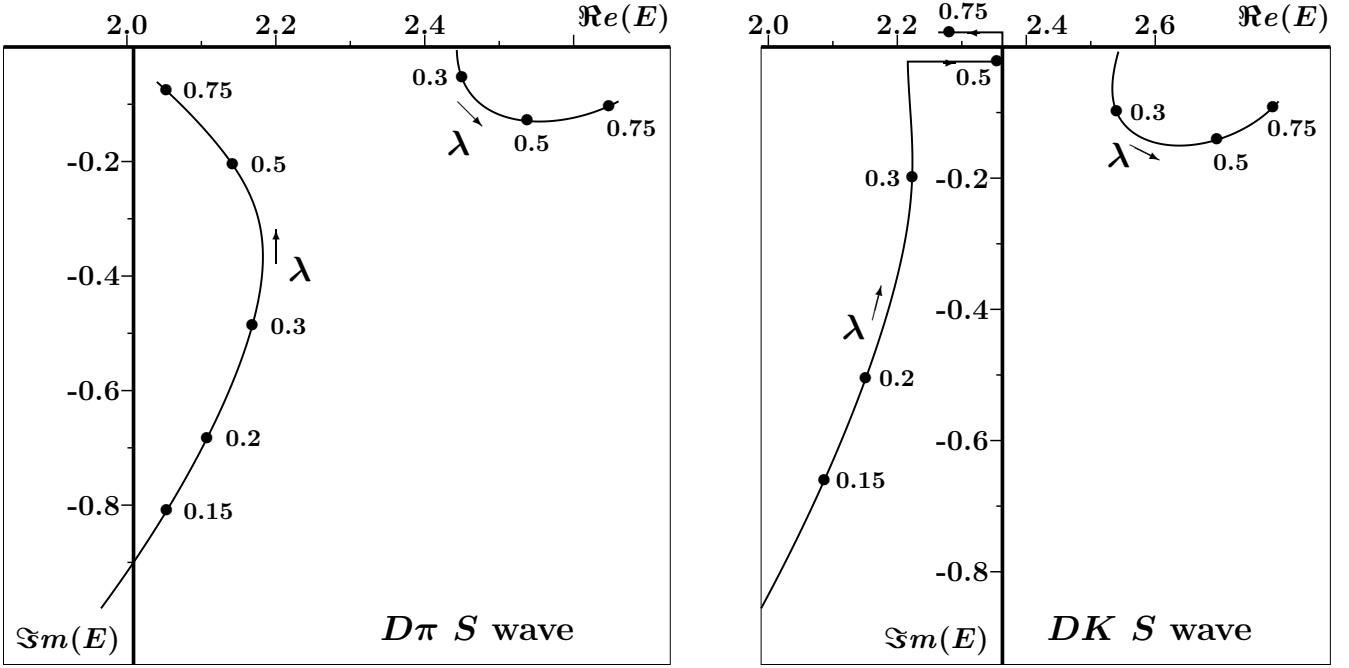}
\caption{$D_0^\star(2300)$ (left) and $D_{s0}^\star(2317)$ (right) pole
trajectories as a function of $\lambda$ (also see text and
Ref.~\cite{beveren03}).}
\end{figure}
in the amplitudes of $S$-wave $D\pi$ (left) and $DK$ (right) scattering
as a function of the coupling $\lambda$, for the simple model of
Ref.~\cite{beveren03}. In both cases we see a pole arising from deep down in
the complex-energy plane for small coupling, but then settling relatively close
to ($D_0^\star(2300)$) or on top of ($D_{s0}^\star(2317)$) the real axis for
the physical value of $\lambda$. On the other hand, for either meson, there is
a pole that starts at the ground-state bare energy level for zero coupling,
moving into the complex plane for increasing $\lambda$. So the former poles are
of a dynamical nature, whereas the latter, intrinsic ones correspond to usual
resonances that can be clearly linked to states in the bare $q\bar{q}$
confinement spectrum. These dynamical poles are of the same nature as the light
scalar mesons \cite{beveren86} and their manifestation as a broad or narrow
resonance, or even a bound state like $D_{s0}^\star(2317)$\footnote
{As already mentioned, $D_{s0}^\star(2317)$ is not really a bound state, 
because it can decay to $D_s\pi$ in an isospin-violating and OZI-suppressed
process, which gives rise to a very small width.},
depends on details of where their most relevant thresholds lie. For the latter
state, whose lowest (OZI-allowed) decay mode is $DK$, the pole ends up below
threshold, owing to the relatively large kaon mass. Note that the pole's
movement, becoming first a virtual bound state and only a true bound state
after passing through the ``eye'' of the $DK$ threshold, is typical for
$S$-waves \cite{taylor}. This is symbolically depicted in
Fig.~6
by displacing the virtual and genuine bound-state pole trajectories slightly
downwards or upwards, respectively. In contrast, the $D_0^\star(2300)$ pole
slows down while approaching the real axis, remaining complex. This is due to
the presence of a kinematical Adler-type zero \cite{rupp05} in the amplitude
just below the $D\pi$ threshold, due to the small pion mass. Similar Adler
zeros are to a large extent responsible \cite{bugg03,beveren06c} for the broad
$\sigma$ and $\kappa$ scalar resonances. 

A final comment is due concerning these dynamical poles. Namely, the only
physically relevant poles are those for parameter values that reproduce the
data and not the trajectories as a function of $\lambda$ that lead to the
eventual pole positions. Now, it turns out that these trajectories are very
sensitive to minor changes in other parameters and also to the included
meson-meson channels. For instance, in Ref.~\cite{beveren06b} it was shown
that very small variations in the delta-shell radius $a$ can make the
dynamical $D_{s0}^\star(2317)$ pole become intrinsic and vice versa (see
Fig.~4a in Ref.~\cite{beveren06b}). In other words, there can be a sudden
crossover of the two pole trajectories for certain parameter values. However,
the final pole positions hardly change with such tiny variations, so the very
definition of dynamical vs.\ intrinsic resonance can be highly questionable for
certain mesons. We shall come back to this issue when dealing with the
axial-vector charmonium state $\chi_{c1}(3872)$ in Subsecs.~\ref{x3872p},
\ref{x3872r1}, and \ref{x3872r2} below.

\subsubsection{Charmed axial-vector mesons
\boldmath$D_{s1}(2460)$, \boldmath$D_{s1}(2536)$,
\boldmath$D_{1}(2430)$, and \boldmath$D_{1}(2420)$}
\label{axials-simple}
Shortly after the discovery of $D_{s0}^\star(2317)$ by BABAR \cite{babar03},
the Belle Collaboration not only announced \cite{belle03} a confirmation of
this charmed-strange scalar meson, but also evidence of a another new 
charmed-strange state, viz.\ $D_{sJ}(2457)$, decaying to $D^\star K$ and
meanwhile confirmed as the axial-vector $D_{s1}(2460)$ \cite{PDG2020}.
Moreover, Belle had already indicated \cite{belle02} a new and very broad
charmed-light axial-vector meson, confirming its observation in
Ref.~\cite{belle04}, which is now listed \cite{PDG2020} as $D_{1}(2430)^0$.
Now, axial-vectors with open charm have no definite $C$-parity, so that
each of them gives rise to two physical $J^P=1^+$ mesons as different and 
orthogonal mixtures of the spectroscopic states \tpo\ and \spo, just like the
two strange mesons $K_1(1270)$ and $K_1(1400)$ \cite{PDG2020}. The two new
states $D_{s1}(2460)$ and $D_{1}(2430)^0$, together with the then
already established mesons $D_{s1}(2536)$ and $D_{1}(2420)$,
unfold a very puzzling pattern of axial-vector masses and widths. On the one
hand, $D_{s1}(2460)$ and $D_{s1}(2536)$ are both very narrow,
but separated in mass by 75~MeV, which is difficult to explain with
standard spin-orbit splittings. On the other hand, $D_{1}(2430)$ and
$D_{1}(2420)$ are almost degenerate in mass, but the former resonance
is extremely broad ($\Gamma\sim 300$--400~MeV), whereas the latter is
relatively narrow ($\Gamma\approx25$~MeV) and all the more so when considering
that it can decay to $D^\star\pi$ in an $S$-wave, with a phase space of about
280~MeV. Furthermore, the approximate mass degeneracy of $D_{1}(2430)$ and
$D_{1}(2420)$ destroys any attempt to explain the mass difference between
$D_{s1}(2460)$ and $D_{s1}(2536)$ with spin-orbit interactions. Clearly,
unitarity must somehow come to the rescue.

In order to describe \cite{beveren04a} these four charmed axial-vector mesons
with the above simple model and its parameters fitted to $K\pi$ $S$-wave phase
shifts, we faced the problem of needing to deal with two instead of one
$q\bar{q}$ channel, viz.\ \tpo\ and \spo. So in order to be able to apply this
model again, just like in the case of $D_{s0}^\star(2317)$ and
$D_0^\star(2300)$, we showed how a Schr\"{o}dinger system of two degenerate
$q\bar{q}$ channels coupled in the same way to one continuum state naturally
gives rise to two orthogonal solutions that either completely decouple from
the continuum or fully couple. Of course, this is an approximation in the
case of the charmed axial-vector mesons, since they couple to several continuum
channels and moreover in a different way. For instance, the \tpo\ and \spo\
$c\bar{n}$ ground states couple to the $S$-wave $D^\star\pi$ channel with
strengths $g^2_{\mbox{\scriptsize\tpo}}=1/36$ and
$g^2_{\mbox{\scriptsize\spo}}=1/72$, respectively \cite{coito11b}. However, for
the corresponding couplings to the $D$-wave $D^\star\pi$ channel, the roles are
reversed, yielding the values $g^2_{\mbox{\scriptsize\tpo}}=5/144$ and
$g^2_{\mbox{\scriptsize\spo}}=5/72$, respectively \cite{coito11b}. The
\tpo\ and \spo\ $c\bar{s}$ ground states have the same kind of relative
couplings to the $S$-wave and $D$-wave $D^\star K$ channel. Also for the other
meson-meson channels, which are all closed, there is a clear tendency for 
the several coupling strengths of the \tpo\ and \spo\ components to average
out. So we expect our argument based on diagonalising the above very simple
Hamiltonian for two degenerate states to largely hold.
The same procedure was employed in Ref.~\cite{beveren04a} to rather
successfully model the tiny mass difference between $K_L$ and $K_S$, besides
predicting axial-vector resonances in the open-bottom sector.

When applying this approach to the pair $D_{s1}(2460)$ and $D_{s1}(2536)$, the
latter state just remains as a continuum bound-state at its bare energy level
of 2545~MeV resulting from the fixed model parameters of Ref.~\cite{nijmegen83},
whereas the pole of the former state first moves downwards into the
complex-energy plane before settling as a bound state at 2446~MeV on the real
axis, below the $D^\star K$ threshold (see Fig.~3 in Ref.~\cite{beveren04a}).
The resulting small deviations of -14~MeV and +9~MeV from the experimental
$D_{s1}(2460)$ and $D_{s1}(2536)$ masses, respectively, are insignificant 
in view of this parameter-free prediction. In the charmed-light case, the
continuum bound state $D_1(2420)$ stays at 2443~MeV, i.e., 20 MeV higher
than in experiment \cite{PDG2020}, while the resonance pole settles with
a real part of about 2.3~GeV and an imaginary part of roughly 0.1~GeV,
being again very sensitive to small parameter changes. Clearly, the used
model is too coarse to describe very broad resonances. In particular,
the inclusion of additional meson-meson channels may significantly alter
the trajectories of such poles. In Subsec.~\ref{axials} we shall come back
in detail to these four charmed axial-vector mesons, in the context of the
full RSE model.
\subsubsection{Replacing the infinite RSE sum by channel recoupling
coefficients}
The way we have so far replaced the infinite sum over the confinement wave
function at the string-breaking distance $a$ in Eq.~(\ref{cotgd}) was useful
in the considered cases, but is not general enough to describe a large
variety of meson resonances. Namely, the few constants replacing the infinite
sum as determined by fitting the $S$-wave $K\pi$ phase shifts only make sense
when dealing with similar situations, namely coupling essentially only one
$q\bar{q}$ channel to one dominant $S$-wave meson-meson channel. On the other
hand, the values of $\lambda$ and $a$ resulting from the $S$-wave $K\pi$ fit
cannot be applied to other systems, as the $P$-wave $K\pi$ already showed,
by leading to a quite different value of $a$. So a scheme is needed to replace
the --- in principle --- infinite number of real constants in the RSE sum by
other numbers that make sense physically. At this point we recall the 
microscopic formalism \cite{beveren84} developed by one of us (EvB), which
amounts to deriving the recoupling constants of a meson with any
quantum numbers, as well as of its radial and angular excitations, to all
OZI-allowed two-meson decay channels via ground-state \tpz\ pair creation,
while conserving total spin $J$, parity $P$, and (when applicable) $C$-parity.
It is based on overlaps of wave functions for the original $q\bar{q}$ pair,
the created \tpz\ pair, and the two resulting mesons, in a harmonic-oscillator
(HO) basis. These recouplings for the lowest radial quantum number have already
been used in the coordinate-space model \cite{beveren86} for the light scalar
mesons. In this calculation, the Schr\"{o}dinger equation ensures that higher
radially excited solutions for the $q\bar{q}$ wave functions have smaller and
smaller effective couplings to the decay channels. However, in the RSE model
formulated in momentum space, these wave functions only enter through the
moduli of their values at the radius $a$, which \em a priori \em \/are not
known. Nevertheless, it makes perfect sense to replace these real numbers by
the recouplings constants determined with the HO overlap method of
Ref.~\cite{beveren84}, which then automatically provides a suppression of
higher radial excitations just like in the coordinate-space approach. Moreover,
the standard normalisation of solutions of the Schr\"{o}dinger equations is
reflected in the normalised HO wave functions used in Ref.~\cite{beveren84},
which ensures that the sum of the squares of the recouplings to all allowed
decay channels is equal to 1. Therefore, mass shifts resulting from coupling
an increasing number of decay channels will never grow indefinitely and will
even tend to rapidly converge in practice.

Thus, we can write down \cite{beveren06b} the partial-wave $T$-matrix for the
still quite simple case of one $q\bar{q}$ channel coupled to $N$ meson-meson
channels yet all with the same radial dependence of the corresponding
recoupling constants as
\begin{equation}
\left[ T_{\ell}\right]_{ij}(p)\; =\;
\frac
{
\lambda^{2}\;
\left\{
\dissum{n=0}{\infty}
\frac{r_{i}(n)r_{j}(n)}{\sqrt{s}-E_{n\ell}}\;
\right\}\;
2a
\sqrt{\frac{\mu_{i}\mu_{j}}{p_{i}p_{j}}}\;
p_{i}p_{j}
j_{\ell_{i}}\left( p_{i}a\right)j_{\ell_{j}}\left( p_{j}a\right)
}
{
1+\lambda^{2}
\dissum{m=1}{N}
\left\{
\dissum{n=0}{\infty}
\frac{\abs{r_{m}(n)}^{2}}{\sqrt{s}-E_{n\ell}}\;
\right\}\;
2ia\mu_{m}p_{m}
j_{\ell_{m}}\left( p_{m}a\right)
h^{(1)}_{\ell_{m}}\left( p_{m}a\right)
}
\; ,
\label{Tij}
\end{equation}
where $j_{\ell_i}$ is the usual spherical Bessel function for channel $i$,
$h^{(1)}_{\ell_i}$ is the corresponding spherical Hankel function of the
first kind, defined by
$h^{(1)}_{\ell_i}=j_{\ell_i}+in_{\ell_i}$, with $n_{\ell_i}$ the spherical
Neumann function, and the coefficient $r_i(n)$ is the recoupling constant of
the $n$-th bare radial $q\bar{q}$ state to the $i$-th decay channel, as
determined with the formalism of Ref.~\cite{beveren84}. As an example,
we show in Table~\ref{scalar_recoupling} the recouplings of
\begin{table}[!h]
\begin{center}
\begin{tabular}{||c|c|c|c||}
\hline\hline
&&& \\[-4mm]
{\bf Meson 1} & {\bf Meson 2} & {\bf Relative} & {\bf Recoupling Coefficients}\\
\hline
&&& \\[-4mm]
$(nj\ell s)_{1}$ $\left( j^{PC}\right)$ & $(nj\ell s)_{2}$
$\left( j^{PC}\right)$ & $LS$ & $\{ r(n)\}^{2}\times 4^{n}$ \\[0.5mm]
\hline\hline & & & \\[-1mm]
$\left( 0,0,0,0\right)$ $\left( 0^{-+}\right)$ &
$\left( 0,0,0,0\right)$ $\left( 0^{-+}\right)$ & $0,0$
& $\frac{1}{24}(n+1)$\\ [10pt]
$\left( 0,0,0,0\right)$ $\left( 0^{-+}\right)$ &
$\left( 1,0,0,0\right)$ $\left( 0^{-+}\right)$ & $0,0$
& $\frac{1}{144}(2n+3)(n-1)^{2}$\\ [10pt]
$\left( 1,0,0,0\right)$ $\left( 0^{-+}\right)$ &
$\left( 1,0,0,0\right)$ $\left( 0^{-+}\right)$ & $0,0$
& $\frac{1}{3456}n(2n+1)(2n+3)(n-3)^{2}$\\ [10pt]
$\left( 0,0,0,0\right)$ $\left( 0^{-+}\right)$ &
$\left( 0,1,1,1\right)$ $\left( 1^{++}\right)$ & $1,1$
& $\frac{1}{6}$\\ [10pt]
$\left( 0,1,0,1\right)$ $\left( 1^{--}\right)$ &
$\left( 0,1,0,1\right)$ $\left( 1^{--}\right)$ & $0,0$
& $\frac{1}{72}(n+1)$\\ [10pt]
$\left( 0,1,0,1\right)$ $\left( 1^{--}\right)$ &
$\left( 0,1,0,1\right)$ $\left( 1^{--}\right)$ & $2,2$
& $\frac{1}{18}(2n+5)$\\ [10pt]
$\left( 0,1,0,1\right)$ $\left( 1^{--}\right)$ &
$\left( 1,1,0,1\right)$ $\left( 1^{--}\right)$ & $0,0$
& $\frac{1}{432}(2n+3)(n-1)^{2}$\\ [10pt]
$\left( 0,1,0,1\right)$ $\left( 1^{--}\right)$ &
$\left( 0,1,2,1\right)$ $\left( 1^{--}\right)$ & $0,0$
& $\frac{1}{540}(2n+3)(2n-5)^{2}$\\ [10pt]
$\left( 0,1,0,1\right)$ $\left( 1^{--}\right)$ &
$\left( 0,1,1,0\right)$ $\left( 1^{+-}\right)$ & $1,1$
& $\frac{1}{6}$\\ [10pt]
$\left( 0,0,1,1\right)$ $\left( 0^{++}\right)$ &
$\left( 0,0,1,1\right)$ $\left( 0^{++}\right)$ & $0,0$
& $\frac{1}{432}(2n+3)(n-3)^{2}$\\ [10pt]
$\left( 0,1,1,1\right)$ $\left( 1^{++}\right)$ &
$\left( 0,1,1,1\right)$ $\left( 1^{++}\right)$ & $0,0$
& $\frac{1}{144}(2n+3)(n-2)^{2}$\\ [10pt]
$\left( 0,1,1,0\right)$ $\left( 1^{+-}\right)$ &
$\left( 0,1,1,0\right)$ $\left( 1^{+-}\right)$ & $0,0$
& $\frac{1}{144}(2n+3)(n-1)^{2}$\\ [10pt]
\hline\hline
\end{tabular}
\end{center}
\caption{Recoupling coefficients as a function of radial excitation $n$,
for scalar ($J=0$, $\ell =1$, $s=1$, $n$) decay into two mesons.
Note that the squared recoupling coefficients for $n=0$ add up to one.
This means that, in the harmonic-oscillator approach,
there are no additional two-meson channels that can couple to
the ground state of the confinement spectrum.
For the higher radial excitations, the table is still very incomplete.}
\label{scalar_recoupling}
\end{table}
generic scalar meson to its allowed meson-meson channels. Note that these
constants still must be multiplied by $SU(3)_{\mbox{\scriptsize flavour}}$
relative couplings. For detailed cases, see Refs.~\cite{beveren99a,beveren99b}.
Let us here just pick two typical examples from Table~\ref{scalar_recoupling},
as illustrations. First, we see in row no.~6 a decay to two ground state vector
mesons in a $D$-wave and with total intrinsic spin $S=2$, though total $J=0$
of course. This process has a relatively large recoupling constant.
Another example is row no.~3, with a decay to two radially excited pseudoscalar
mesons ($n=1$) with a very small recoupling. Note that all coefficients in the
fourth column still must be divided by $4^n$, with $n$ the radial quantum number
of the bare scalar state.

\subsubsection{Light scalar mesons in a multichannel RSE approach}
\label{light-scalars}
The first detailed application \cite{beveren06c} of the multichannel RSE
formalism as given by Eq.~(\ref{Tij}) was a fit to $S$-wave $\pi\pi$ and $K\pi$
phase shifts as well as $f_0(980)$ and $a_0(980)$ line shapes, by including
all pseudoscalar-pseudoscalar channels for the different scalar mesons. A
further extension of the model was required, in order to deal with the coupled
$\sigma$-$f_0(980)$ system by taking into account two instead of only one
quark-antiquark channels, viz.\ $n\bar{n}$ and $s\bar{s}$. The corresponding
expressions for the $T$-matrix became much more complicated than in
Eq.~(\ref{Tij}) (see Eqs.~(4,5) in Ref.~\cite{beveren06c}), but
exact $T$-matrix unitarity is still guaranteed, as the dependence of the
recouplings on the radial quantum number is the same in all cases. Since the
only free fit parameters were $\lambda$ and $a$, with the quark masses and bare
energy levels fixed at the values from Ref.~\cite{nijmegen83}, a very accurate
description of the experimental data was not to be expected. Nevertheless,
all essential features of the mentioned phase shifts and line shapes were
reproduced for $\lambda$ and $a$ values varying at most about 10\% about
the averages for the cases $\sigma$ only, $\kappa$, $\sigma$ coupled to
$f_0(980)$, and $a_0(980)$. Also, the found scalar-meson poles came out fully
compatible with contemporary and also present-day PDG \cite{PDG2020} limits. 

In Table~\ref{scalar-poles} we illustrate the dynamical nature of these
\begin{table}[!t]
\begin {center}
\begin{tabular}{|r|c|c|c|c||}
\hline\hline\\[-4mm]
$\lambda$ & $\sigma$ & $\kappa$ & $f_0(980)$ & $a_0(980)$\\
\hline\hline \\[-4mm]
 1.5 &  942 - i794 &     ---     &     ---     &     ---      \\
 2.0 &  798 - i507 &     ---     &     ---     &     ---      \\
 2.2 &  738 - i429 &  791 - i545 &     ---     & 1081 - i8.0  \\
 2.4 &  682 - i368 &  778 - i472 &     ---     & 1051 - i25   \\
 2.6 &  633 - i319 &  766 - i409 & 1041 - i13  & 1024 - i45   \\
 2.8 &  589 - i278 &  754 - i355 & 1028 - i26  &  998 - i61   \\
 3.0 &  549 - i243 &  743 - i309 & 1015 - i35  &  978 - i60   \\
 3.5 &  468 - i174 &  717 - i219 &  976 - i37  &  896 - i142  \\
 4.0 &  404 - i123 &  693 - i155 &  948 - i38  &  802 - i103  \\
 5.0 &  308 - i50  &  651 - i69  &  889 - i34  &  711 - i40   \\
 7.5 &  216 + i0   &  610 + i0   &  752 - i25  &  632 + i0    \\
10.0 &  142 + i0   &  560 + i0   &  633 - i17  &  577 + i0    \\
\hline\hline
\end{tabular}
\end {center}
\caption{Movement of the $\sigma$, $\kappa$, $f_0(980)$, and $a_0(980)$
poles as the coupling constant $\lambda$ is varied. Bound states are
indicated by ``+i0''. Units are MeV for the poles and GeV$^{-3/2}$ for
$\lambda$. For further details, see Ref.~\cite{beveren06c}.}
\label{scalar-poles}
\end{table}
scalar-meson poles by presenting a number of pole positions for small, medium,
and large couplings. The physical poles lie at the complex energies for
$\lambda$ values close to 3.0~GeV$^{-3/2}$ in all cases. The $\sigma$ and
$\kappa$ poles clearly originate in the scattering continuum and only turn 
into bound-state poles for very large values of $\lambda$, below the $\pi\pi$
and $K\pi$ thresholds, respectively. This slowing
down of either pole is due to the kinematical Adler-type zero just below
the corresponding lowest threshold, as we have seen above in the case of
the $D_0^\star(2300)$ resonance. As for the $a_0(980)$ second-sheet pole,
it appears to be strongly attracted by the $K\bar{K}$ threshold, but becoming
again a bound state below the $\eta\pi$ threshold for very large coupling.
For small values of $\lambda$, it disengages from the $K\bar{K}$ threshold,
moving further upwards as a more and more unphysical second-sheet pole.
The behaviour of the $f_0(980)$ pole is similar, although now an extremely
large value of $\lambda$ is needed to make it turn it into a bound-state pole
below the $\pi\pi$ threshold. This is probably due to repulsion by the
$\sigma$ pole.

\subsubsection{Multichannel RSE prediction of radial excitations of
\boldmath$D_{s0}^\star(2317)$ and \boldmath$D_0^\star(2300)$}
In our simple RSE description \cite{beveren03} of $D_{s0}^\star(2317)$ and
$D_0^\star(2300)$, we observed how the lowest confinement states turned into
relatively broad ($\Gamma\sim200$~MeV) resonances as poles moving upwards in
the complex-energy plane (see
Fig.~6).
However, in the corresponding
energy region, several other meson-meson channels become relevant and may even
allow decay to occur. Therefore, a more realistic description at higher
energies requires the inclusion of several additional channels. In
Ref.~\cite{beveren06d} such a calculation was carried out, employing
Eq.~(\ref{Tij}) like in Subsubsec.~\ref{light-scalars} for $K_0^\star(700)$
and $a_0(980)$, but now also including $S$-wave and $D$-wave vector-vector
(VV) channels, besides the $S$-wave pseudoscalar-pseudoscalar (PP) channels.
In order to allow reliable predictions at higher energies, the parameter
$\lambda$ is calibrated so as to reproduce the $D_{s0}^\star(2317)$ mass,
with the parameter $a$ in the $c\bar{s}$ and $c\bar{n}$ cases separately 
related to the value 3.2~GeV$^{-1}$ found for $K_0^\star(700)$ via flavour
symmetry of the equations. Moreover, a subthreshold suppression of closed
meson-meson channels is included via the same form factor as used in 
Ref.~\cite{beveren06c}. For further details, see Ref.~\cite{beveren06d}.
Thus, apart from the $D_{s0}^\star(2317)$ bound-state, the following
resonance poles are found (all in MeV):
\begin{equation}
\mbox{\boldmath$c\bar{s}$}: (2779-i233), \; (2842-i23.6)\;; \;\;\;
\mbox{\boldmath$c\bar{n}$}: (2174-i96.4),\; (2703-i228), \; (2737-i24.0) \;.
\label{cscn-radial}
\end{equation}
Starting with the charmed-strange scalars, the calibrated $D_{s0}^\star(2317)$
pole now originates in the lowest $c\bar{s}$ state at 2545~MeV, as was
checked by slowly increasing $\lambda$ from 0 to its final value. This is 
to be contrasted with the simple RSE model calculation \cite{beveren03} in
which $D_{s0}^\star(2317)$ showed up as a dynamical pole. Conversely, the
broad $c\bar{s}$ resonance found here at $(2779-i233)$~MeV now turns out
to be a dynamical state, whereas the nearby resonance at a mass of about 
2.79~GeV seen in Ref.~\cite{beveren03} corresponded to the ground state
of the confinement spectrum but pushed to higher energies (see Fig.~2 in
Ref.~\cite{beveren03}). As for the $c\bar{s}$ pole at $(2842-i23.6)$~MeV,
it is simply linked to the first radial excitation in the bare
spectrum.

Coming now to the $c\bar{n}$ resonances, we see that the
$D_0^\star(2300)$ pole, with peak mass near 2.19~GeV
\cite{beveren06d}, improves as compared to the values found in
Refs.~\cite{beveren03,beveren04b}, albeit still lying roughly 100~MeV
too low as for its real part. This is not very worrying in view of the
sensitivity of this pole to details of the model's parameters, besides
the problem of reliably extracting mass and width of a very broad
resonance from the data employing Breit-Wigner parametrisations (see our
above discussion on this issue). Here, it is worthwhile to note that this
resonance is now called $D_0^\star(2300)$ by the PDG \cite{PDG2020},
whereas in prior PDG editions it was designated as $D_0^\star(2400)$.
The next $c\bar{n}$ resonance, with pole at $(2703-i228)$~MeV, 
is the bare confinement ground state yet displaced by 
the dynamics to much higher energies and acquiring a large imaginary 
part. This pole can be compared to the one found in Ref.~\cite{beveren03}
at about 2.64~GeV, which was also of an intrinsic nature. Finally, the
pole at $(2737-i24.0)$~MeV comes again from the first radial excitation
in the bare scalar $c\bar{n}$ spectrum.

Focusing now on the $c\bar{s}$ pole at $(2842-i23.6)$~MeV, when we
presented \cite{rupp07} our above findings, in particular on this
$D_{s0}^\star$ resonance with a peak mass of 2847 and a width of 47~MeV,
we were informed about a new charmed-strange state
decaying to $DK$, called $D_{sJ}(2860)^+$, with mass 2857~MeV and width
47~MeV, observed by the BABAR Collaboration and later published in
Ref.~\cite{babar06}. Naturally, we associated our prediction with the
BABAR observation. However, a few years later the BABAR Collaboration
published \cite{babar09b} new results on $D_{sJ}(2860)$, most notably the
observation of $D^\star K$ decays, which would exclude a scalar assignment.
In a Comment \cite{beveren10a} on this paper, we argued on the basis of
e.g.\ branching ratios of \ttpt\ and \otft\ $c\bar{s}$ states decaying to $DK$
and $D^\star K$, as well as an inevitable mixing of \ttpt\ with \otft, that
the observed $D_{sJ}(2860)$ structure may correspond to overlapping
$J^P=0^+$ and $J^P=2^+$ resonances. Then, in 2012, the LHCb Collaboration
again confirmed \cite{lhcb12} $D_{sJ}(2860)$, with unknown $J^P$, as decaying
to $DK$ and $D^\star K$, but now with a slightly increased mass of 2866~MeV and
a very significantly larger width of 70~MeV. Two years later, the same LHCb
Collaboration claimed \cite{lhcb14}, from a Dalitz-plot analysis, that
the $D_{sJ}(2860)$ structure is after all composed of two overlapping
resonances, namely a $D_{s1}^\star$ $J^P=1^-$ state with mass 2859~MeV and
width 159~MeV and a $D_{s3}^\star$ $J^P=3^-$ state with mass 2860.5~MeV and
width 53~MeV. These are the states listed as such in the PDG tables
\cite{PDG2020}. In Ref.~\cite{beveren15a} we argued, on the basis of
structures resulting from threshold enhancements due to the opening of the
$DK^\star$ and $D^\star K^\star$ channels, that the $D_{sJ}(2860)$ 
bump may hide several much narrower states than the extracted \cite{lhcb14}
surprisingly broad ($\Gamma=159$~MeV) $D_{s1}^\star$ resonance and possibly
with completely different $J^P$ assignments. We shall come back to threshold
enhancements in Sec.~\ref{production}.

\subsection{Fully unitary multichannel RSE model in momentum space}
\label{rse2}
The most straightforward way to derive the general RSE expression for
the $T$-matrix describing a system of several quark-antiquark channels
coupled to an arbitrary number of meson-meson (MM) channels is by realising
that the thus constructed effective meson-meson potential is separable. Namely,
we do not consider $t$-channel exchanges between the two scattered mesons, 
only $s$-channel exchanges amounting to towers of bare $q\bar{q}$ states,
which resemble \cite{beveren09a} Regge propagators. Because of the separability
of the MM potential, the Lippmann-Schwinger equation for the $T$-matrix can be
solved in closed form, even for energy-dependent potentials, which is the case
here. Graphically we can depict the Born series for the $T$-matrix as
\cite{beveren09a}
\begin{center}
\begin{tabular}{ccccc}
\raisebox{8.5mm}{$T\;\;\;=$} &
\includegraphics[trim = 15mm 0mm 0mm 0mm,clip,height=2.0cm,angle=0]
{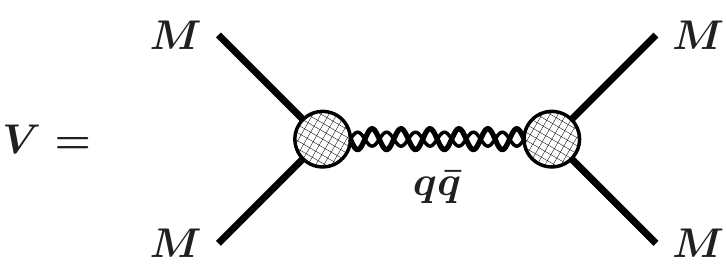} &
\raisebox{8.5mm}{$+$} &
\includegraphics[trim = 24mm 0mm 0mm 0mm,clip,height=2.0cm,angle=0]
{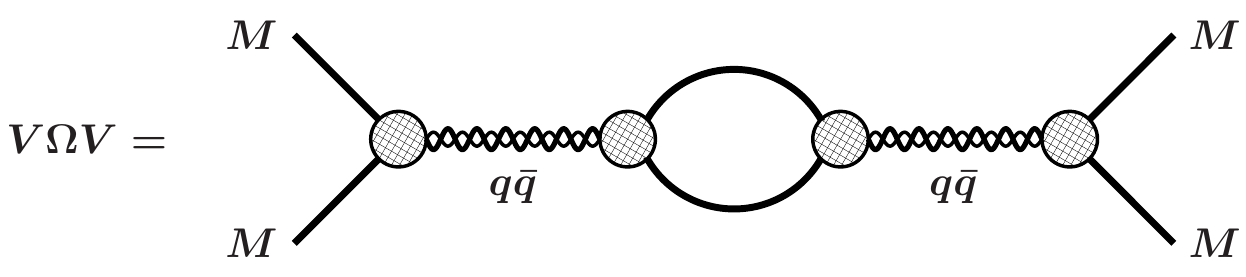} &
\raisebox{8.5mm}{$+\;\;\;\ldots$ \,.} 
\end{tabular} 
\end{center}
Here, the first diagram on the right-hand side stands for the effective MM
potential $V$ generated by the MM$q\bar{q}$ vertices and the RSE $q\bar{q}$
propagator in the intermediate state. The wiggly lines indicate that this
is not just one $q\bar{q}$ state but a whole tower, consisting in principle
of an infinite number
of $q\bar{q}$ states with the same quantum numbers, The second diagram is
the once-iterated $V$, with an MM loop in between, and so the second term
in the Born series. The whole series can be summed up easily, giving rise
to a closed-form expression for the $T$-matrix. The explicit momentum-space
formulae for $V$ read \cite{coito09}
\begin{equation}
V_{ij}^{L_i,L_j}(p_i,p'_j;E)=
\lambda^2j^i_{L_i}(p_ia)\,\mathcal{R}_{ij}(E)\,j^j_{L_j}(p'_ja) \;,
\label{inter}
\end{equation}
\begin{equation}
\mathcal{R}_{ij}(E)=\sum_{i_{q\bar{q}}=1}^{N_{q\bar{q}}}\sum_{n=0}^{\infty}
\frac{g^i_{(i_{q\bar{q}},n)}g^j_{(i_{q\bar{q}},n)}}{E-E_n^{(i_{q\bar{q}})}}\;,
\label{rsev}
\end{equation}
where the RSE propagator $\mathcal{R}$ contains an infinite tower of
$s$-channel bare $q\bar{q}$ states, corresponding to the discrete spectrum
of an arbitrary confining potential. Here, $E_n^{(i_{q\bar{q}})}$ is the energy
level of the $n$-th recurrence in the $i_{q\bar{q}}$-th $q\bar{q}$ channel,
with $N_{q\bar{q}}$ the number of $q\bar{q}$ channels having the same quantum
numbers, and $g^i_{(i_{q\bar{q}},n)}$ is the corresponding coupling to the
$i$-th MM channel. Furthermore, in Eq.~(\ref{inter}), $\lambda$ is the overall
coupling constant for \tpz\ decay (now with dimensions GeV$^{1/2}$), and
$\bes{i}(p_i)$ and $p_i$ are the
$L_i$-th order spherical Bessel function and the (relativistically defined)
off-shell relative momentum in MM channel $i$, respectively. The spherical
Bessel function originates in our string-breaking picture of OZI-allowed decay,
being just the Fourier transform of a spherical delta-shell of radius $a$.
The channel couplings $g^i_{(i_{q\bar{q}},n)}$ in Eq.~(\ref{rsev}) are computed
following the formalism developed in Ref.\cite{beveren84}, namely from
overlaps of HO wave functions for the original $q\bar{q}$ pair, the created
\tpz\ pair, and the quark-antiquark states corresponding to the outgoing
two mesons. In most cases, this method produces the same couplings for
ground-state mesons as the usual point-particle approaches, but also
provides a clear prescription for excited mesons, with the additional
advantage of always resulting in a finite number of non-vanishing couplings.
Because of their fast decrease for increasing radial quantum number $n$,
practical convergence of the infinite sum in Eq.~(\ref{rsev}) is achieved by
truncating it after at most 20 terms.

With this effective energy-dependent MM potential, the $T$-matrix reads
explicitly\cite{coito09}
\begin{eqnarray}
\lefteqn{\tmat{i}{j}(p_i,p'_j;E)=-2a\lambda^2\sqrt{\mu_ip_i}\,\bes{i}(p_ia)
\times} \nonumber \\
&&\hspace*{-1pt}\sum_{m=1}^{N}\rse_{im}\left\{[\One-\Omega\,\mathcal{R}]^{-1}
\right\}_{\!mj}\bes{j}(p'_ja)\,\sqrt{\mu_jp'_j} \; ,
\label{tmatrix}
\end{eqnarray}
with the loop function
\begin{equation}
\Omega_{ij}(k_j)=-2ia\lambda^2\mu_jk_j\,\bes{j}(k_ja)\,\han{j}(k_ja)\,
\delta_{ij}\;, 
\label{loop}
\end{equation}
where $\han{j}(k_ja)$ is the spherical Hankel function of the first kind,
$k_j$ and $\mu_j$ are the on-shell relative momentum and reduced mass in
MM channel $j$, respectively, and the matrix $\mathcal{R}(E)$ is
given by Eq.~(\ref{rsev}). Note that no regularisation is needed in this
all-orders model, since the Bessel functions at the vertices make the meson
loops finite.
The fully on-shell and unitary $S$-matrix is then given by
\begin{equation}
S_{ij}^{(L_i,L_j)}(k_i,k_j;E)\;=\;\delta_{ij}\:+\:2i
\tmat{i}{j}(k_i,k_j;E)\;.
\label{Smatrix}
\end{equation}
This RSE $S$-matrix can be used to describe a large variety of non-exotic
mesons coupling to several open and closed two-meson decay channels, even
when some of these involve resonances themselves, by applying a mathematical
procedure (see Appendix B of Ref.~\cite{coito11a}) to re-unitarise the
resulting non-unitary $S$-matrix.

\section{Understanding the charmed axial-vector mesons and
\boldmath$\chi_{c1}(3872)$}
\label{enigmatic}
\subsection{Revisiting
\boldmath$D_{s1}(2460)$, \boldmath$D_{s1}(2536)$,
\boldmath$D_{1}(2430)$, and \boldmath$D_{1}(2420)$}
\label{axials}
When describing the charmed axial-vector mesons with the basic RSE model
of Subsubsec.~\ref{axials-simple}, we were limited to only one quark-antiquark
channel. We then argued that diagonalising a simple Hamiltonian for two
degenerate $q\bar{q}$ states coupled in the same way to one continuum channel
would give rise to one eigensolution that decouples and an orthogonal one
that fully couples. This allowed to study \cite{beveren04a} $D_{s1}(2460)$
and $D_{1}(2430)$ by using Eq.~(\ref{cotgd}). Here we review results
\cite{coito11b} obtained by employing the full RSE formalism as given in
Eqs.~(\ref{rsev}--\ref{loop}). So in both the $c\bar{s}$ and $c\bar{n}$ cases,
we take two $q\bar{q}$ channels with $\ell=1$, viz.\ for the \tpo\ and \spo\
spectroscopic states, which are degenerate in energy without explicit spin-orbit
interactions.

Taking for the bare $q\bar{q}$ state the usual values of quark masses and
HO radial levels from Ref.~\cite{nijmegen83}, we optimise $\lambda$ and $a$
for the broad $D_{s1}(2430)$ resonance, whose pole is extremely sensitive to
especially the decay radius (see below). In the $c\bar{s}$ case, these
$\lambda$ and $a$ values are then scaled with the reduced quark masses, as
successfully done before in Refs.~\cite{beveren04b,beveren06d} for the charmed
scalar mesons. So we have only two relatively free parameters, albeit within
the ranges expected \cite{coito11b} from previous work, to fit the four masses
as well as the four widths of the ground-state charmed axial-vectors. The
corresponding pole positions \cite{coito11b} are given in
Table~\ref{axialcharm}, together with the 
\begin{table}[ht]
\centering
\begin{tabular}{|c|c|l|}
\hline \hline && \\[-11pt]
Quark Content & Radial Excitation & Pole in MeV \\[0.5ex]
\hline
\mbox{ }&& \\[-11pt]
$c\bar{q}$  & $0$ & $2439 - i\times3.5$     \\
$c\bar{q}$  & $0$ & $2430 - i\times191$     \\
$c\bar{s}$  & $0$ & $2540 - i\times0.7$     \\
$c\bar{s}$  & $0$ & $2452 + i\times0.0$     \\
$c\bar{q}$  & $1$ & $2814 - i\times7.8$     \\
$c\bar{q}$  & $1$ & $2754 - i\times47.2$    \\
$c\bar{s}$  & $1$ & $2915 - i\times6.7$     \\
$c\bar{s}$  & $1$ & $2862 - i\times25.7$    \\
\hline \hline
\end{tabular}
\caption{Poles of ground-state ($n\!=\!0$) and first radially excited
($n\!=\!1$) charmed axial-vector mesons. Parameters: $\lambda=$ 1.30 (1.19) 
GeV$^{1/2}$ and $r_0=$ 3.40 (3.12) GeV$^{-1}$ for $c\bar{q}$ ($c\bar{s}$)
states. For further details, see Ref.~\cite{coito11b}.}
\label{axialcharm}
\end{table}
predictions for the first radial excitations. We immediately see that the
dynamics of the RSE coupled-channel equations generates the right mixture
of the included \tpo\ and \spo\ $q\bar{q}$ components, which is simply imposed
in other approaches (see discussion in Ref.~\cite{coito11b}). Thus, on the one
hand we end up with two resonances, i.e., $D_{s1}(2536)$ and $D_1(2420)$, that
are much narrower than what could be expected from their $S$-wave decays and
available phase space, becoming quasi-bound states in the continuum. On the
other hand, we find a strongly coupling state, i.e., $D_{s1}(2460)$, shifting
so much downwards that it settles as a bound-state pole on the real axis, below
\begin{figure}[!hb]
\begin{tabular}{cc}
\includegraphics[trim = 23mm 120mm 35mm 23mm,clip,width=8.75cm,angle=0]
{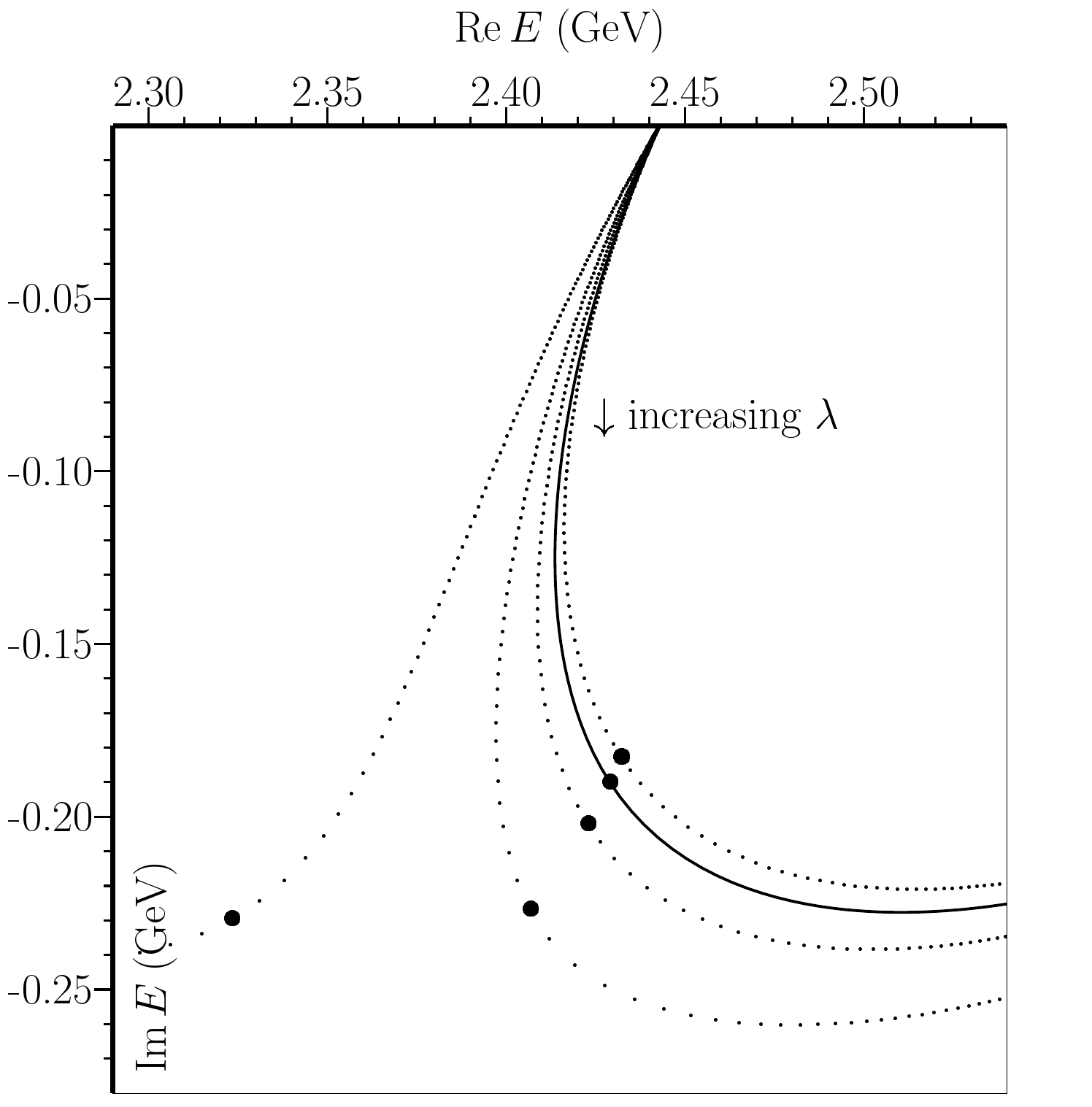}
&
\includegraphics[trim = 21mm 120mm 35mm 6mm,clip,width=9.2cm,angle=0]
{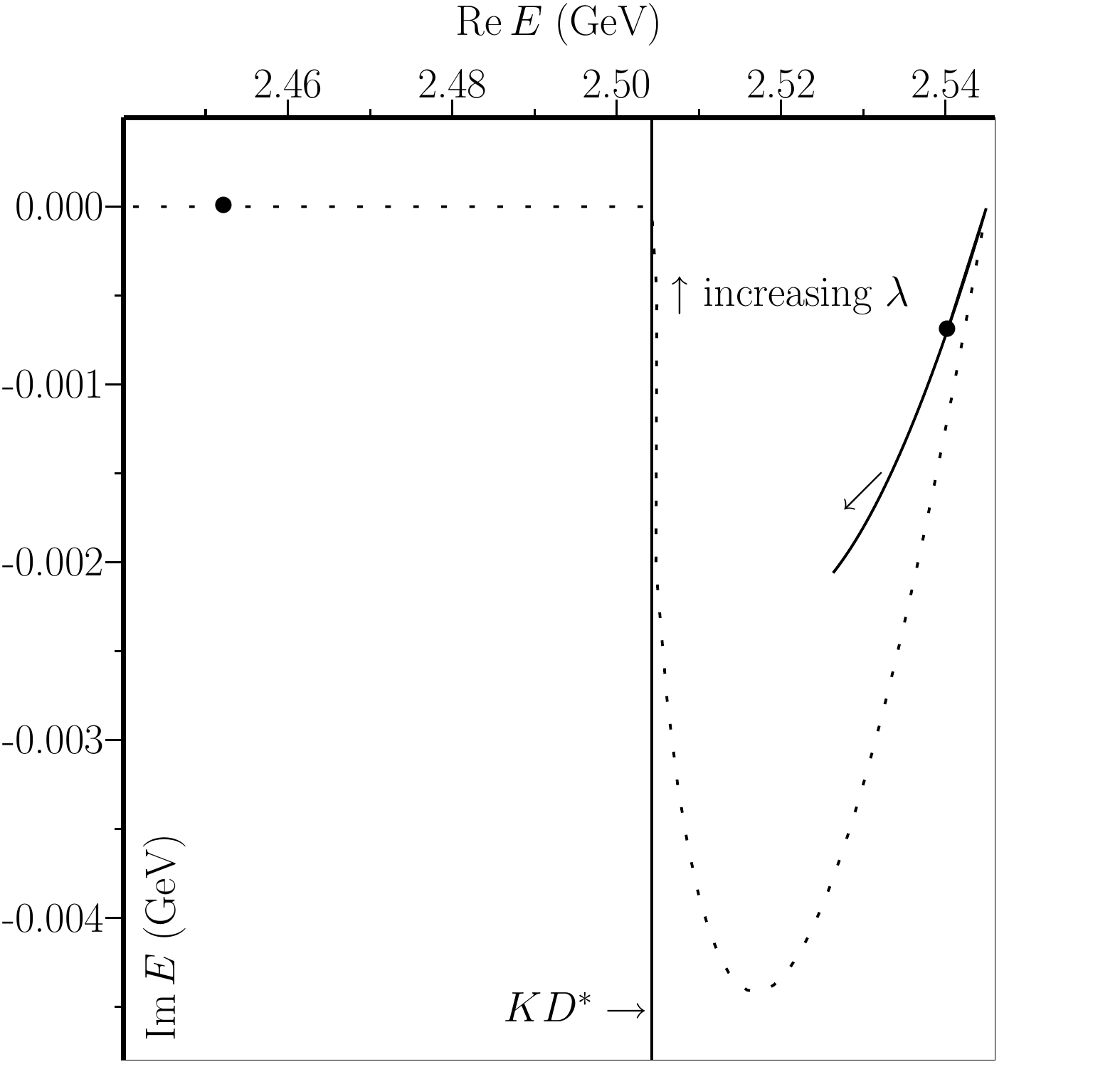} 
\end{tabular}
\caption{Left-hand plot: $D_1(2430)$ pole trajectories as a function of
$\lambda$, for $r=$ 3.2, 3.3, 3.4, and 3.5 GeV$^{-1}$ (left to right);
solid curve and bullets correspond to $r=$ 3.4~GeV$^{-1}$ and $\lambda=1.30$,
respectively.
Right-hand plot: $D_{s1}(2460)$ (left) and $D_{s1}(2536)$ (right) pole
trajectories as a function of $\lambda$ for $r=3.4$~GeV$^{-1}$; bullets
correspond to $\lambda=1.19$ (also see text and Ref.~\cite{coito11b}).}
\label{d1ds1poles}
\end{figure}
the $D^\star K$ threshold, in much the same way as $D_{s0}^\star(2317)$ with
respect to the $DK$ threshold. Finally, the other strongly coupling state,
i.e., $D_1(2430)$, becomes a very broad resonance and in a highly non-linear
fashion, with its real part first decreasing and then increasing again, at
least for the optimum parameter values. The corresponding pole trajectory as a
function of $\lambda$ is depicted in Fig.~\ref{d1ds1poles}, together with those
of $D_{s1}(2460)$ and $D_{s1}(2536)$. The behaviour of the $D_1(2430)$ pole is
again related to the Adler-type zero just below the $D^\star\pi$ threshold,
which prevents the latter channel from attracting the pole so as to become a
bound state. To conclude this topic, it is most remarkable that with only two
quite restricted parameters the very unusual pattern of masses and widths of
the charmed axial-vector mesons can be reproduced, with rather insignificant
deviations from the experimentally observed \cite{PDG2020} values. This lends
further support to the predictive power of the RSE model.

\subsection{Describing \boldmath$\chi_{c1}(3872)$ in momentum and coordinate
space}
\label{X3872}
Hundreds of papers have been published on the axial-vector charmonium-like
state $\chi_{c1}(3872)$, after its first observation \cite{x3872} in 2003 by
the Belle Collaboration, having been called $X(3872)$ for many years. The main
reason for this excitement is the extreme closeness of its mass to the
$\bar{D}^{\star0}D^0$ threshold, now within 0.01~MeV according to the
PDG average \cite{PDG2020}. Moreover, if confirmed as a $J^{PC}=1^{++}$
$c\bar{c}$ state, its mass would be about 80~MeV lower than in the GI
\cite{GI85} and similar static quark models, which is much more than any other
discrepancy found for established charmonium states in most of such models.
Another coincidence is that the $\chi_{c1}(3872)$ mass lies less than 1~MeV
below the central $\rho^0J/\psi$ threshold, which is one of the main observed
hadronic decay modes, besides e.g.\ $\omega J/\psi$ and
$\bar{D}^{\star0}D^0$. Since the $\rho^0J/\psi$ decay of $\chi_{c1}(3872)$
is both isospin violating and OZI-forbidden, its contribution to the width is
small despite the available phase space owing to the large $\rho$ width.
For a summary of different model approaches to $\chi_{c1}(3872)$, see the
more general review on hidden-charm pentaquark and tetraquark states
\cite{xiang16}.

Next we shall briefly review three different model calculation of
$\chi_{c1}(3872)$, with each one focusing on specific aspects.
\subsection{RSE modelling of \boldmath$\chi_{c1}(3872)$}
\label{x3872p}
\begin{figure}[!t]
\begin{tabular}{cc}
\includegraphics[trim = 25mm 108mm 45mm 5mm,clip,width=8.75cm,angle=0]
{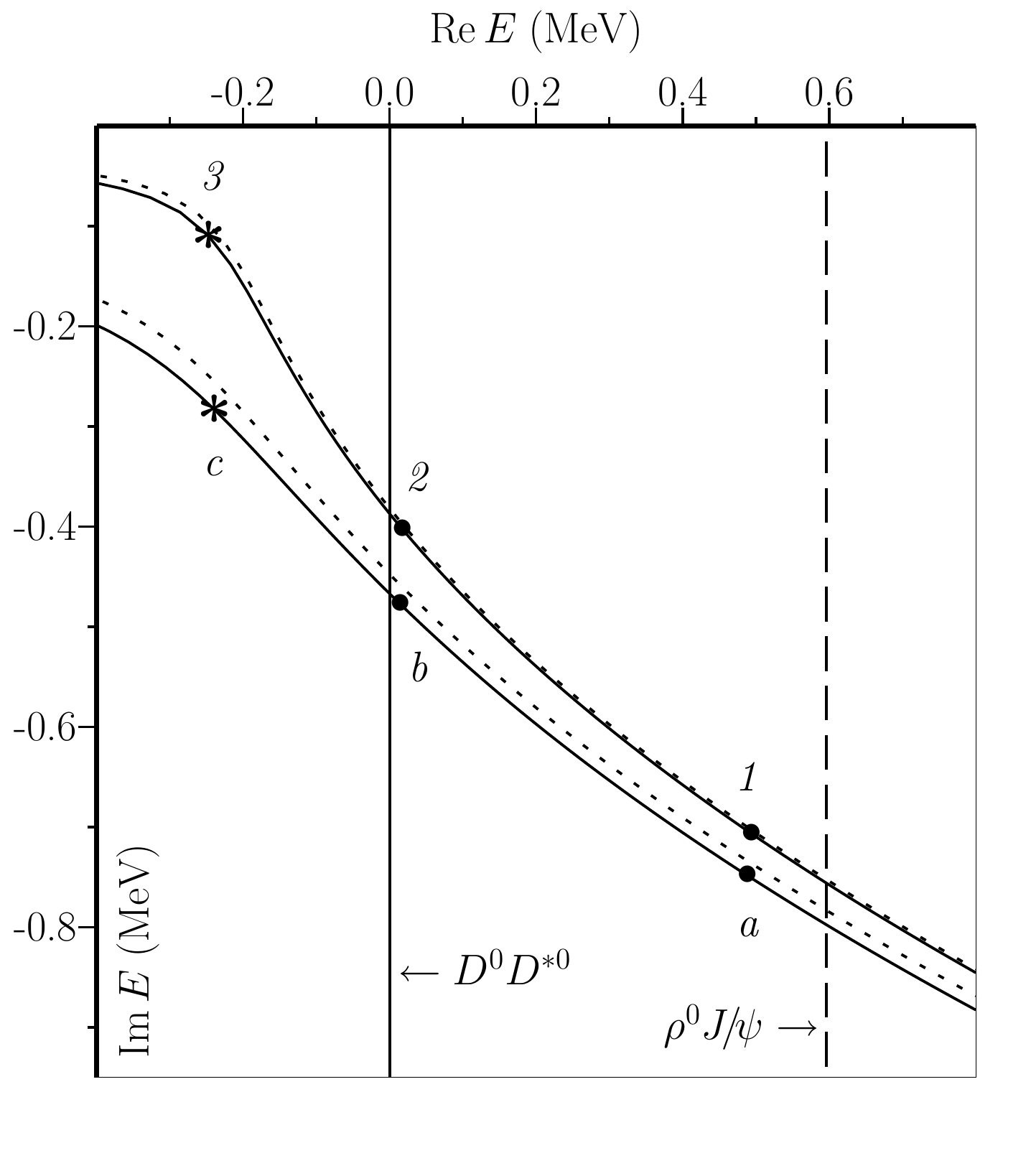}
&
\includegraphics[trim = 25mm 120mm 50mm 0mm,clip,width=8.25cm,angle=0]
{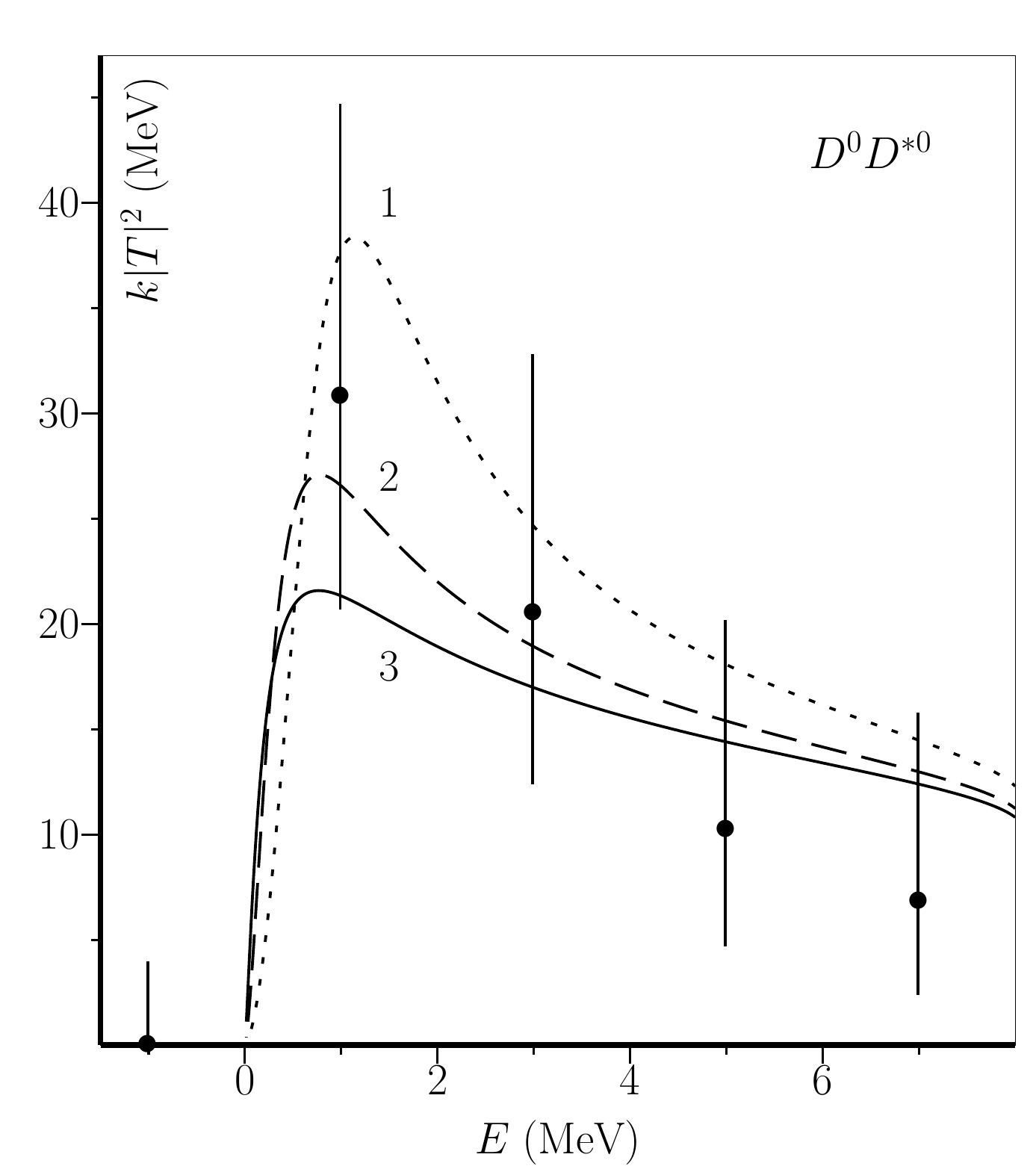}
\end{tabular}
\caption{Left: Pole trajectories of $\chi_{c1}(3872)$ as a function of
overall coupling $\lambda$; pole positions $1,2,3$ and $a,b,c$ for two
different values of OZI-forbidden coupling (see text); dotted curves -
different decay radius.
Right: corresponding $(1,2,3)$ elastic $\bar{D}^{\star0}D^0$ amplitudes with
arbitrarily normalised data.
See Ref.~\cite{coito11a} for more details.}
\label{x3872-poles}
\end{figure}

In Ref.~\cite{coito11a} the $\chi_{c1}(3872)$ state was studied by employing
the full RSE formalism, with an emphasis on the behaviour of the \ttpo\
$c\bar{c}$ pole and $\bar{D}^{\star0}D^0$ amplitude in the vicinity of the
latter threshold, and also on the effect of the OZI-forbidden decays
$\rho^0J/\psi$ and $\omega J/\psi$. So the goal was not to predict the precise
mass of $\chi_{c1}(3872)$, which would anyhow be impossible due to the needed
fine-tuning of parameters, but rather to demonstrate that this meson can be
understood as a strongly unitarised \ttpo\ charmonium state. Thus, the bare
RSE $J^{PC}=1^{++}$ spectrum is coupled to the OZI-allowed vector-pseudoscalar
channels $\bar{D}^{\star0}D^0$, $D^{\star\mp}D^\pm$, and
$D_s^{\star\mp}D_s^\pm$, allowed in both $S$- and $D$-waves, as well as the
charge-averaged vector-vector channel $\bar{D}^\star D^\star$ in $D$-waves
only. Furthermore, the already mentioned OZI-forbidden channels $\rho^0J/\psi$
(also isospin violating) and $\omega J/\psi$ are coupled as well. Upon 
increasing the overall coupling $\lambda$, the bare \ttpo\ $c\bar{c}$ pole at
3979~MeV (for $m_c=1562$~MeV and HO frequency $\omega=190$~MeV
\cite{nijmegen83}) moves about 100~MeV downwards in the complex-energy plane,
crossing the $\rho^0 J/\psi$ and $\bar{D}^{\star0}D^0$ thresholds as depicted
on the left-hand plot of Fig.~\ref{x3872-poles} for different sets of
parameters. Note that in this calculation the $\chi_{c1}(3872)$ pole will not
end up as a bound state on the real axis below the $\bar{D}^{\star0}D^0$
threshold, because the $\rho^0 J/\psi$ channel is included with a complex
$\rho^0$ mass in order to mimic the broadness of this resonance. Nevertheless,
the resulting $S$-matrix has been made unitary again with an empirical yet 
mathematical transformation \cite{coito11a} using the fact that $S$ is always
a symmetric matrix. For further details, see Ref.~\cite{coito11a}.

The pole trajectories as well as the amplitudes shown in Fig.~\ref{x3872-poles}
support the interpretation of $\chi_{c1}(3872)$ as a regular \ttpo\ $c\bar{c}$
state, but strongly influenced by the $S$-wave $\bar{D}^{\star0}D^0$ decay
channel and --- to a lesser extent --- by the OZI- and isospin-violating
$\rho^0 J/\psi$ channel.

\subsection{Coordinate-space modelling of \boldmath$\chi_{c1}(3872)$
wave function and pole}
\label{x3872r1}
In Ref.~\cite{coito13} the simple coordinate-space model of
Ref.~\cite{beveren83} was employed to find out whether it is reasonable
to call $\chi_{c1}(3872)$ a $\bar{D}^{\star0}D^0$ molecule. Thus, a
harmonically confined \tpo\ $c\bar{c}$ state is coupled to the
$\bar{D}^{\star0}D^0$ channel through a delta-shell interaction at a radius $a$
and with coupling constant $g$, similarly to the RSE modelling. The advantage
of the $r$-space formulation is that one can easily obtain the bound-state
wave function for a pole on the real axis below threshold. One can then study
how variations in the binding energy with respect to this threshold affects
the $c\bar{c}$ and $\bar{D}^{\star0}D^0$ wave-function components, in
particular their relative probabilities, besides computing the r.m.s.\ radius
of the system in the different situations.

As one of the main results, we find that for the then experimentally reported
average binding energy of 0.16~MeV, the $c\bar{c}$ probability varies as
7.48\%--11.18\% for $a=2.0$--3.0~GeV$^{-1}$, and the corresponding 
$\bar{D}^{\star0}D^0$ probability as 92.52\%--88.82\%. Nevertheless, the
$c\bar{c}$ and $\bar{D}^{\star0}D^0$ wave functions are of comparable size
in the inner region, with the $\bar{D}^{\star0}D^0$ probability only being so
large because of its very long tail, owing to the small binding energy. 
As for the r.m.s.\ radius, it is quite stable at almost 8~fm for either
value of $a$ and again a binding of 0.16~MeV. For the present PDG value of
the binding energy, the $\bar{D}^{\star0}D^0$ probability would be larger
than 99\% and the r.m.s.\ radius of the order of 100~fm, but the two
wave-function components would still be of similar size in the interior.
For further variations, see Ref.~\cite{coito13}. A more complete study
of the $\chi_{c1}(3872)$ wave function will be presented in
Subsec.~\ref{x3872r2}.

\begin{figure}[!t]
\begin{tabular}{cc}
\includegraphics[trim = 25mm 175mm 55mm 2mm,clip,width=8.75cm,angle=0]
{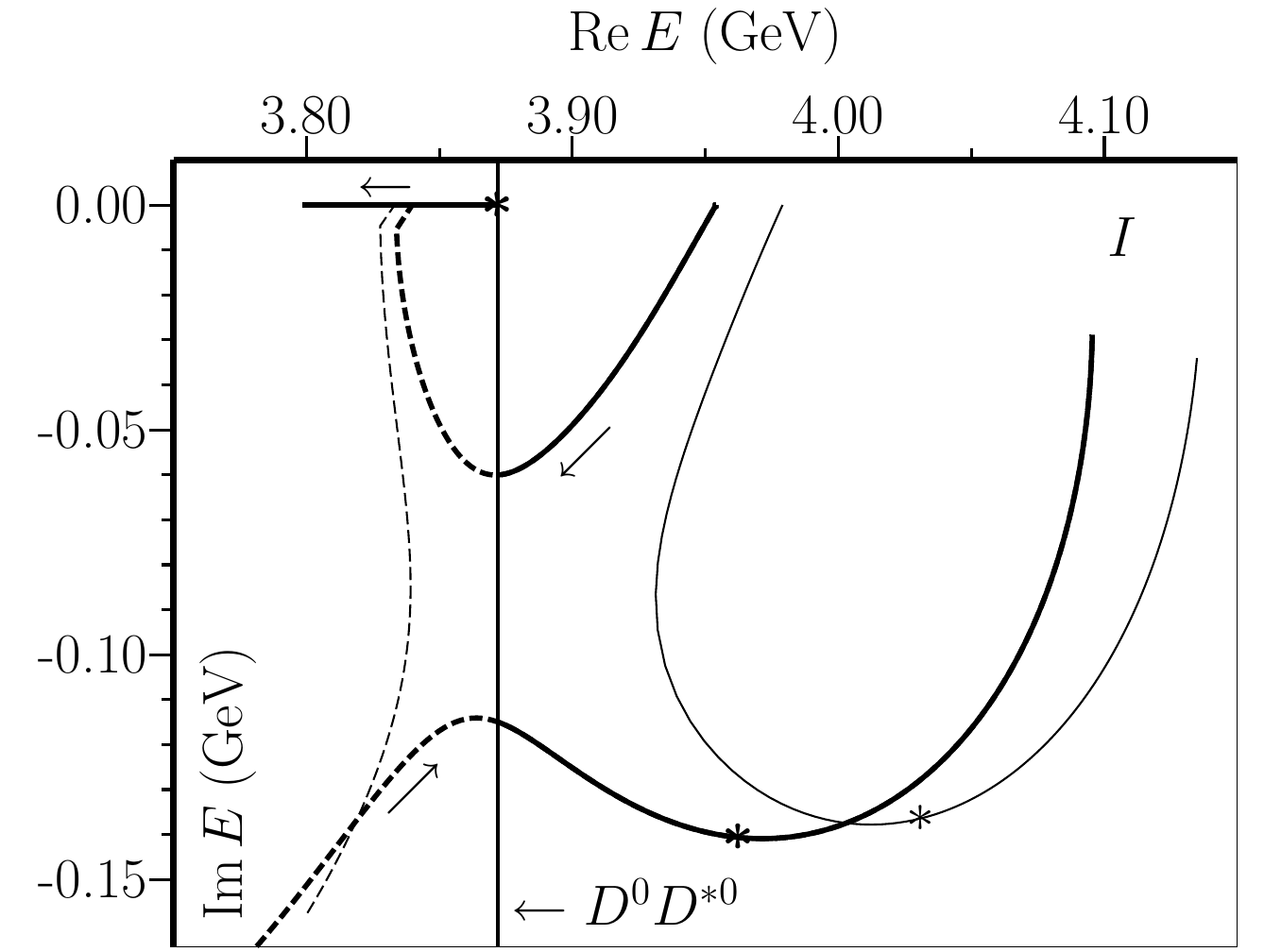}
&
\includegraphics[trim = 25mm 175mm 55mm 2mm,clip,width=8.75cm,angle=0]
{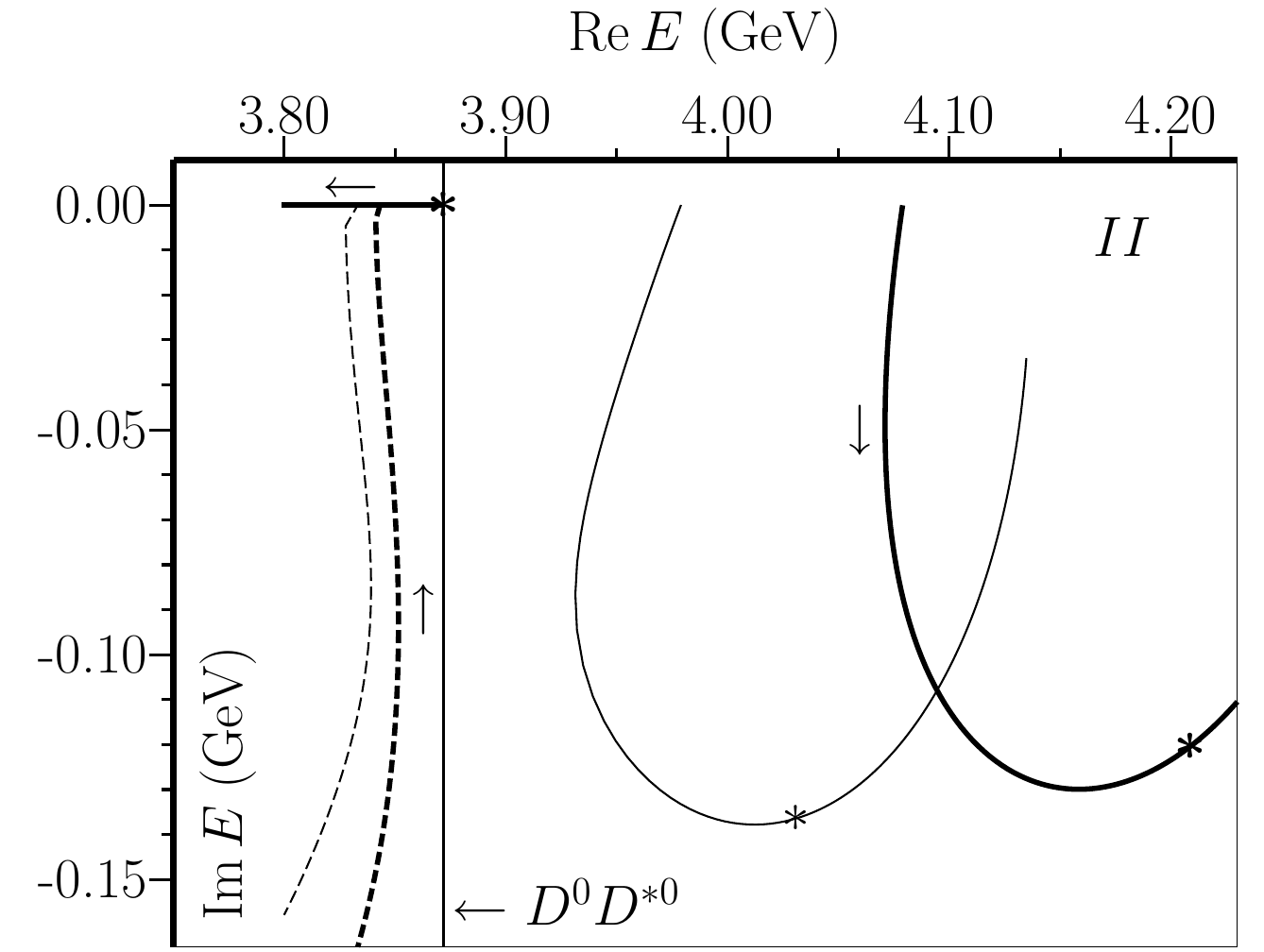}
\end{tabular}
\caption{Left: $\chi_{c1}(3872)$ pole trajectories for \ttpo\ bare $c\bar{c}$
mass at 3954~MeV (boldface) and 3979~MeV (non-boldface). Right: same but now
for bare mass at 4079~MeV (boldface) and again 3979~MeV (non-boldface). Also
see text and Ref.~\cite{coito13}.}
\label{x3872-wf_poles}
\end{figure}

The coordinate-space formalism also allows to search for $S$-matrix poles and
Fig.~\ref{x3872-wf_poles} shows how a small change in the energy of the bare
state can affect the pole trajectories. If we take the bare state at 3979~MeV
just like in Ref.~\cite{coito11a}, the $\chi_{c1}(3872)$ pole now turns out to
be dynamical instead of intrinsic (cf.\ non-boldface dashed trajectories on
both plots). At first sight, this qualitative difference might be due to our
much simpler modelling here, with only one meson-meson channel. However, if we
lower the \ttpo\ bare $c\bar{c}$ mass by only 25~MeV to 3954~MeV, which is 
almost exactly the value predicted in the GI model \cite{GI85}, the
$\chi_{c1}(3872)$ pole trajectory suddenly connects to this bare energy
level and so becomes of an intrinsic nature (upper boldface trajectory on
left-hand plot). The corresponding dynamical pole then moves close to the
intrinsic pole in the standard situation (lower boldface trajectory and
non-boldface solid trajectory, respectively, on the same plot). So we
witness a crossover of poles for small parameter variations, just as in
the case of $D_{s0}^\star(2317)$ \cite{beveren06b} and discussed above.
We must conclude that also for  $\chi_{c1}(3872)$ its assignment
as either a dynamical or intrinsic state is very problematic. The right-hand
plot in Fig.~\ref{x3872-wf_poles} displays the situation when the \ttpo\
bare $c\bar{c}$ mass is increased by 100~MeV. Although this does not represent
a very likely scenario compared to other models, it does show how insensitive
the $\chi_{c1}(3872)$ pole is to very significant changes in the confinement
spectrum (cf.\ the non-bold and bold dashed trajectories on the right-hand
plot). So our findings appear to be  quite model independent.

From the studied $\chi_{c1}(3872)$ wave function and pole trajectories we
conclude \cite{coito13} that this enigmatic meson is not a
$\bar{D}^{\star0}D^0$ molecule.
\subsection{Multichannel coordinate-space modelling of
\boldmath$\chi_{c1}(3872)$}
\label{x3872r2}
In Ref.~\cite{cardoso15} electromagnetic transition rates of $\chi_{c1}(3872)$
to $J/\psi$ and $\psi(2S)$ were calculated in a multichannel coordinate-space
model like the one employed in Refs.~\cite{nijmegen80,beveren04c}, which is
a generalisation of the model used in Subsec.~\ref{x3872r1}. In order
to ensure the derivation of realistic wave functions for the involved three
charmonium states, several open-charm decay channels that should acquire
appreciable probabilities in the total wave function are coupled to the
$c\bar{c}$ channel(s) in each case. Using an abbreviated notation in which
$D$ represents $D^0$, $D^+$, or $D_s^+$, we get for the vectors $J/\psi$ and
$\psi(2S)$ the channels $\bar{D}D$, $\bar{D}^\star D$, and
$\bar{D}^\star D^\star$ coupling in $P$-waves to the \tso\ $c\bar{c}$
wave-function component, and $\bar{D}^\star D^\star$ coupling only
in an $F$-wave to the \tdo\ $c\bar{c}$ component. In the $\chi_{c1}(3872)$
case, we must couple the $\bar{D}^{\star0}D^0$ and $D^{\star\mp} D^\pm$
channels separately, with the correct mass splitting and relative couplings, in
view of the closeness of the pole to the former threshold. Moreover, there is
now only one $c\bar{c}$ component, viz.\ \tpo, to which we couple the
mentioned $\bar{D}^{\star0}D^0$ and $D^{\star\mp} D^\pm$ channels,
as well as $D_s^{\star\mp}D_s^\pm$, in both $S$- and $D$-waves, and finally
$\bar{D}^\star D^\star$ (with $D=D^0,D^\pm,D_s$) only in $D$-waves.

\begin{figure}[!t]
\begin{center}
\includegraphics[trim = 0mm 0mm 0mm 0mm,clip,width=18cm,angle=0]
{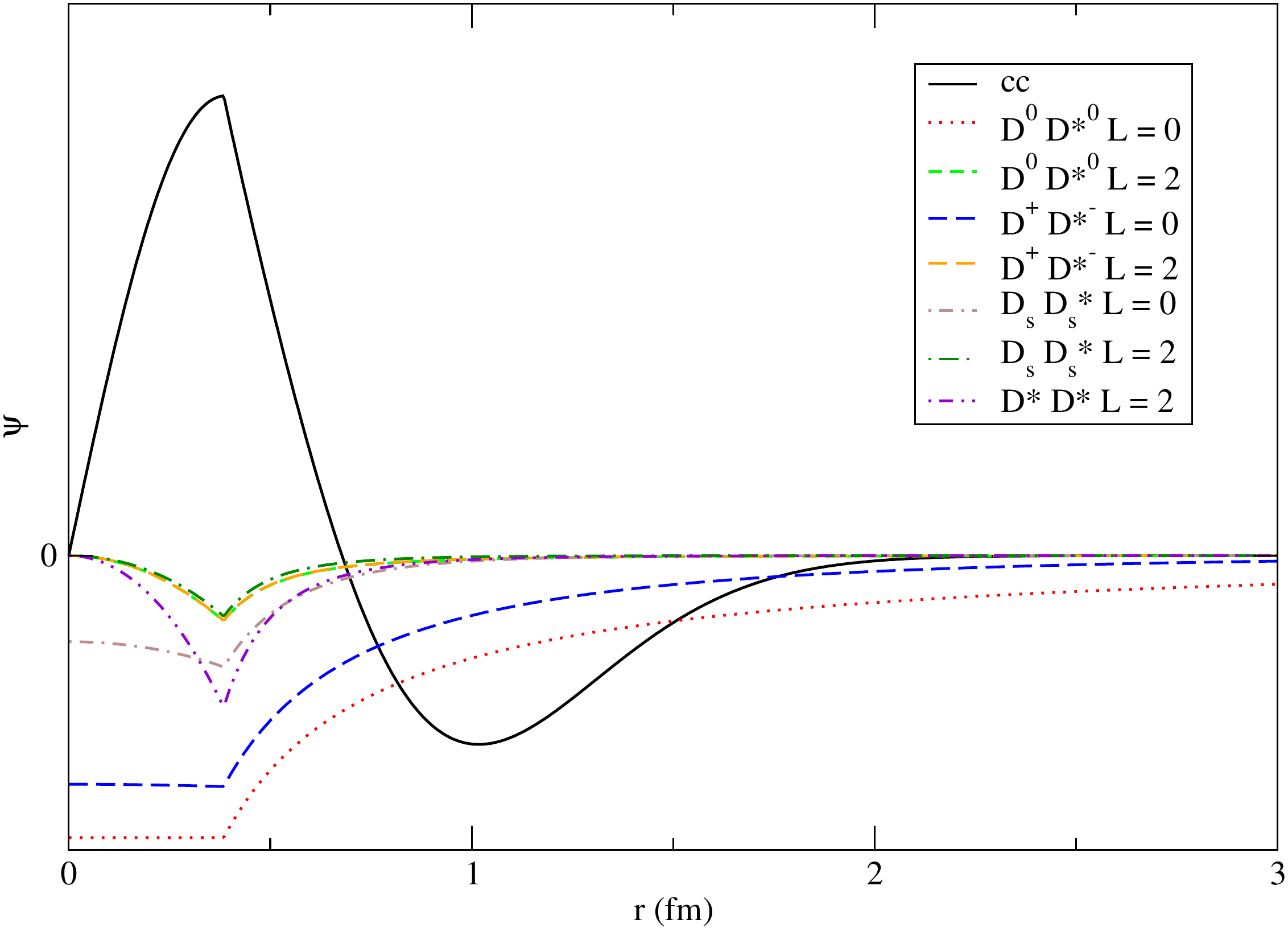}
\end{center}
\caption{Radial wave function of $\chi_{c1}(3872)$;
see text and Ref.~\cite{cardoso15}.}
\label{x3872wf}
\end{figure}

The parameters $\lambda$ and $a$ are fixed such that the masses of the three
charmonia are exactly reproduced. This can be done with the same $a$, but two
$\lambda$ values are needed, differing by about 15\% between the two vectors
and the axial-vector. This is perfectly reasonable, as the included meson-meson
channels have different orbital angular momenta (also see the discussion in
Ref.~\cite{cardoso15}). The resulting multicomponent radial wave function of
$\chi_{c1}(3872)$ is depicted in Fig.~\ref{x3872wf}. For clarity,
the meson-meson components are plotted with a negative sign. We see that now
the $c\bar{c}$ channel is clearly dominant in the interior region, followed
by the $S$-wave channels $\bar{D}^{\star0}D^0$ and $D^{\star\mp}D^\pm$. This
is logical, since their thresholds lie much closer to the $\chi_{c1}(3872)$
mass and so are less suppressed kinematically.

With the $\chi_{c1}(3872)$ wave function at hand, we can again compute the
r.m.s.\ radius of the system as well as the $c\bar{c}$ and
$\bar{D}^{\star0}D^0$ probabilities, which come out at 6.57~fm, 26.76\%, and
65.03\%, respectively. The at first sight surprisingly large $c\bar{c}$
percentage as compared to the value found in the simple model of
Subsec.~\ref{x3872r1} can be understood by realising that all channels 
contribute to the negative mass shift of the bare \ttpo\ $c\bar{c}$ state at
3979~MeV, and in particular the now also quite nearby $D^{\star\mp}D^\pm$
threshold. This reduces the relative importance of the $\bar{D}^{\star0}D^0$
component in the wave function, which is precisely the one with by far the
longest tail, and so its probability is significantly reduced. This benefits
primarily the $c\bar{c}$ component, which is most prominent in the interior
region. For further details, see Ref.~\cite{cardoso15}.

After also solving for the radial wave functions of $J/\psi$ and $\psi(2S)$,
the formalism developed in Ref.~\cite{verschuren91} allows to compute
\cite{cardoso15} the electromagnetic branching ratio
\begin{equation}
\mathcal{R}_{\gamma\psi}=\frac
{\Gamma\left(X(3872)\to\gamma\psi(2S)\right)}
{\Gamma\left(X(3872)\to\gamma J/\psi\right)} = 1.17 \; .
\label{x3872em}
\end{equation}
This value is too low as compared to the PDG \cite{PDG2020} average of
$2.6\pm 0.6$, but a huge improvement due to unitarisation as compared to a
static calculation with the same confinement mechanism. For further discussion
on possible theoretical and experimental improvements, see
Ref.~\cite{cardoso15}.

This concludes our extensive discussion of this fascinating meson. We are very
well aware that there are many other descriptions of
$\chi_{c1}(3872)$, as e.g.\ a dynamical pole in a coupled-channel model
\cite{danilkin10}, a purely molecular state from $t$-channel meson exchanges
\cite{lee09}, or using effective field theory \cite{wang13} (see, however,
Ref.~\cite{baru15}). For many more references, 
see Ref.~\cite{xiang16}. Nevertheless, we believe a unitarised quark-model
formulation is the most promising one in the context of general meson
spectroscopy. The lattice appears to confirm this (see Sec.~\ref{lattice}).

\section{Recent Lattice Results on Meson Resonances}
\label{lattice}
Lattice QCD is a non-perturbative approach to the non-Abelian gauge theory
of quarks and gluons, which uses Monte-Carlo techniques to numerically
simulate gauge configurations in discretised Euclidean space-time. Originally,
quarks were taken as static sources of the colour fields, in view of the
enormous numerical effort needed in treating also the quarks dynamically, i.e.,
via a fermion determinant. This approximative handling of the theory is
generally called quenched lattice QCD. However, present-day computer power
allows to go beyond the quenched approximation, doing realistic calculations
of mesons in 3+1 dimensions. Now, in unquenched lattice QCD, effects of
$q\bar{q}$ loops are fully taken into account, by including dynamical quark
degrees of freedom. Nevertheless, allowing for virtual $q\bar{q}$
pairs does not paint a complete picture, as the created quark and antiquark
may recombine with the original (anti)quarks so as to form two new
colourless mesons. Even if the mass of the initial meson is smaller
than the sum of the new mesons' masses, so that no real decay can take
place, the corresponding virtual processes via meson-meson
loops will contribute to the total mass. This is expected to be all the more
significant according as the decay-threshold energy lies closer to the
original meson's mass. On the other hand, if the latter mass is above
threshold, the meson actually becomes a resonance, whose properties are
determined by $S$-matrix analyticity and unitarity. 

In recent years, different lattice groups employing L\"{u}scher's method
\cite{luscher91} or extensions thereof (see Refs.~\cite{dudek18a,dudek18b} for
a list of references) have
managed to extract unitary scattering phase shifts and/or resonance properties
from unquenched finite-volume simulations that include meson-meson or
meson-baryon interpolating fields, besides the usual $q\bar{q}$ or $qqq$ ones,
respectively (see Ref.~\cite{dudek18a} for a recent review).
Some of these works on mesonic resonances \cite{PDG2020} show that
sizeable mass shifts may result from unitarisation, even when analytically
continued to underneath the lowest strong-decay threshold. On the other hand,
dynamical resonances, not present in the quenched meson spectrum, may show
up as well. Finally, there is even an indication that radial level spacings
can be affected considerably. In the following we shall briefly review a
typical selection of such recent lattice applications.

\subsection{\boldmath$\chi_{c1}(3872)$ as a $\bar{c}c+\bar{D}^\star D$ state}
In Ref.~\cite{lang15} a lattice study of charmonium-like mesons with
$J^{PC}=1^{++}$ was performed, considering three types of quark contents, viz.\
$\bar{c}c\bar{d}u$, $\bar{c}c(\bar{u}u+\bar{d}d)$, and $\bar{c}c\bar{s}s$,
where the latter two can mix with $\bar{c}c$. The corresponding simulation with
$N_f=2$ and $m_\pi=266$~MeV aimed at finding possible signatures of exotic
tetraquark states. A large basis of $\bar{c}c$, two-meson, and
diquark-antidiquark
interpolating fields was employed, with diquarks in both antitriplet and sextet
colour representations. Thus, a lattice candidate for $X(3872)$ (alias
$\chi_{c1}(3872)$) with $I=0$ is observed very close to the experimental state,
but only if both $\bar{c}c$ and $\bar{D}^\star D$ interpolators are included.
The candidate is not found if diquark-antidiquark and $\bar{D}^\star D$ are
used without a $\bar{c}c$ interpolating field. Furthermore, no candidate for a
neutral or charged $X(3872)$ or any other exotic candidates are found in the
$I=1$ channel. Also no signatures of exotic $\bar{c}c\bar{s}s$ candidates are
found below 4.2~GeV, such as $Y(4140)$ (alias $X(4140)$ or $\chi_{c1}(4140)$
\cite{PDG2020}).

Besides these very significant results concerning $\chi_{c1}(3872)$ in
particular, it is worthwhile to pay attention to the following quote from
Ref.~\cite{lang15}:
\begin{quote}  \em
``In the physical world with $N_c=3$, it is argued that tetraquarks could
exist at subleading orders \cite{weinberg13}
of large $N_c$ QCD. However, in the presence of the leading order two-meson
terms, one should take caution in interpreting the nature of the levels purely
based on their overlap factors onto various four-quark interpolators.'' \em
\end{quote}
The latter warning about not jumping to conclusions concerning evidence of
tetraquarks also applies to other approaches involving four-quark components
in the employed formalism. For a discussion focusing on the light scalar
mesons, see Ref.~\cite{rupp17}.

\subsection{\boldmath$D_{s0}^\star(2317)$ as a $c\bar{s}+DK$ state}
In the lattice calculation of Ref.~\cite{mohler13b}, the charmed-strange
scalar meson $D_{s0}^\star(2317)$ is found $37\pm17$~MeV below the $DK$
threshold, in a simulation of the $J^P=0^+$ channel using, for the first time,
both $DK$ and $\bar{s}c$ interpolating fields. The simulation is done on
$N_f=2+1$ gauge configurations with $m_\pi\simeq 156~$MeV, and the resulting
$M_{D_{s0}^\star}-\frac{1}{4}(M_{D_s}+3M_{D_s^\star})=(266\pm16)$~MeV is close
to the experimental value $(241.5\pm0.8$~MeV. The
energy level related to the scalar meson is accompanied by additional discrete
levels due to $DK$ scattering states. The levels near threshold lead to the
negative $DK$ scattering length $a_0=-(1.33\pm0.20)$~fm that indicates the
presence of a state below threshold. 

These results were published with more details in Ref.~\cite{lang14}, with
additionally a lattice confirmation of the $J^P=1^+$ charmed-strange meson
$D_{s1}(2460)$ as a unitarised $\bar{s}c$ state below the $D^\star K$
threshold, similarly to $D_{s0}^\star(2317)$ below the $DK$ threshold. (Also
see Ref.~\cite{mohler13a} for charmed-light mesons with $J^P=0^+$ or
$J^P=1^+$.) Very recently, another lattice collaboration \cite{wagner20}, by
including quark-antiquark, tetraquark, and two-meson interpolators in their
computation, obtained essentially the same results for $D_{s0}^\star(2317)$,
while even concluding:
\begin{quote} \em
``The coupling to the tetraquark interpolating fields is essentially zero,
  rendering a tetraquark interpretation of the $D_{s0}^\star(2317)$ meson rather
  unlikely.'' \em
\end{quote}

\subsection{Light scalar mesons as $q\bar{q}$ states with meson-meson components}
The light scalar mesons $f_0(500)$, $f_0(980)$, $K_0^\star(700)$, and
$a_0(980)$ were studied by the same lattice collaboration in
Refs.~\cite{dudek17,dudek18b,dudek15,dudek16}, respectively. In the $f_0(500)$
($\sigma$) case, two types of interpolating fields were included in the
simulation \cite{dudek17}, namely single-meson-like operators resembling
$q\bar{q}$ constructions of both $(u\bar{u}+d\bar{d})$ and $s\bar{s}$ flavours,
as well as operators resembling a pair of pions $\pi\pi$ with definite relative
and total
momentum, projected onto isospin $I=0$. The calculations were done for two
$u,d$ quark masses, corresponding to pion masses of 261 and 391~MeV,
respectively. The resulting amplitudes are described in terms of a $\sigma$
meson which evolves from a bound state below the $\pi\pi$ threshold at the
heavier quark mass to a broad resonance at the lighter quark mass. A precise
determination of the $\sigma$'s pole position is not possible yet, because
the employed parametrisations, while maintaining elastic unitarity, do not
necessarily respect the analytical constraints placed on them by causality
and crossing symmetry, apart from the need to extrapolate to the
physical pion mass. So for future simulations, adaptation dispersive
approaches and the use of smaller $u,d$ quark masses are planned
\cite{dudek17}.

In Ref.~\cite{dudek18b} the first lattice-QCD study of $S$-wave and $D$-wave
scattering in the coupled isoscalar $\pi\pi$, $K\bar{K}$, and $\eta\eta$
channels was carried out from discrete finite-volume spectra computed on
lattices for a light-quark mass corresponding to $m_\pi\sim391$~MeV. In the
$J^P=0^+$ sector analogues of the experimental $\sigma$ and $f_0(980)$ states
are found, where the $\sigma$ appears as a stable bound-state below the
$\pi\pi$ threshold and $f_0(980)$ manifests itself as a dip in the $\pi\pi$
cross section in the vicinity of the $K\bar{K}$ threshold, as also seen in 
experiment. For $J^P=2^+$ two states resembling $f_2(1270)$ and
$f_2^\prime(1525)$, observed as narrow peaks, with the lighter state dominantly
decaying to $\pi\pi$ and the heavier state to $K\bar{K}$, in agreement with
experiment \cite{PDG2020}. The presence of all these states is determined
rigorously by finding the pole singularity content of scattering amplitudes,
and their couplings to decay channels are established using the residues of
the poles. As an extension of L\"{u}scher's method \cite{luscher91}, an
approach which has proven successful proceeds by parametrising the
energy dependence of coupled-channel amplitudes and fitting a large set of
energy levels, from one or more lattice volumes, within a kinematic window.
A dense spectrum of energy levels will tightly constrain the possible
energy dependence of the scattering $t$-matrix, and to acquire as many energy
levels as possible, systems with various total momenta may be considered.
For further details and references, see Ref.~\cite{dudek18b}.

Coupled-channel $\pi K$ and $\eta K$ scattering amplitudes for $J^P=0^+$,
$1^-$, and $2^+$ were determined in Ref.~\cite{dudek15} by studying the
finite-volume energy spectra obtained from dynamical lattice-QCD calculations.
Using a large basis of interpolating operators, including both those resembling
a $q\bar{q}$  construction and those resembling a pair of mesons with definite
relative momentum, a reliable excited-state spectrum can be obtained. Working at
$m_\pi=391$~MeV, a gradual increase in the $J^P=0^+$ $\pi K$ phase shift is found,
which may be identified with a broad scalar resonance that couples strongly to
$\pi K$ and weakly to $\eta K$. The low-energy behaviour of this amplitude
suggests a virtual bound state that may be related to the $\kappa$
($K_0^\star(700)$) resonance. At higher energies, a broad resonance is found
that couples dominantly to $\pi K$ and not $\eta K$, with a pole mass
of $m=(1370\pm45)$~MeV and width of $\Gamma=(530\pm45)$~MeV. This should
represent the $K_0^\star(1430)$ \cite{PDG2020} resonance, although the here
found width is clearly too large. The (approximate) decoupling of the $\eta K$
channel is in agreement with what was observed in Ref.~\cite{beveren07}. For
a more definite determination of the $K_0^\star(700)$ resonance, the use of
considerably smaller pion masses will be necessary.

In Ref.~\cite{dudek16} the first lattice-QCD calculation of coupled-channel
meson-meson scattering in the $I^G=1^-$ sector was carried out, with the
channels $\pi\eta$, $K\bar{K}$, and $\pi\eta^\prime$, for a pion mass of 391~MeV
and a strange-quark mass approximately tuned to its physical value. The energy
dependence of the $S$-matrix is determined and a prominent cusp-like structure
in the $S$-wave $\pi\eta\to\pi\eta$ amplitude close to the $K\bar{K}$ threshold
is observed, coupled with a rapid turn-on of amplitudes leading to the
$K\bar{K}$ final state. This behaviour is traced to an $a_0(980)$-like
resonance, strongly coupled to both $\pi\eta$ and $K\bar{K}$, which is
identified with a pole in the complex energy plane, appearing on only a single
unphysical Riemann sheet. Explicit lattice calculations at a smaller quark
mass will be needed in order to confirm an $a_0(980)$ resonance in agreement
with experiment.

\subsection{\boldmath$K^\star(1410)$ and the issue of radial splittings}
\begin{figure}[!t]
\begin{tabular}{cc}
\includegraphics[trim = 0mm 7mm 0mm 5mm,clip,width=8.75cm,angle=0]
{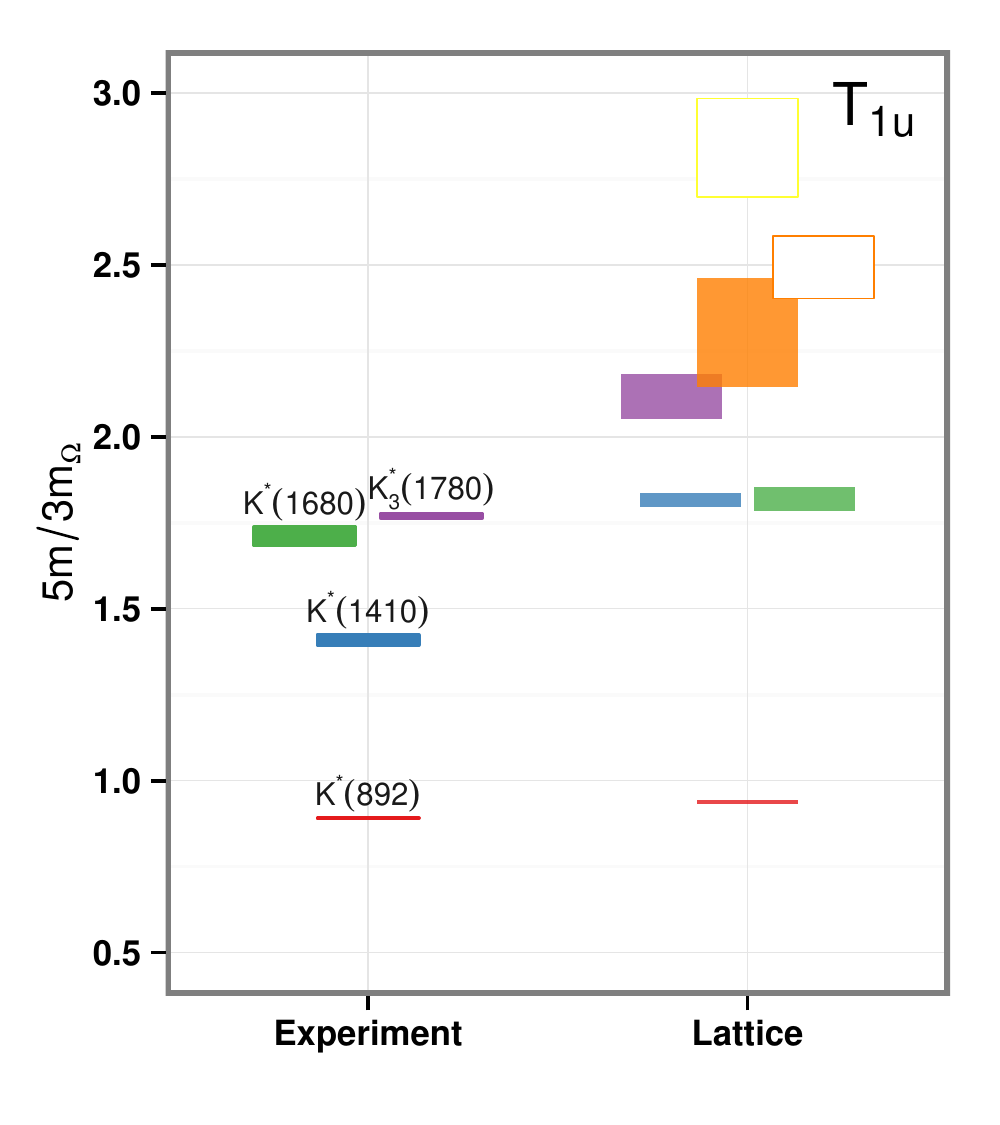}
&
\includegraphics[trim = 0mm 7mm 0mm 5mm,clip,width=8.75cm,angle=0]
{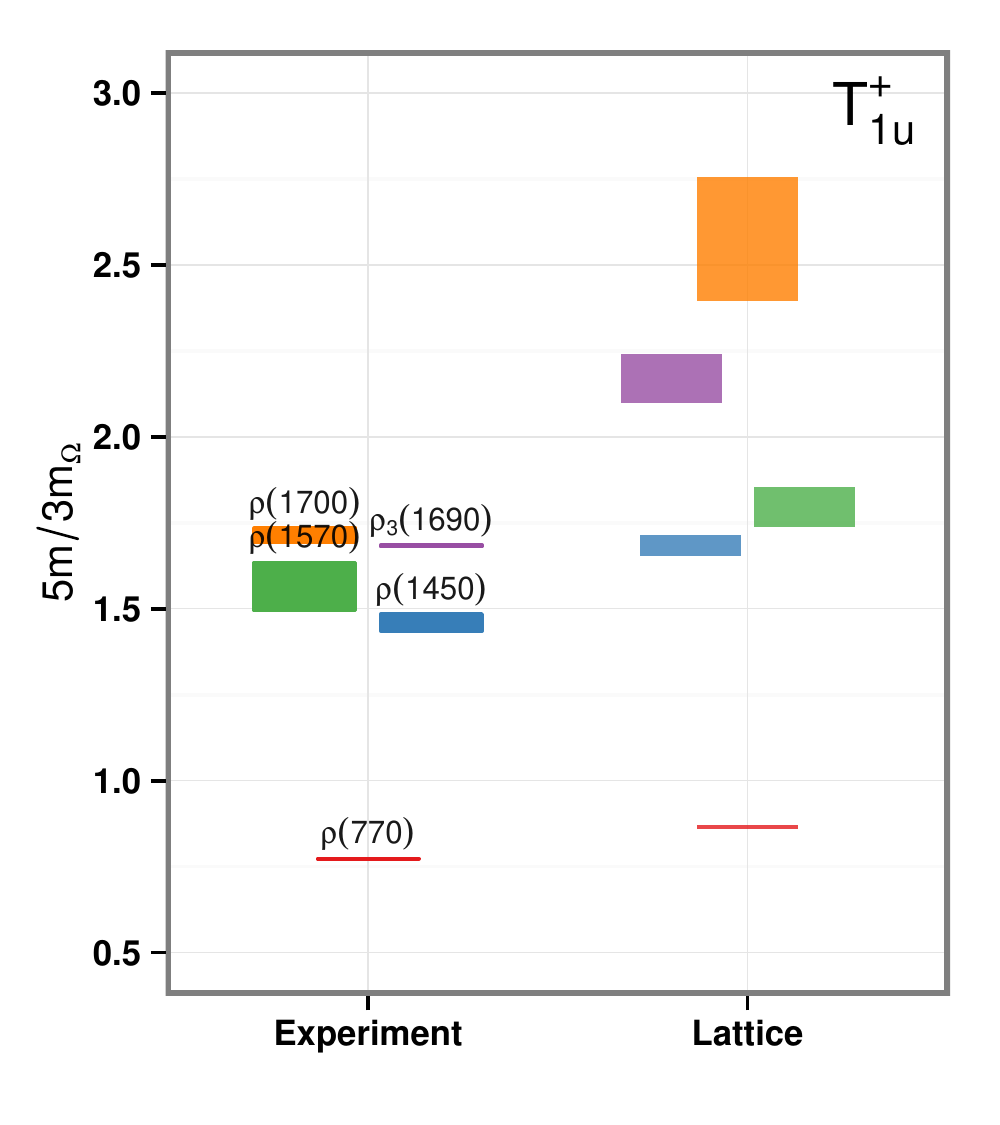}
\end{tabular}
\caption{Left: excited strange-meson masses, as ratios with respect to 3/5
of the $\Omega$ baryon mass $m_\Omega$, for stationary states expected to
evolve into the single-meson resonances in infinite volume. The height of
each box indicates its statistical uncertainty. The hollow boxes at the
top show higher-lying states extracted with less certainty due to the
expected presence of lower-lying two-meson states that have not been taken
into account. Right: the analogous plot for the isovector channel, the
superscript in $T^+_{1u}$ indicating $G$-parity. See Ref.~\cite{morningstar14}
for further details.}
\label{morningstar}
\end{figure}

As already shown above in our discussion of the GI model \cite{GI85}, the
$K^\star(1410)$ first radial excitation of lowest strange vector meson
$K^\star(892)$ is one of the many examples of a considerably smaller radial
mass splitting than predicted in the GI and similar static quark models.
This makes any possible lattice prediction of the $K^\star(1410)$ all
the more interesting. Now, in Ref.~\cite{morningstar14} excited states
in unquenched lattice QCD were computed, though without meson-meson
interpolators, with the resulting bound-state spectra for strange
and isovector mesons shown in Fig.~\ref{morningstar}. In both cases,
the radial splitting between the ground-state vector meson and its first
radial excitation is much too large, even with the proviso that the used pion
mass is around 390 MeV and that the states have been extracted in a finite
volume. Therefore, it is most opportune to check similar predictions by 
a different lattice group.
\begin{figure}[!hb]
\begin{tabular}{cc}
\includegraphics[trim = 0mm 0mm 0mm 0mm,clip,width=8.75cm,angle=0]
{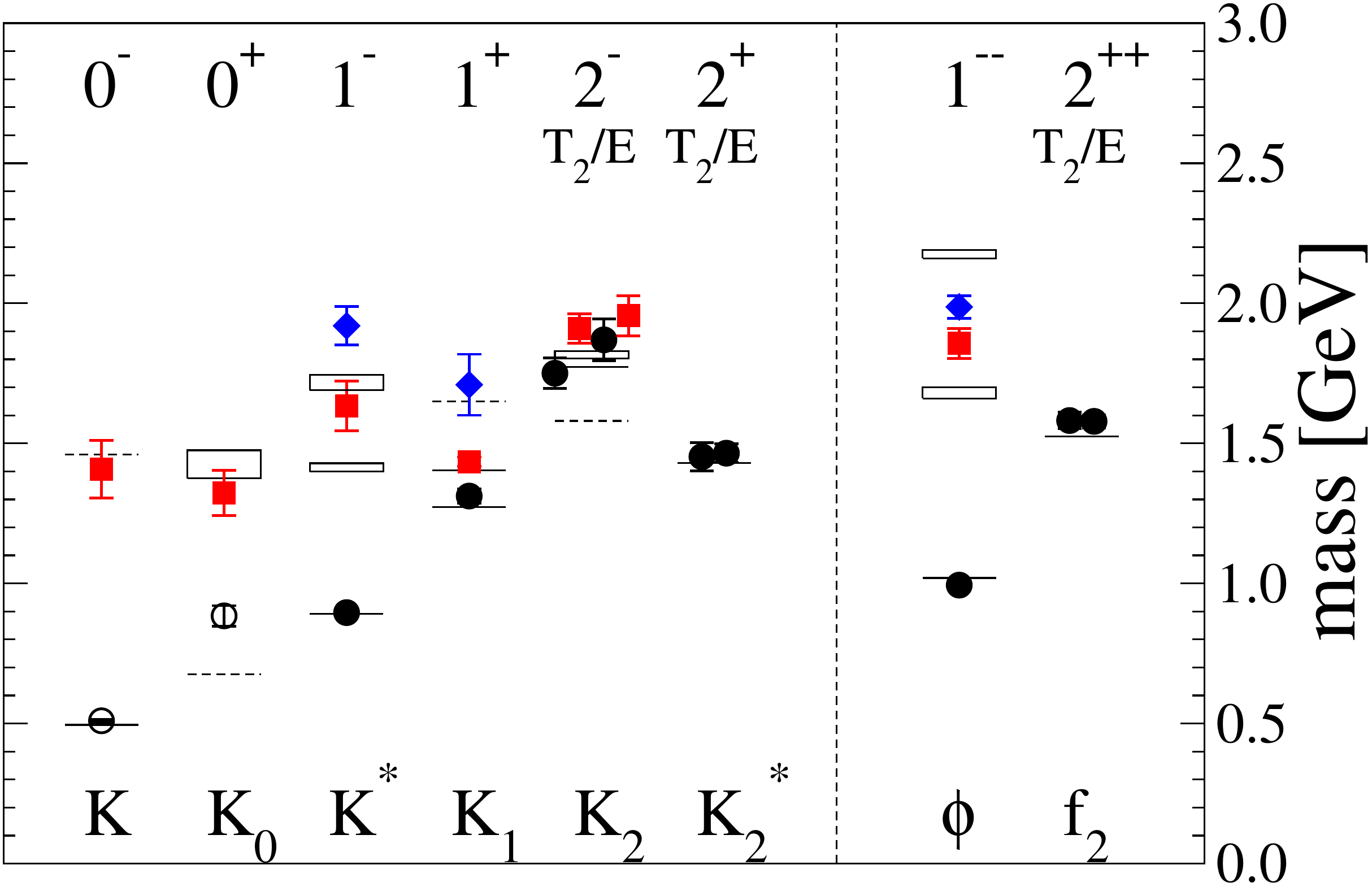}
&
\parbox[t]{8.75cm}{\mbox{ } \\[-5.6cm]
\includegraphics[trim = 0mm 0mm 0mm 0mm,clip,width=8.75cm,angle=0]
{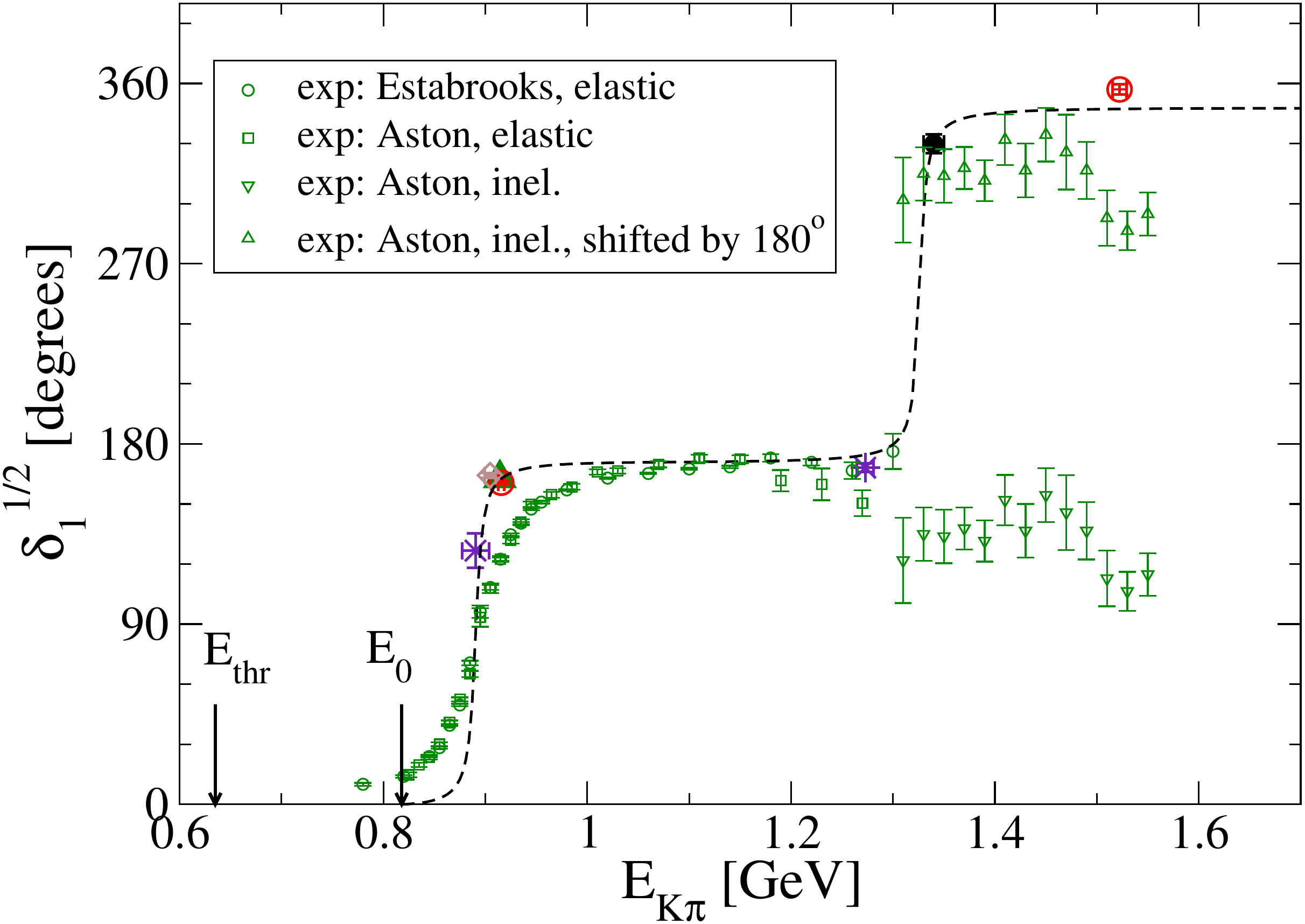}
}
\end{tabular}
\caption{Left: radial excitation spectrum of several mesons including
$K^\star(892)$ from the lattice calculation in Refs.~\cite{lang12,lang13c}.
Right: $P$-wave $K\pi$ phase shifts from the lattice calculation in
Refs.~\cite{lang13a,lang13b}.
Figures reprinted from Refs.~\cite{lang13c} (left) and \cite{lang13b} (right)
by kind permission of the authors.}
\label{kstar1410}
\end{figure}

In the unquenched lattice calculation of Refs.~\cite{lang12,lang13c}, yet again
without including two-meson interpolators, results were presented for radial
excitation spectra of light and strange mesons as well as positive- and
negative-parity baryons. On the left-hand plot of Fig.~\ref{kstar1410} we
display the
corresponding radial excitations of, among other strange mesons, $K^\star(892)$
and also two isoscalar mesons. We again see a too large first radial
splitting in the isodoublet vector case, i.e., roughly 200~MeV more than in
experiment, though not as huge as in Ref.~\cite{morningstar14}. The big
surprise, though, we see on the right-hand plot of Fig.~\ref{kstar1410},
taken from Refs.~\cite{lang13a,lang13b}, in which $P$-wave $K\pi$ phase shifts
were calculated by members of the same lattice group as in
Refs.~\cite{lang12,lang13c}. This more realistic simulation includes $K\pi$
two-meson interpolators, which allows to extract phase shifts using
L\"{u}scher's method \cite{luscher91}. As expected, the phase passes through
90$^\circ$ more or less at the mass of the $K^\star(892)$ resonance. However,
there is a second phase-shift jump of 180$^\circ$ right above 1.3~GeV,
corresponding to the first radially excited $K^\star$ resonance, with an
extracted mass of $(1.33\pm0.20)$~GeV, which is about 300~MeV lower than in
the lattice calculation --- without considering decay --- of
Refs.~\cite{lang12,lang13c}. Admittedly, the simulation in
Refs.~\cite{lang13a,lang13b} is an approximation in the $K^{\star\prime}$ case,
because it ignores the important \cite{PDG2020} $K^\star(892)\pi$ decay mode
and also $K\rho$. However, it is hardly conceivable that the inclusion of these
channels will drastically change the $K^{\star\prime}$ mass, and certainly not
shift it to much higher energies. So the discrepancy of roughly 300~MeV between
the mass of an excited meson calculated with or without two-meson interpolators
in a lattice simulation poses an enormous challenge to lattice practitioners as
well as quark-model builders.

\subsection{Conclusions on lattice results for some puzzling mesons}
\label{lattice-conclusions}
\indent
Summarising, the above lattice results in
Refs.~\cite{dudek18b,dudek17,dudek15,dudek16,wagner13} largely confirm our
earlier interpretation and modelling of the light scalar mesons as dynamical
$q\bar{q}$ resonances in Refs.~\cite{beveren86,beveren01,beveren06c}. Moreover,
the lattice descriptions of $D_{s0}^\star(2317)$ in
Refs.~\cite{mohler13b,lang14,wagner20} support our former unitarised model
for this state tuned to an $S$-wave $K\pi$ phase-shift fit. Similarly,
the lattice results for the charmed axial-vector mesons in
Refs.~\cite{lang14,mohler13a} are in agreement with our prior modelling of
these resonances in Refs.\cite{beveren04a,coito11b}. Furthermore, the
lattice approach to $\chi_{c1}(3872)$ with quark-antiquark, tetraquark, and
meson-meson interpolators in Ref.~\cite{lang15} corroborates our
preceding description of this meson as a non-exotic yet strongly unitarised
\ttpo\ $c\bar{c}$ state in Refs.~\cite{coito11a,coito13,cardoso15}. Finally,
the lattice calculation of $P$-wave $K\pi$ phase shifts in Ref.~\cite{lang13a}
lends indirect support to the $\rho(1250)$ resonance resulting from the
analysis in Ref.~\cite{hammoud20}, as a consequence of the low-lying
$K^{\star\prime}$ found on the lattice.

To conclude this discussion, there has been spectacular progress over the past
decade in describing physical meson resonances through QCD simulations on the
lattice, which promise to become very helpful in the future to further reduce
model bias in determining resonance pole positions from experiment. Even
certain data analyses might need to be reassessed in view of clear conflicts
with solid lattice predictions.
\section{Mesonic Enhancements and Production Processes}
\label{production}
\begin{figure}[!h]
\begin{center}
\includegraphics[trim = 0mm 0mm 0mm 0mm,clip,width=15cm,angle=0]
{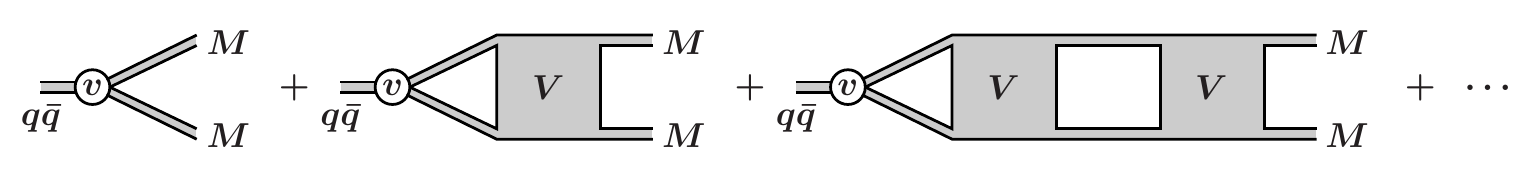}
\end{center}
\caption{Schematic production process followed by rescattering; see text and
Ref.~\cite{beveren08a}.}
\label{inhomogeneous}
\end{figure}

So far we have only considered a unitary $S$-matrix for meson-meson scattering
in order to search for poles describing non-exotic meson resonances. However,
because direct meson-meson scattering is not feasible experimentally, such data
long ago used to be extracted from meson scattering off nucleons. For example,
the still often cited and even used LASS \cite{lass88} phase shifts for
$K^-\pi^+$ scattering were obtained from the reaction $K^-p\to K^-\pi^+n$.
However, nowadays mesonic resonances are mostly observed in $e^+e^-$ collisions
and decays of the resulting vector states, or in multiparticle decays of
open-bottom and open-charm mesons. One may argue that, from a theoretical point
of view, there is no fundamental difference between the various mechanisms to
produce meson resonances, as their pole positions are generally accepted to be
universal and so independent of the process. However, resonance line shapes may
very well depend on the production mechanism, which makes it crucial
to have an adequate formalism at hand for a reliable analysis of the
experimental data. But still more importantly, the common feature of production
processes is an initial state with only one quark-antiquark pair
and not a two-meson system, as shown in Fig.~\ref{inhomogeneous}
\cite{beveren08a}.
Here, the first diagram shows an initial $q\bar{q}$ pair (resulting from
$e^+e^-$ annihilation or heavy-meson decay) producing two mesons e.g.\ via
\tpz\ pair creation at the vertex $v$ followed by an OZI-allowed decay. These
two mesons can and will then rescatter through iterations of the effective
potential $V$, and so the two-body $T$-matrix. Now there are two essential
remarks to be made. In the first place, the amplitude for the production
process depicted in Fig.~\ref{inhomogeneous} will have the same resonance poles
as $T$, provided the first diagram is not singular. Secondly, if one assumes
that the effective potential $V$ involves $q\bar{q}$ pair annihilation
followed by a new creation, then the first term is of leading order in the
strength of the creation/annihilation process, as it only amounts to one
$q\bar{q}$ creation.

In Ref.~\cite{beveren08a} we derived the formalism of production processes in
the context of the RSE model. The resulting relation between the production
amplitude for an initial quark-antiquark state labelled $\alpha$ to a final
meson-meson state labelled $i$ reads
\begin{equation}
P_i^{(\alpha)} \; \propto \; \lambda\sum_{L,M}\,(-i)^{L}\,
Y^{(L)}_{M}(p_{i})\,
Q^{(\alpha )}_{\ell_{q\bar{q}}}(E)
\left\{ g_{\alpha i}\, j_{L}(p_{i}a)
-i\sum_{\nu}\,\mu_{\nu}\,p_{\nu}\,h^{(1)}_{L}(p_{\nu}a)\,
g_{\alpha\nu}\,T^{(L)}_{i\nu}
\right\} \; ,
\label{prodscat}
\end{equation}
where $T^{(\ell)}_{i\nu}$ is a partial-wave $T$-matrix
element for transitions between the two-meson channels $i$ and $\nu$. 
Furthermore, $j_L(p_ia)$ and $h^{(1)}_L(p_ia)$ are the usual spherical
Bessel and Hankel-1 functions, and the $g_{\alpha i}$ are the coupling
constants for transitions between the initial $q\bar{q}$ pair, with all its
quantum numbers symbolically abbreviated as $\alpha$, and the $i$-th 
meson-meson channel, following again the recoupling scheme of
Ref.~\cite{beveren84}. The largely unknown function
$Q^{(\alpha)}_{\ell_{q\bar{q}}}(E)$ describes the initial quark-antiquark
state $\alpha$ with energy E, which must be determined by the precise
production mechanism. Note that indeed $P$ and $T$ share
the same resonance poles, as the spherical Bessel function in the lead term
of Eq.~(\ref{prodscat}) is smooth. Furthermore, the production amplitude
$P$ manifestly satisfies \cite{beveren08a,beveren08b}
the so-called extended-unitarity relation
\begin{equation}
\Imag{P}\; =\; T^{\star}\, P \; ,
\label{extended}
\end{equation}
despite the fact that the coefficients in front of the $T$-matrix
elements in Eq.~(\ref{prodscat}) are complex \cite{beveren08b,beveren08c}
due to the spherical Hankel function $h^{(1)}_L$. In Ref.~\cite{pennington08}
it was argued that these coefficients should be real, but we showed
\cite{beveren08c} this to imply that the coefficients themselves must contain
$T$-matrix elements, whereas our definition in Eq.~(\ref{prodscat})
involves purely kinematical coefficients, proportional to $h^{(1)}_L$. Besides
this clear advantage, Eq.~(\ref{prodscat}) explicitly displays the mentioned
lead term in the production amplitude, which is linearly dependent on the
coupling $g_{\alpha\nu}$ between the initial $q\bar{q}$ state and meson-meson
channel $\nu$, and has the shape of a spherical Bessel function. This term will
generally result in an enhancement in the cross section right above a threshold
opening, which may be mistaken for a genuine resonance or distort the signal
when indeed a true resonance pole is nearby. In Ref.~\cite{beveren08a}
we also showed the connection between our production formalism and the
Breit-Wigner approximation as well as the $K$-matrix approach.

Notwithstanding these nice properties of the formalism, one should realise
that actually reproducing experimental production data is much more difficult
than finding resonance poles or fitting scattering phase shifts with the RSE
$T$-matrix. Namely, in order to fit production cross sections, one would need
accurate knowledge of the largely unknown function
$Q^{(\alpha)}_{\ell_{q\bar{q}}}(E)$ in the lead term of Eq.~\ref{prodscat}, as
well as of the generally multichannel meson-meson $T$-matrix. Therefore, we
have limited ourselves to more empirical applications in a number of concrete
cases involving puzzling resonances. Three typical examples will be dealt with
next, viz.\ the enigmatic vector charmonium states $\psi(4260)$ and
$\psi(4660)$, as well as the usually uncontroversial \cite{PDG2020} bottomonium
resonance $\Upsilon(10580)$. For a more general discussion of the former two
candidates for exotic charmonia, see the review in Ref.~\cite{xiang16}.
We also mention here Ref.~\cite{beveren09c} for our production description of
several threshold and resonance structures in $e^+e^-$ annihilation at the
mass scales of strangeonium, charmonium, and bottomonium.

\subsection{Non-resonant vector charmonium state \boldmath$\psi(4260)$}
The $\psi(4260)$ vector charmonium state, previously called $X(4260)$,
has 51 listed \cite{PDG2020} decay modes, of which 39 are \em ``not seen'', \em
one is \em ``possibly seen'', \em and five are just mentioned. None of the 
\em ``(possibly) seen'' \em \/modes correspond to decays to open-charm mesons.
It lies outside the scope of the present paper to mention all the published
explanations of this most peculiar feature of $\psi(4260)$, usually in terms
of (crypto-)exotic non-$c\bar{c}$ configurations, so we just refer to the
dedicated review of exotic candidates in Ref.~\cite{xiang16}. Instead, we 
\begin{figure}[!b]
\includegraphics[trim = 48mm 63mm 8mm 65mm,clip,width=16cm,angle=0]
{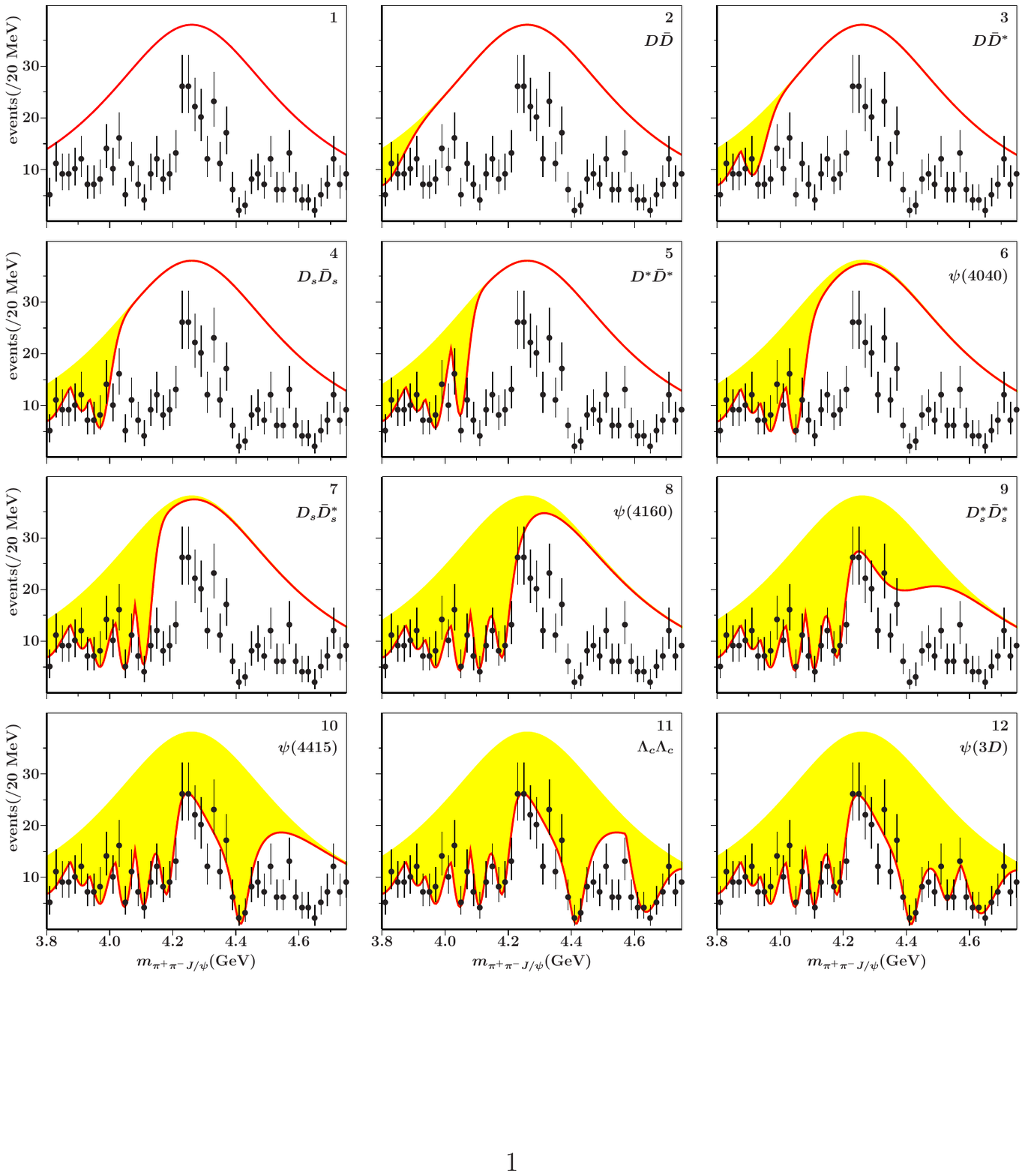}
\caption{
Reconstruction of $\psi(4260)$ data in
$e^{+}e^{-}\to\pi^{+}\pi^{-}J/\psi$ \cite{babar05} via a stepwise depletion
of a presumed broad threshold structure by OZI-allowed inelastic processes
(also see text).
From upper left to lower right: broad structure (1), depletion by respectively
$D\bar{D}$ (2), $D\bar{D}^{\ast}$ (3), $D_{s}\bar{D}_{s}$ (4),
$D^{\ast}\bar{D}^{\ast}$ (5), $\psi (4040)$ (6), $D_{s}\bar{D}_{s}^{\ast}$ (7),
$\psi (4160)$ (8), $D_{s}^{\ast}\bar{D}_{s}^{\ast}$ (9), $\psi (4415)$ (10),
$\Lambda_{c}\bar{\Lambda}_{c}$ (11), and $\psi (3D)$ (12) (also see text
and Ref.~\cite{beveren10b}.
}
\label{octopsi}
\end{figure}
briefly summarise here our non-resonant interpretation of $\psi(4260)$, as
published in Ref.~\cite{beveren10b}. Assuming a very broad threshold
structure in the $J/\psi\pi^+\pi^-$ channel dominated by $J/\psi f_0(500)$,
we interpret the various dips in the $J/\psi\pi^+\pi^-$ event distribution
as strong inelasticity effects from the opening of OZI-allowed channels with
pairs of charmed mesons as well as established vector charmonia in these
channels. This also helps to explain the very pronounced and puzzling dip
precisely at the mass of $\psi(4415)$. Thereabove, the opening of the
$\Lambda_c\bar{\Lambda}_c$ threshold and a tentative, so far unlisted
$\psi(3D)$ resonance are fundamental to understand the data. These large
inelasticity effects we called \cite{beveren10b} \em depletion, \em which we
believe to give rise to a non-resonant, apparent $\psi(4260)$ enhancement.

In Fig.~\ref{octopsi} we show how the data can be explained stepwise with the
mentioned broad structure, thresholds of open-charm meson pairs, known
charmonium resonances, the $\Lambda_c\bar{\Lambda}_c$ baryon-antibaryon
threshold, and finally a proposed, so far unlisted, $\psi(3D)$ resonance.
Note that already in Ref.~\cite{beveren09b} we had found indirect indications
of a $\psi(3D)$ resonance not far underneath the $\Lambda_c\bar{\Lambda}_c$
threshold. The here extracted $\psi(3D)$ mass and width are 4.53~GeV and
80~MeV, respectively. In the GI model, this vector charmonium state was
predicted \cite{GI85} at 4.52~GeV.

To conclude our discussion on $\psi(4260)$, let us just mention a very recent
calculation \cite{coito20} from a unitarised effective Lagrangian, which also
supports the interpretation of this state as a non-resonant enhancement, due
to the proximity of the $D_s^\star\bar{D}_s^\star$ threshold and the
$\psi(4160)$ pole.

\subsection{Vector charmonium \boldmath$\psi(4660)$ as a 
\boldmath$\Lambda_c\bar{\Lambda}_c$ threshold enhancement}
The vector charmonium $\psi(4660)$, previously called $X(4660)$ and also
known as $Y(4660)$, is listed \cite{PDG2020} with an average mass and width of
$(4633\pm7$)~MeV and $(64\pm9)$~MeV, respectively. These small errors as given
by the PDG are surprising in view of the published ranges of mass and
width values of 4626--4669~MeV and 42--104~MeV, respectively. Also puzzling
is that the only \em ``seen'' \em \/open-charm decay mode of $\psi(4660)$ is 
$D_s^+D_{s1}(2536)^-$.

\begin{figure}[!t]
\begin{tabular}{cc}
\includegraphics[trim = 0mm 0mm 0mm 0mm,clip,width=8.75cm,angle=0]
{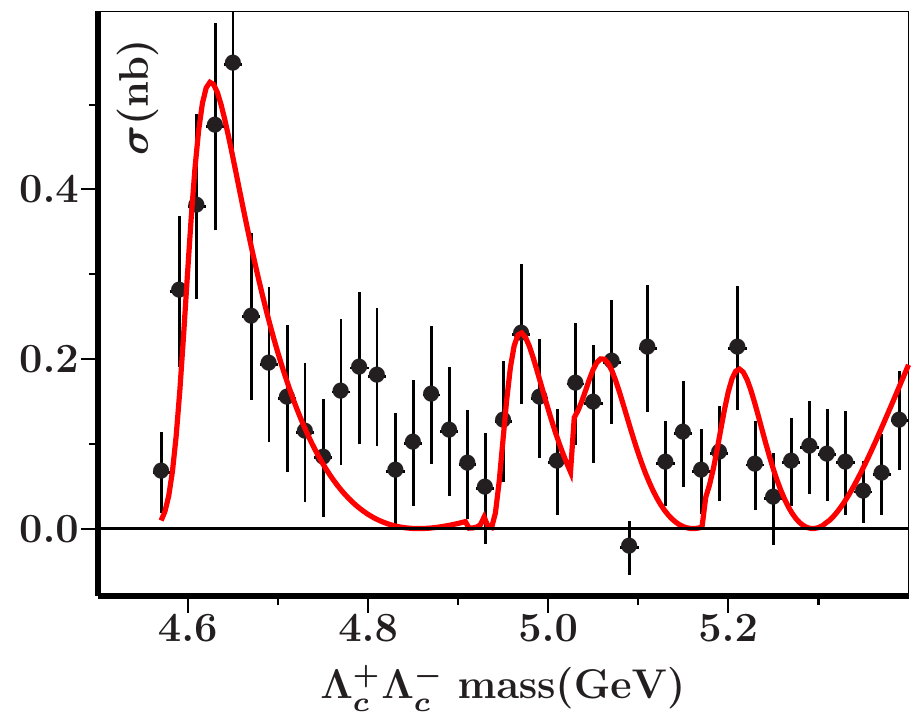}
&
\includegraphics[trim = 0mm 0mm 0mm 0mm,clip,width=8.75cm,angle=0]
{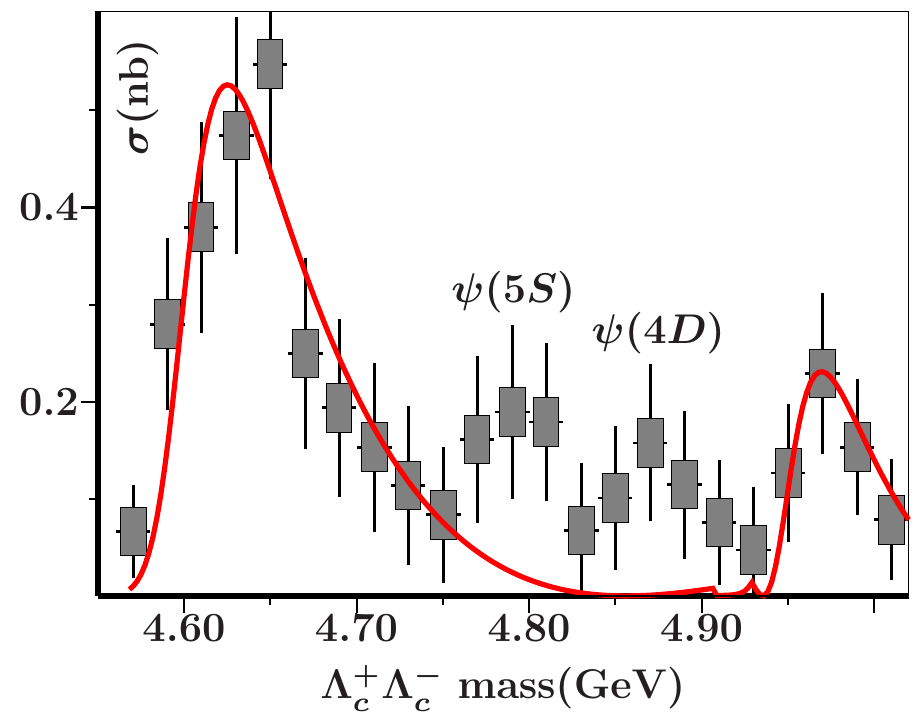}
\end{tabular}
\caption{Left: experimental data \cite{belle08} in the energy region
4.57--5.40~GeV. Right: same data with the proposed $\psi(5S)$ and
$\psi(4D)$ resonances. Solid curves on both plots are from the fit
in Ref.~\cite{beveren09b}.}
\label{psi5s4d}
\end{figure}

In Ref.~\cite{beveren09b} we analysed data on the
$e^+e^-\to\Lambda_c\bar{\Lambda}_c$ cross section published \cite{belle08} by
the Belle Collaboration, employing a phenomenological ansatz based on our
production formula in Eq.~(\ref{prodscat}). From our fit we concluded that the
reported structure ``$X(4630)$''at 4634~MeV is a non-resonant threshold
enhancement due to the opening of the $\Lambda_c\bar{\Lambda}_c$ channel (also
see Ref.~\cite{beveren10b}). Moreover, we found indications of the
higher vector charmonia $\psi(5S)$, $\psi(4D)$, $\psi(6S)$, and
$\psi(5D)$, as well as an indirect indication of $\psi(3D)$, later supported by
the analysis in Ref.~\cite{beveren10b}. In Fig.~\ref{psi5s4d} we show the
experimental data \cite{belle08} in the energy region 4.57--5.40~GeV
(left-hand plot), and the data with the suggested
$\psi(5S)$ and $\psi(4D)$ resonances\footnote
{This way of displaying the data and the proposed two new $\psi$ states only
appears in the arXiv version 0809.1151v3 [hep-ph] of Ref.~\cite{beveren09b}.}
in the interval 4.57--5.10~GeV (right-hand plot). The solid curves on both
plots represent our fit based on the opening of the $\Lambda_c\bar{\Lambda}_c$
and three charmed-$\Sigma$-baryon thresholds. For details, see
Ref.~\cite{beveren09b}.
\subsection{\boldmath$\Upsilon(10580)$ as a non-resonant enhancement
\label{Ups10580}
between the $B\bar{B}$ and $B^\star \bar{B}$ thresholds}
Finally, we discuss a bottomonium state that is generally believed to be very
well established, namely $\Upsilon(10580)$, which has been included in the PDG
tables for many years and is listed \cite{PDG2020} as $\Upsilon(4S)$. However,
we interpreted \cite{beveren09d,beveren10d,rupp19} it rather as a non-resonant
bump right between the $BB$ and $BB^\star$ thresholds. The crucial point is
that the data \cite{babar09a} show a small yet clear enhancement on top of the
$B_s\bar{B_s}$ threshold, whereas there are unmistakable dips at the openings
of the $B\bar{B^\star}$ and $B^\star\bar{B^\star}$ thresholds (see 
Fig.~16). This pattern can be understood by assuming an
\begin{figure}[t]
\begin{center}
\includegraphics[trim = 0mm 0mm 0mm 0mm,clip,width=18cm]
{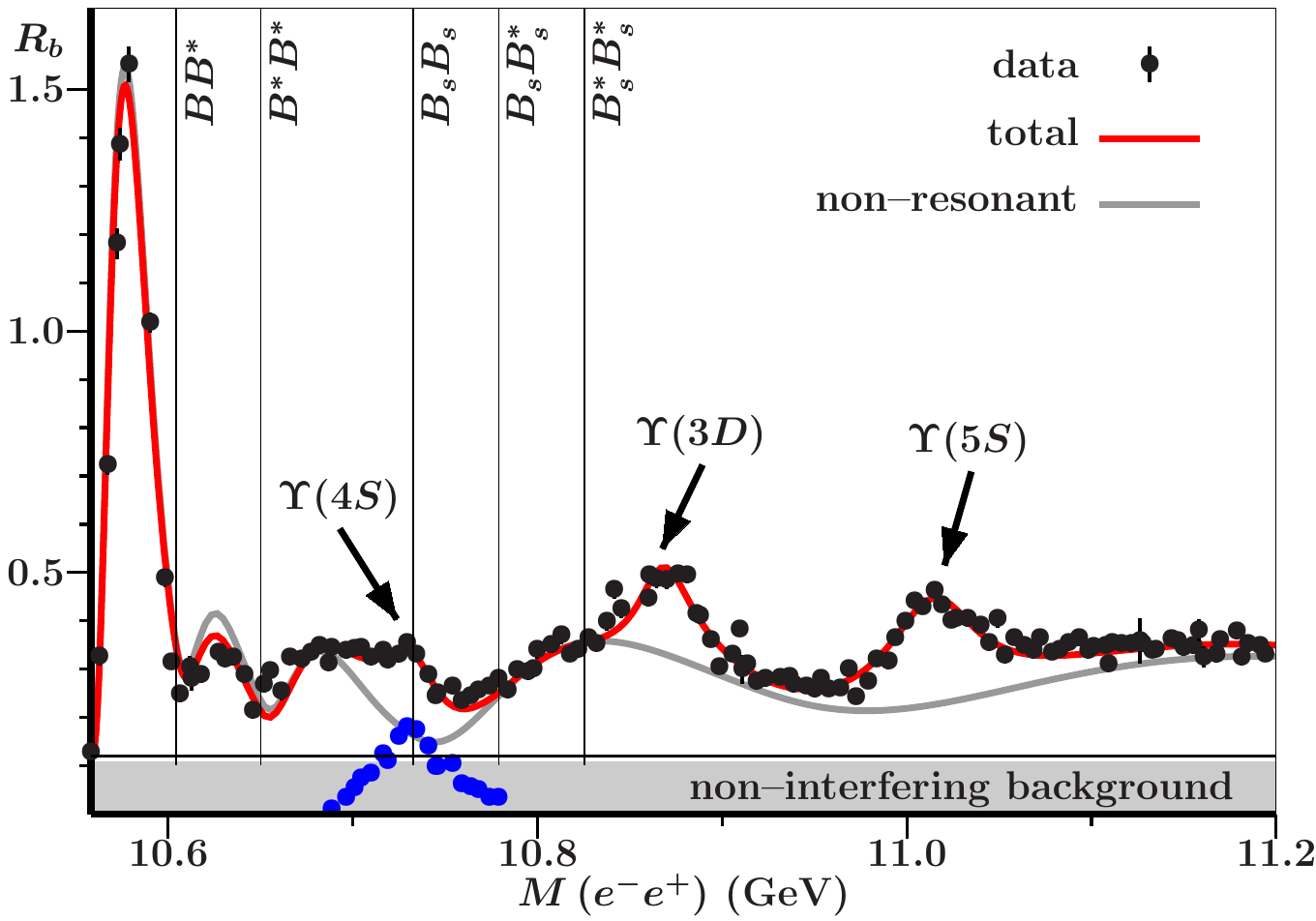}
\caption{Fit \cite{beveren09d} to bottomonium vector states, with 
open-bottom thresholds marked and locations of $\Upsilon(4S)$,
$\Upsilon(3D)$, and $\Upsilon(5S)$ in the data \cite{babar09a}.
The blue dots indicate the difference between the data and the
non-resonant contribution disregarding interference.
For further details, see text and Refs.~\cite{beveren09d,beveren10d}.}
\end{center}
\label{upsilon10580}
\end{figure}
$\Upsilon(4S)$ resonance somewhat above the $B_s\bar{B_s}$ threshold, which
also allows to identify $\Upsilon(10860)$ as $\Upsilon(3D)$ and
$\Upsilon(11020)$ as $\Upsilon(5S)$. From our phenomenological fit to the data,
based on threshold enhancements as following from our production formalism
 described above, we extracted \cite{beveren09d,beveren10d} an $\Upsilon$
resonance with a mass of $10.735$~GeV and a width of 38~MeV, which we interpret
as the true $\Upsilon(4S)$ state. Most interestingly, the Belle Collaboration
very recently observed \cite{belle19,PDG2020} a new vector bottomonium
state, with a mass of 10.753~GeV and a width of 35.5~MeV.

As for our non-resonant interpretation of $\Upsilon(10580)$, also very
recently $\Upsilon$ resonances above the open-bottom threshold were studied
\cite{oset20} in a simple effective model based on the \tpz\ mechanism. In
this paper, the vector bottomonium resonances listed in the PDG tables are
described via a propagator dressed with loops of $B$, $B^\star$, $B_s$,
and $B_s^\star$ mesons. From this dressed propagator, a
wave-function-renormalisation constant $Z$ is obtained, which should be
close to one for a state that is mostly $b\bar{b}$. This condition is
strongly fulfilled by all $\Upsilon$ resonances except for $\Upsilon(10580)$,
which
shows a large deviation. Moreover, the high peak and relatively large width of
$\Upsilon(10580)$ is argued \cite{oset20} to be incompatible with a vector
$b\bar{b}$ resonance decaying only to $B\bar{B}$ and with little phase space.
We believe that these results lend support to considering $\Upsilon(10580)$
a non-resonant $B\bar{B}$ threshold enhancement, amplified \cite{beveren10d}
by the pole of a so far unlisted (yet also see Ref.~\cite{beveren10c})
$\Upsilon(2D)$ state not far below the $B\bar{B}$ threshold.

\section{Summary and Conclusions}
\label{conclusions}
In this review we aimed at making the case for a systematic treatment of
meson spectroscopy based on the quark model for $q\bar{q}$ states only, yet
imposing the requirements of $S$-matrix unitarity. Thus, in
Sec.~\ref{intro} we started with a brief introduction to mainstream
quark models of mesons using a Coulomb-plus-linear confining potential, 
and mentioned the inevitable problem with radial spacings in the spectra of
especially mesons made of light and strange quarks. In Sec.~\ref{unitarity}
we employed a simple unitary single-channel model for the $S$-matrix in order
to show discrepancies that may arise when using standard Breit-Wigner
parametrisations, in particular when applied to very broad resonances not far
above the lowest threshold, like in the case of the light scalar mesons
$f_0(500)$ and $K_0^\star(700)$. Section~\ref{static} was devoted to a detailed
discussion of static quark models, in which the dynamical effects of strong
decay or virtual meson loops on the spectra are ignored. The shortcomings of
the relativised meson model of Godfrey and Isgur \cite{GI85} were illustrated
with many examples from particularly light-meson spectra. Furthermore, two
fully relativistic static quark models were reviewed as well and shown to have
similar or even worse problems. In Sec.~\ref{coupled} we briefly reviewed the
very disparate predictions for meson mass shifts, some of them really huge,
due to unitarisation or coupled channels in a series of old and more recent
models, discussing their differences. Section~\ref{rse} treated a simple
unitarised model in momentum space, called Resonance Spectrum Expansion (RSE)
and inspired by the unitarised quark-meson model in coordinate space developed
by the Nijmegen group. Its predictive power was demonstrated by successfully
describing e.g.\ the $K_0^\star(700)$ resonance, the charmed scalar meson
$D_{s0}^\star(2317)$, the charmed $J^P=1^+$ mesons, and --- in a multichannel
extension of the model --- even the whole light scalar-meson nonet.
Furthermore, the most general RSE model, applicable to systems with various
quark-antiquark channels coupled to an arbitrary number of meson-meson 
channels, was shown to be exactly solvable, both algebraically and
analytically, owing to the separability of the effective meson-meson
interaction and the employed string-breaking mechanism. In
Sec.~\ref{enigmatic} the latter general RSE model was used to analyse
again the charmed $J^P=1^+$ mesons, thus allowing to dynamically produce the 
physical states as orthogonal mixtures of the \tpo\ and \spo\ quark-antiquark
components. This gives rise to two quasi-bound states in the continuum and two
strongly shifted states, thus reproducing the observed disparate pattern of
masses and widths with remarkable accuracy. Moreover, the same full RSE model
as well as its multichannel coordinate space version were employed to describe
the axial-vector charmonium state $\chi_{c1}(3872)$, modelling it as a
unitarised \ttpo\ charmonium state. The resulting pole trajectories, wave
function, and electromagnetic transitions support our
interpretation of this very enigmatic meson. Section~\ref{lattice} was
devoted to several recent lattice calculations of controversial meson
resonances that include meson-meson interpolating fields in the simulations,
in order to allow for the computation of phase shifts and extract the
corresponding resonance or bound-state pole positions. The results for 
$\chi_{c1}(3872)$, $D_{s0}^\star(2317)$, the charmed $J^P=1^+$ mesons, and
the light scalars largely confirm our description of these mesons. In
Sec.~\ref{production} we presented a formalism for production processes
that is strongly related to our RSE model and satisfies the
extended-unitary condition $\Imag{P}=T^{\star}\,P$, with $T$ the RSE
$T$-matrix. The general expression features a purely
kinematical, non-resonant real lead term, plus a combination of two-body
$T$-matrix elements with also kinematical yet complex coefficients.
Fully empirical applications of the formalism to the controversial vector
charmonium states $\psi(4260)$ and $\psi(4660)$ as well as the established
$\Upsilon(10580)$ bottomonium state allowed to fit all three resonance-like
structures as non-resonant threshold enhancements.

Let us repeat that we did not aspire to carry out a comprehensive review of
general meson spectroscopy. Therefore, several alternative descriptions of
mesons with very interesting results, like e.g.\ unitarised chiral models
\cite{oller00,oset10}, the generalised Nambu--Jona-Lasinio model
\cite{osipov07}, or the quark-level linear $\sigma$ model \cite{scadron13},
have not been dealt with here at all. Nevertheless, we believe that these
approaches have a more restricted applicability to meson spectroscopy, being
usually limited to specific resonances or ground states only. In order to be
able to infer information on the confining potential, it is necessary to be
able to calculate radially excited states without introducing new parameters.
We have also not paid attention to truly exotic meson candidates, as e.g.\
the charmed charmonium-like and bottomonium-like states $Z_c^\pm(3900)$,
$Z_c^\pm(4430)$, $Z_b(10610)$, and $Z_b(10650)$ \cite{PDG2020}. For a
discussion of such exotics, we refer again to Ref.~\cite{xiang16}.
Nevertheless, in this context it is worthwhile to mention a very recent paper
by the COMPASS Collaboration \cite{compass20}, in which for the first time
a triangle-singularity model is fitted directly to partial-wave data, viz.\
for the controversial $a_1(1420)$ state reported \cite{compass15} by COMPASS
itself five years ago. The conclusion of this fit is that including
the triangle singularity allows for a better fit to the data with fewer
parameters, so that after all there is no need to introduce the new $a_1(1420)$
resonance \cite{compass20}. This result may have far-reaching consequences for
exotic spectroscopy, in view of the increasing number of observed enhancements
in the data that cannot be accommodated as $q\bar{q}$ mesons. Clearly, all such
controversial states will have to be refitted in a similar fashion. For a
detailed discussion of triangle singularities, see Ref.~\cite{xiang16}.

To conclude, we recall the mentioned email exchange with a co-spokesperson
\cite{appel01} of the E791 Collaboration about the need for easy formulae to
fit the data and a related discussion at the LHCb workshop \em `Multibody
decays of $D$ and $B$ mesons'', \em in Rio de Janeiro, 2015 \cite{lhcb16}. The
latter meeting focused on alternatives to the usual Breit-Wigner (BW) and
Flatt\'{e} parametrisations that guarantee multichannel unitarity, even in the
case of overlapping broad resonances. In that spirit, we proposed (see
Ref.~\cite{lhcb16}, pages 36--39) our general RSE formalism
(cf.\ Eqs.~(\ref{rsev}--\ref{Smatrix})), yet with the bare HO energies 
replaced by a few to-be-fitted real energies and possibly also the Bessel and
Hankel-1 functions by more flexible expressions, thus allowing much more
accurate fits to the data. Apart from thus guaranteeing manifest multichannel
unitarity, the usual two BW parameters for each resonance could then be
replaced by only one real energy. Finally, a similar generalisation of
our production formalism should also be possible.

\section{Acknowledgement}
We thank R.~Kami\'{n}ski for demonstrating the importance of a fully unitary
$S$-matrix description instead of the usual Breit-Wigner parametrisation even
in the simplest case of elastic resonances, which was one of our motivations
to write the present review.

\end{document}